\documentclass[a4paper,11pt]{article}

\usepackage{amssymb,amsmath,epsfig,graphicx,psfrag}

\numberwithin{equation}{section}

\newcommand{\per}{\,}
\newcommand{\pp}[2]{p_{#1}\hspace{-0.ex} p_{#2}}
\newcommand{\pq}[2]{p_{#1}\hspace{-0.ex} q_{#2}}
\newcommand{\qq}[2]{q_{#1}\hspace{-0.ex} q_{#2}}
\newcommand{\st}{\bigg(} 
\newcommand{\dt}{\bigg)} 
\newcommand{\sq}{\bigg[} 
\newcommand{\dq}{\bigg]} 
\newcommand{\sg}{\bigg\{} 
\newcommand{\dg}{\bigg\}} 
\newcommand{\pst}{\Big(} 
\newcommand{\pdt}{\Big)} 
\newcommand{\psq}{\Big[} 
\newcommand{\pdq}{\Big]} 
\newcommand{\psg}{\Big\{} 
\newcommand{\pdg}{\Big\}} 

\newcommand{\as}{\alpha_{\mathrm{S}}}
\newcommand{\bra}[1]{\langle #1 |}

\newcommand{\Caption}[1]{\caption{\it #1}}
\newcommand{\e}{\varepsilon}
\newcommand{\J}{\boldsymbol{J}}             
\newcommand{\ket}[1]{| #1 \rangle}

\newcommand{\M}{{\cal M}}               

\newcommand{\pol}{\varepsilon}

\newcommand{\qu}[1]{{Q}_{#1}}   
\newcommand{\cS}{{\cal S}}               
\newcommand{\sy}[2]{\bigl(#1\, #2\bigr)_{sym}}
\newcommand{\sym}[3]{\bigl(#1\, #2\, #3\bigr)_{sym}}
\newcommand{\symm}[4]{\bigl(#1\, #2\, #3\, #4\bigr)_{sym}}
\newcommand{\T}{\boldsymbol{T}}                   
\newcommand{\ui}{\mathrm{i}}             

\def\nn{\nonumber}
\def\beq{\begin{equation}}
\def\eeq{\end{equation}}
\def\beeq{\begin{eqnarray}}
\def\eeeq{\end{eqnarray}}
\newcommand\g{g_{\mathrm{S}}}
\newcommand\eqcs{\raisebox{-2mm}{\rlap{\,{\rm cs}\,}} \raisebox{.0ex}{$\,=\,$}}
\newcommand\G{\boldsymbol{\Gamma}}
\newcommand{\oS}{{\overline {\cal S}}} 
\newcommand{\w}{w}

\title{{\bf Triple (and quadruple) soft-gluon radiation\\ in QCD hard scattering}}

\author{
   {\bf Stefano Catani, Dimitri Colferai} and {\bf Alessandro Torrini}\\[1ex]
   {\sl INFN, Sezione di Firenze and Dipartimento di Fisica e Astronomia,}\\
   {\sl Universit\`a di Firenze, I-50019 Sesto Fiorentino, Florence, Italy}\\[10mm]
}

\date{}

\begin{document}

\maketitle

\begin{abstract}
We consider the radiation of three soft gluons in a generic process for multiparton hard scattering in QCD.
In the soft limit the corresponding scattering amplitude has a singular behaviour that is factorized and controlled by a colorful soft current.
We compute the tree-level current for triple soft-gluon emission from both massless
and massive hard partons.
The three-gluon current is expressed in terms of maximally non-abelian irreducible correlations. 
We compute the soft behaviour of squared amplitudes and the colour correlations produced by the squared current. 
The radiation of one and two soft gluons leads to colour dipole correlations. 
Triple soft-gluon radiation produces in addition colour quadrupole correlations
between the hard partons.
We examine the soft and collinear singularities of the squared current in various
energy ordered and angular ordered regions.
We discuss some features of soft radiation to all-loop orders for processes with two and three hard partons. 
Considering triple soft-gluon radiation from three hard partons, colour quadrupole interactions break the Casimir scaling symmetry between
quarks and gluons. 
We also present some results on the radiation of four soft gluons from two hard partons,
and we discuss the colour monster contribution and its relation with the violation
(and generalization) of Casimir scaling.
We also compute the first correction of ${\cal O}(1/N_c^2)$ to the eikonal formula for multiple soft-gluon radiation with strong energy ordering from two hard gluons.

\end{abstract}

\vskip 1cm

\vspace*{\fill}
\begin{flushleft}
     August 2019 

\end{flushleft}
\newpage

\tableofcontents

\setcounter{footnote}{2}
\newpage
\section{Introduction\label{s:intro}}

The first two operational runs of proton--proton collisions at the LHC have produced a large amount of high-precision data on hard-scattering final states. Similar data
are expected in the next phases of the LHC. The high-precision LHC data demand for a corresponding accuracy in theoretical predictions. Such theoretical accuracy is required both to test our present understanding of the Standard Model and to discover
and investigate (probably tiny) signals of new physics phenomena.

In the context of QCD, one way to increase the theoretical accuracy consists in performing calculations at higher perturbative orders in the QCD coupling $\as$.
The LHC physics program has moved the present frontier of perturbative calculations to the next-to-next-to-next-to-leading order (N$^3$LO). During the last few years, much effort has been devoted to high-order perturbative computations and much progress
has been already achieved at the N$^3$LO frontier. We limit ourselves to explicitly mentioning few examples among very many others.  
The total (partonic) cross section for Higgs boson production in hadron--hadron collisions is known up to N$^3$LO 
\cite{Anastasiou:2014vaa, Anastasiou:2015ema, Mistlberger:2018etf}.
Substantial advances have been achieved toward the complete N$^3$LO calculation 
\cite{Moch:2017uml, Moch:2018wjh} of the evolution kernels of the parton distribution functions. The structure of the infrared (IR) divergences of multileg QCD scattering amplitudes has been explicitly computed at the three-loop level
\cite{Almelid:2015jia}.

A relevant feature of QCD scattering amplitudes is the presence of singularities in soft and collinear regions of the phase space, and the corresponding presence of IR divergences in virtual radiative corrections at the loop level. The theoretical study of these aspects of the scattering amplitudes is relevant `per se' in QCD and, more generally, in perturbative gauge field theories. We know that soft/collinear singularities and IR divergences
have a process-independent structure, and they are controlled by universal factorization formulae.

In the computation of physical observables, phase space and loop singularities 
cancel between themselves, but much technical effort is required to achieve the cancellations, and the effort highly increases at higher perturbative orders. 
The explicit knowledge of soft/collinear factorization of scattering amplitudes at
${\cal O}(\as)$ has been essential to devise fully general (observable-independent 
and process-independent) methods to carry out next-to-leading order (NLO) QCD calculations (see, e.g., Refs.~\cite{Frixione:1995ms, csdip}).
Analogously, the knowledge of soft/collinear factorization at 
${\cal O}(\as^2)$
\cite{Bern:1999ry}--\cite{Becher:2009qa}
is exploited to develop methods
(see, e.g., a list of references in Sect.~1.2.4 of Ref.~\cite{Bendavid:2018nar})
at the next-to-next-to-leading order (NNLO). Soft/collinear factorization formulae at 
${\cal O}(\as^3)$ can be used in the context of N$^3$LO calculations.

The perturbative radiative corrections to hard-scattering observables in kinematical regions close to the exclusive boundary of the phase space are affected by large 
logarithmic contributions. These large contributions have to be computed at sufficiently-high perturbative orders and, possibly, resummed to all orders
(see, e.g., the list of references in Refs.~\cite{Luisoni:2015xha, Becher:2014oda}).
The large logarithms arise from the unbalance between loop and real radiative corrections in the soft and collinear regions of the phase space.
The explicit knowledge of soft/collinear factorization at ${\cal O}(\as^3)$ 
gives information that is necessary in resummed calculations at the 
next-to-next-to-next-to-leading logarithmic accuracy. Independently of resummation,
the explicit calculation of large logarithmic terms can be used to obtain
approximated fixed-order results. Indeed, we note that the first approximated 
N$^3$LO result for Higgs boson production \cite{Anastasiou:2014vaa}
was obtained by computing soft (and virtual) contributions at relative order
$\as^3$. We also recall that soft/collinear factorization of scattering amplitudes provides the theoretical basis of parton shower algorithms for Monte Carlo event generators (see, e.g., Ref.~\cite{Buckley:2011ms}) for high-energy particle collisions.

This paper is devoted to soft emission in QCD scattering amplitudes at 
${\cal O}(\as^3)$. More specifically, we consider triple soft-gluon radiation at the tree level in QCD hard scattering. Partial results on triple soft-gluon radiation
are available in the literature 
\cite{Bassetto:1984ik}--\cite{Delenda:2015tbo}
and we comment on them throughout the paper. These results are limited to specific processes and to energy ordering approximations.

In this paper we use soft factorization of QCD scattering amplitudes and we explicitly compute the singular factor due to radiation of three soft gluons.
The three soft gluons have `arbitrary' energies (i.e., there are no restrictions on their relative energies) and they are radiated in a generic multiparton 
hard-scattering process with massless and massive partons. We discuss several features of triple soft-gluon radiation at the level of both scattering amplitudes and squared amplitudes. 
Single and double soft-gluon radiation leads to colour dipole correlations between
the hard partons in the squared amplitude.
We find that triple soft-gluon radiation also produces non-abelian colour quadrupole
correlations between the hard partons in generic processes with three or more hard partons.
We apply in details our results to processes with two and three hard partons,
and we also highlight some all-order features of soft-gluon radiation in these processes. 

Exploiting our analysis of triple soft-gluon radiation, we also present some results on the tree-level radiation of four soft gluons in scattering processes with two hard partons. In particular, we discuss the violation (and generalization) of Casimir scaling between quark and gluon hard scattering at ${\cal O}(\as^4)$.

The outline of the paper is as follows.
In Sect.~\ref{s:sc} we introduce our notation, and we recall the soft-gluon factorization formula for scattering amplitudes and the known results
on the tree-level currents for single and double soft-gluon emission.
In Sect.~\ref{s:sc3} we derive the tree-level current for triple soft-gluon radiation, and we present its expression in colour space in terms of
irreducible correlations that are maximally non-abelian.
We also compute the corresponding colour stripped current \cite{BeGi89}
for colour-ordered multigluon amplitudes.
In Sect.~\ref{s:square} we consider squared amplitudes, and we recall the known results for the one-gluon and two-gluon squared currents.
The tree-level squared current for triple soft-gluon radiation is presented
in Sect.~\ref{s:sq123}. We express the result in terms of irreducible correlations that are controlled by colour dipole and colour quadrupole interactions.
We derive some simplified expressions that are valid within energy ordering approximations. In Sect.~\ref{s:coll} we discuss the collinear singularities of the three-gluon squared current.
Soft-gluon radiation in the specific processes with three and two hard partons is considered in Sect.~\ref{sec:3hard} and \ref{sec:2hard}, respectively.
In Sect.~\ref{sec:3hardgeneral} we consider soft-gluon emission at arbitrary loop orders and we discuss the simplified colour structure for processes with three hard partons. Section~\ref{sec:3hardtree} is devoted to soft radiation at the tree level,
and we present the explicit results of the squared currents for emission of one, two and three soft gluons. We also derive a results that is valid for emission of an arbitrary number of soft gluons with strong ordering in energy.
In Sect.~\ref{sec:2hardgeneral} we discuss some all-order features of soft-gluon radiation from two hard partons. The squared currents for tree-level emission of one, two and three soft gluons are presented in Sect.~\ref{sec:3from2}.
Section~\ref {sec:quadruple} is devoted to present some results on tree-level radiation of four soft gluons from two hard partons. We obtain the general colour structure of the squared current, and we explicitly compute the four-gluon irreducible correlation by using some energy ordering approximations.
We discuss the colour monster contribution \cite{Dokshitzer:1991wu}, which is suppressed by a relative factor of ${\cal O}(1/N_c^2)$ in the limit of a large number $N_c$ of colours, and its relation with the violation (and generalization)
of quadratic Casimir scaling at ${\cal O}(\as^4)$. We also explicitly compute the first correction to the multi-eikonal formula \cite{Bassetto:1984ik}
for multiple soft-gluon radiation from two hard gluons.
In Sect.~\ref{s:exp} we present the exponentiation structure of the generating functional for tree-level soft-gluon radiation in generic hard-scattering processes.
A brief summary of our results is presented in Sect.~\ref{sec:sum}.
In Appendix~\ref{a:quad} we discuss the properties of the colour quadrupole
operators and their colour algebra.
The Appendix~\ref{a:ee} includes the large explicit expressions of the dipole and quadrupole correlation functions for triple soft-gluon radiation.

\section{Soft factorization and soft-gluon currents\label{s:sc}}

In this section we introduce our notation. Then we briefly recall the factorization properties of scattering amplitudes in the soft limit and the known results for
the emission of one and two soft gluons at the tree level.
The new results on the soft current for triple gluon emission at the tree level are presented and discussed in Sect.~\ref{s:sc3}.

\subsection{Soft factorization of scattering amplitudes\label{s:sc12}}

We consider the amplitude $\M$ of a generic 
scattering process whose external particles (the external legs of $\M$)
are QCD partons (quarks, antiquarks and gluons) and, possibly, additional 
non-QCD particles (i.e. partons with no colour
such as leptons, photons, electroweak vector bosons, Higgs bosons and so forth).
As is well known, the non-QCD particles in $\M$ play no relevant active role in the context of soft-gluon factorization. Throughout the paper we make no distinctions between massless and massive quarks and antiquarks, and we treat them on equal footing. Soft-gluon factorization can also be extended in a straightforward way to scattering amplitudes with other types of particles that carry QCD colour charge
(such as, for instance, squarks and gluinos in supersymmetric theories).

The external QCD
partons are
on-shell with physical spin polarizations (thus, $\M$
includes the corresponding spin wave functions). All external particles 
of $\M$ are treated as `outgoing' particles, with corresponding outgoing momenta and quantum numbers (e.g., colour and spin). 
Note, however, that we do no restrict our treatment to scattering processes with physical partons in the final state. 
In particular, the time component $p^0$ (i.e.,
the `energy' $E=p^0$) of the outgoing momentum $p^\mu$ ($\mu=0, 1, \dots, d-1$
in $d$ space-time dimensions) of an external particle is {\em not} (necessarily)
positive definite. 
Different types of physical processes with initial- and final-state particles
are described
by considering different kinematical regions of the parton momenta and
by simply applying crossing symmetry relations to the spin wave functions and quantum numbers of the same scattering amplitude $\M$.
According to our definition, an outgoing particle $A$ in $\M$ describes two different physical processes: the production of particle $A$ in the final state if its momentum has positive `energy', and the collision of the {\em antiparticle} ${\overline A}$
in the initial state if the momentum has negative `energy'.

The scattering amplitude $\M$ can be evaluated perturbatively as a power series
expansion (i.e., loop expansion) in the QCD coupling $\g$, with $\as=\g^2/(4\pi)$
(and other couplings, such as electroweak couplings, of the theory). The loop expansion produces IR and ultraviolet (UV) divergences in the physical 
four-dimensional Minkowsky space.
We regularize the divergences by applying dimensional regularization through analytic continuation in $d=4-2\epsilon$ space-time dimensions.
Throughout the paper we always use (unless otherwise explicitly stated) 
the customary
procedure of conventional dimensional regularization (CDR) \cite{cdr}
with $d-2$ physical spin polarization states for on-shell gluons.

Soft-gluon radiation produces colour correlations. 
To take into account the colour structure we use the colour (+ spin) space formalism of Ref.~\cite{csdip}. The scattering amplitude 
$\M_{\sigma_1 \sigma_2 \dots}^{c_1 c_2\dots}$ 
depends on the colour ($c_i$) and spin ($\sigma_i$) indices of its external-leg partons. This dependence is embodied in a vector $\ket{\M}$ in colour+spin space through the definition (notation)
\begin{equation}\label{Mstate}
  \M_{\sigma_1 \sigma_2 \dots}^{c_1 c_2 \dots} \equiv
  \big(\bra{c_1,c_2,\cdots}\otimes\bra{\sigma_1,\sigma_2,\cdots}\big) \;
  \ket{\M} \;\;,
\end{equation}
where $\{ \,\ket{c_1,c_2,\cdots}\otimes\ket{\sigma_1,\sigma_2,\cdots} \} =
\{ \, \ket{c_1,\sigma_1;c_2,\sigma_2,\cdots} \}$ is an orthonormal basis of abstract vectors in colour+spin space.

In colour space the colour correlations produced by soft-gluon emission are represented by associating a colour charge operator $\T_i$ to the emission of a gluon from each parton $i$. If the emitted gluon has colour index $a$ ($a=1,\dots,N_c^2-1$,
for $SU(N_c)$ QCD with $N_c$ colours) in the adjoint representation, the colour charge operator is $\T_i \equiv \bra{a} \,T_i^a$
and its action onto the colour space is defined by
\begin{equation}\label{defT}
  \bra{a,c_1,\cdots,c_i,\cdots,c_m}\,\T_i\,\ket{b_1,\cdots,b_i,\cdots,b_m} \equiv
  \delta_{c_1 b_1} \cdots (T^a)_{c_i b_i} \cdots \delta_{c_m b_m} \;,
\end{equation}
where the explicit form of the colour matrices $T^a_{c_i b_i}$ depends on the
colour representation of the parton $i$, and we have
\begin{align}
 (T^a)_{b c} &= \ui f^{bac} && \text{(adjoint representation) 
if $i$ is a gluon,} &&\nonumber \\
 (T^a)_{\alpha\beta} &=  t^a_{\alpha\beta} && \text{(fundamental representation with $\alpha,\beta=1,\dots,N_c$) if $i$ is a quark,}  &&\nonumber \\
 (T^a)_{\alpha\beta} &= -t^a_{\beta\alpha}   && \text{if $i$ is an antiquark.} &&\nonumber &&\label{Tcs}
\end{align}
We also use the notation $T_i^a T_k^a \equiv \T_i \cdot \T_k$ and $\T_i^2 = C_i$,
where $C_i$ is the quadratic Casimir coefficient of the colour representation, with the normalization $C_i=C_A=N_c$ if $i$ is a gluon and $C_i=C_F=(N_c^2-1)/(2N_c)$
if $i$ is a quark or antiquark.

Note that each `amplitude vector' $\ket{\M}$ is an overall colour-singlet state.
Therefore, colour conservation is simply expressed by the relation
\beq
\label{colcons}
\sum_i \;\T_i \;\ket{\M} = 0 \;\;,
\eeq
where the sum extends over all the external-leg partons $i$ of the amplitude $\M$. 
For subsequent use, we also introduce the shorthand notation
\beq
\label{csnotation}
\sum_i \;\T_i \;\eqcs \;0 \;\;,
\eeq
where the subscript CS in the symbol $\eqcs$ means that the equality between the terms in the left-hand and right-hand sides of the equation is valid if these (colour operator) terms act (either on the left or on the right) onto colour-singlet states.

We are interested in the behaviour of the scattering amplitude $\M$
in the kinematical configuration where 
{\em one} or {\em more} of the  momenta of the external-leg gluons become soft
(formally vanish). To make the notation
more explicit, the soft-gluon momenta are denoted by $q_\ell^\mu$
($\ell=1,\dots,N$, and $N$ is the total number of soft gluons), while the momenta
of the hard partons in $\M$ are denoted by $p_i^\mu$.
In this kinematical configuration, $\M(\{q_\ell\}, \{p_i\})$ becomes singular. The dominant singular behaviour is given by the following soft-gluon factorization formula in colour space:
\begin{equation}\label{1gfact}
  \ket{\M(\{q_\ell\}, \{p_i\})} \simeq
  (\g \,\mu_0^\epsilon)^N \J(q_1,\cdots,q_N) \; \ket{\M (\{p_i\})} \;,
\end{equation}
where $\mu_0$ is the dimensional regularization scale.
Here $\M (\{p_i\})$ is the scattering amplitude that is obtained from the original
amplitude $\M(\{q_\ell\}, \{p_i\})$ by simply removing the soft-gluon external legs.
The factor $\J$ is the soft current for multigluon radiation from the scattering amplitude.

At the formal level the soft behaviour of $\M(\{q_\ell\}, \{p_i\})$ is specified
by performing an overall rescaling of all soft-gluon momenta as 
$q_\ell \to \xi q_l$ (the rescaling parameter $\xi$ is the same for each soft
momentum $q_\ell$) and by considering the limit $\xi\to 0$.
In this limit, the amplitude is singular and it behaves as $(1/\xi)^N$
(modulo powers of $\ln \xi$ from loop corrections). This dominant singular behaviour is embodied in the soft current $\J$ on the right-hand side of Eq.~(\ref{1gfact}).
In this equation the symbol $\simeq$ means that on the right-hand side we have
neglected contributions that are less singular than $(1/\xi)^N$ in the limit
$\xi\to 0$. 

The soft current $\J(q_1,\cdots,q_N)$ in Eq.~(\ref{1gfact}) depends on the momenta,
colours and spins of both the soft and hard partons in the scattering amplitude
(although, the hard-parton dependence is not explicitly denoted in the argument of
$\J$). However this dependence entirely follows from the external-leg content
of $\M$, and the soft current is completely independent of the internal structure
of the scattering amplitude. In particular, we remark that the factorization in 
Eq.~(\ref{1gfact}) is valid \cite{Bern:1999ry, Catani:2000pi, Feige:2014wja} at arbitrary orders in the loop expansion
of the scattering amplitude. Correspondingly, we have $\J= \J^{(0)}+\J^{(1)}+\dots$,
where $\J^{(n)}$ is the contribution to $\J$ at the $n$-th loop accuracy.
In most of the following sections 
of this paper we limit ourselves to considering only
the soft current $\J^{(0)}$ at the tree level\footnote{Throughout the paper we explicitly remark on features and results (such as Eq.~(\ref{1gfact}))
that are valid at arbitrary orders in the loop expansion for both the soft current and the scattering amplitudes.}
and, for the sake of simplicity,
we simply denotes it by $\J$ (removing the explicit superscript $(0)$).
Owing to the all-order validity of Eq.~(\ref{1gfact}), the tree-level soft current
is {\em universal}, since it
equally contributes to soft-gluon factorization of scattering amplitudes $\M^{(0)}$
at the tree level and $\M^{(n)}$ at the $n$-th loop order.

The all-loop soft current $\J$ in Eq.~(\ref{1gfact}) is an operator that acts from the colour+spin space of $\M(\{p_i\})$ to the enlarged space of 
$\M(\{q_\ell\}, \{p_i\})$. Owing to Bose symmetry, the multi-gluon current
$\J(q_1,\cdots,q_N)$ has a fully symmetric dependence on the $N$ soft gluons.
The dependence on the colour ($a_\ell$) and spin ($\sigma_\ell$)
indices of the soft gluons can be made explicit by projecting $\J$ onto a basis vector:
\beeq
\label{colspinsoft}
J^{a_1 \dots a_N}_{\sigma_1 \dots \sigma_N}(q_1,\cdots,q_N) &=&
\bra{a_1,\sigma_1;\cdots;a_N,\sigma_N} \; \J(q_1,\cdots, q_N) \nn \\
&\equiv&
\e_{(\sigma_1)}^{\mu_1} \cdots \e_{(\sigma_N)}^{\mu_N}
\;J^{a_1 \dots a_N}_{\mu_1 \dots \mu_N}(q_1,\cdots,q_N) \;\;,
\eeeq
where $\e_{(\sigma_\ell)}^{\mu_\ell}=\e_{(\sigma_\ell)}^{\mu_\ell}(q_\ell)$ is the 
physical polarization vector of the gluon with momentum $q_\ell$ and spin component
$\sigma_\ell$.
The current 
$J^{a_1 \dots a_N}_{\mu_1 \dots \mu_N}$ 
is still an operator in the space of the hard partons. Since soft-gluon radiation does not change the spin polarization state 
of the radiating parton, the current is simply proportional to the unit matrix
in the spin space of the hard partons. QCD radiation, no matter how soft it is,
always carries away colour, and the soft current produces correlations
in the colour space of the hard partons. These colour correlations are embodies
in the dependence of 
$J^{a_1 \dots a_N}_{\mu_1 \dots \mu_N}$
on the colour charges
$\T_i$ of the hard partons.

Owing to gauge invariance of the amplitude $\M(\{q_\ell\}, \{p_i\})$, the soft-gluon current $\J$ fulfils the following relation:
\beq
\label{wardid}
q_\ell^{\mu_\ell} \;\left(\prod_{\ell^\prime \neq \ell} 
\e_{(\sigma_{\ell^\prime})}^{\mu_{\ell^\prime}}(q_{\ell^\prime}) \right)
J^{a_1 \dots a_\ell \dots a_N}_{\mu_1 \dots \mu_\ell  \dots \mu_N}(q_1,\cdots,q_N) \;\eqcs  \;0 \;\;,
\eeq
which can be regarded as an on-shell non-abelian Ward identity.
We can also consider current conservation in the following abelian-like form:
\beq
\label{curcons}
q_\ell^{\mu_\ell} \;
J^{a_1 \dots a_\ell \dots a_N}_{\mu_1 \dots \mu_\ell \dots \mu_N}(q_1,\cdots,q_N) \; \eqcs \; 0 
\;\; \quad \quad (\ell=1,\dots,N)
\;\;.
\eeq
In the left-hand side of Eq.~(\ref{curcons}), the current is not multiplied by the
physical polarization vectors of the $N-1$ gluons with $\ell^\prime \neq \ell$
and, therefore, Eq.~(\ref{curcons}) represents a stronger version of gauge invariance
in the multigluon case with $N \geq 2$ gluons (Eq.~(\ref{curcons}) implies Eq.~(\ref{wardid}), whereas the opposite is not true).
Owing to the arbitrariness of the physical polarization vectors and to colour conservation,
the multigluon current $\J$ of Eq.~(\ref{1gfact}) con be expressed in different
forms. We state that it is possible to find an explicit expression of $\J$
that fulfils current conservation as in Eq.~(\ref{curcons}). 
This statement is confirmed by the results of Ref.~\cite{CaGr99} for 
$N=2$ soft gluons and by the current that we present in Sect.~\ref{s:sc3} for 
$N=3$ soft gluons. A general (e.g., to all orders in the loop expansion) and formal
proof of Eq.~(\ref{curcons}) is presented in Appendix~\ref{a:cons}.

\subsection{Tree-level current: single and double gluon emission\label{s:cur12}}

The well-known \cite{Bassetto:1984ik} tree-level current $\J(q)$ for emission of a single soft gluon with momentum $q$ is
\begin{equation}\label{J1}
  J^{a,\mu}(q) = \sum_i \,T^a_i \frac{p_i^\mu}{p_i\cdot q} \;\;.
\end{equation}
This expression can be simply obtained by inserting the soft gluon onto the external lines of $\M(\{p_i\})$ and by using vertices and propagators in the eikonal approximation.

From the expression in Eq.~(\ref{J1}) we have
\begin{equation}\label{consJ1}
q_\mu \,J^{a,\mu}(q) = \sum_i \,T^a_i \;\;,
\end{equation}
and, therefore, by using colour conservation as in Eq.~(\ref{colcons}),
the current conservation relation in Eq.~(\ref{curcons}) (or Eq.~(\ref{wardid}))
is directly fulfilled.

The general expression of the current for double soft-gluon radiation at the tree level was presented in Ref.~\cite{CaGr99}. The current can be
expressed in various (though equivalent) forms. We use the form in Eq.~(102) of 
Ref.~\cite{CaGr99} and we write the current as follows:
\beq
\label{J12}
J^{a_1 a_2}_{\mu_1 \mu_2}(q_1,q_2) = \frac{1}{2} \left\{ J^{a_1}_{\mu_1}(q_1) \,,
J^{a_2}_{\mu_2}(q_2)\right\} + \Gamma^{(2) a_1 a_2}_{\mu_1 \mu_2}(q_1,q_2) \;\;,
\eeq
where the first contribution on the right-hand side is the colour commutator of the single-gluon currents of Eq.~(\ref{J1}). The second contribution is
\beq
\label{gamma2}
\Gamma^{(2) a_1 a_2}_{\mu_1 \mu_2}(q_1,q_2) = \ui f^{a_1 a_2\, b} \sum_i T_i^b 
\;\gamma_{i, \mu_1 \mu_2}(q_1,q_2) \;\;,
\eeq
where
\beq
\label{gammaC}
\gamma_i^{\mu_1 \mu_2}(q_1,q_2) = \frac1{p_i\cdot(q_1+q_2)} \left\{
    \frac{p_i^{\mu_1}  p_i^{\mu_2}}{2\;p_i\cdot q_1}
    + \frac1{q_1\cdot q_2}
    \left(p_i^{\mu_1} q_1^{\mu_2}+\frac12 g^{\mu_1\mu_2}p_i\cdot q_2\right) \right\}
    - \bigl( 1 \leftrightarrow 2 \bigr) \;\;,
\eeq
and the notation $\left( 1 \leftrightarrow 2 \right)$ denotes the exchange of the two gluons. Therefore, the kinematical function $\gamma_i^{\mu_1 \mu_2}(q_1,q_2)= 
-\gamma_i^{\mu_2 \mu_1}(q_2,q_1)$ is antisymmetric with respect to the exchange of the two gluons, whereas $\Gamma^{(2) a_1 a_2}_{\mu_1 \mu_2}(q_1,q_2)$ turns out
to be symmetric because of the antisymmetry 
of its colour coefficient ($f^{a_1 a_2\, b}= - f^{a_2 a_1\, b}$).

The two-gluon current in Eqs.~(\ref{J12})--(\ref{gammaC}) fulfils the current conservation relation in Eq.~(\ref{curcons}) (see Eqs.~(104) and (105) in 
Ref.~\cite{CaGr99}).

The first term on the right-hand side of Eq.~(\ref{J12}) is the only contribution that survives in the abelian case (double soft-photon emission), where it reduces itself to the product of two {\em independent} single-emission currents.
In the two-gluon case, the current embodies the intrinsically non-abelian term
$\G^{(2)}$, which produces two-gluon correlations. Therefore, the form of the
two-gluon current in the right-hand side of
Eq.~(\ref{J12}) can formally be regarded as an expression in terms of an independent-emission contribution and an irreducible-correlation contribution. We note, however, that the term $\{ J^{a_1}(q_1), J^{a_1}(q_2) \}$ does not lead to independent emission
from a physical viewpoint. Indeed, the colour charges in the single-gluon currents
$J^{a_\ell}(q_\ell)$ produce colour-correlations with the hard partons in the scattering amplitude $\M(\{p_i\}$
and, moreover, the currents do not commute ($[J^{a_1}(q_1) , J^{a_1}(q_2)] \neq 0$).

\begin{figure}[ht]
  \centering
  \includegraphics[width=0.7\textwidth]{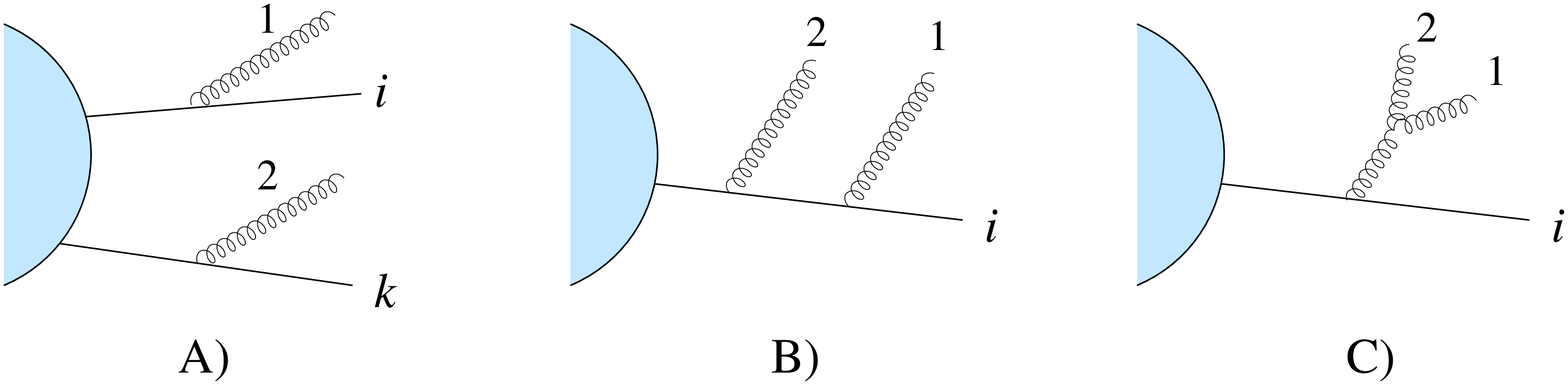}
  \Caption{Diagrams contributing to the two-gluon soft current.}
  \label{f:2s}
\end{figure}

As described in Ref.~\cite{CaGr99}, the result in Eqs.~(\ref{J12})--(\ref{gammaC})
can be obtained 
by inserting the two gluons with momenta $q_1$ and $q_2$ onto the external legs of 
$\M(\{p_i\})$. The relevant Feynman diagrams are presented in 
Fig.~\ref{f:2s}. Here the coupling of the soft gluons to the hard partons $i$ and $k$
is treated by using the eikonal approximation, whereas the soft-gluon propagator
and vertex (Fig.~\ref{f:2s}C) are treated without introducing any soft approximation.


\section{Tree-level soft current for triple gluon emission\label{s:sc3}}

\begin{figure}[ht]
  \centering
  \includegraphics[width=0.7\textwidth]{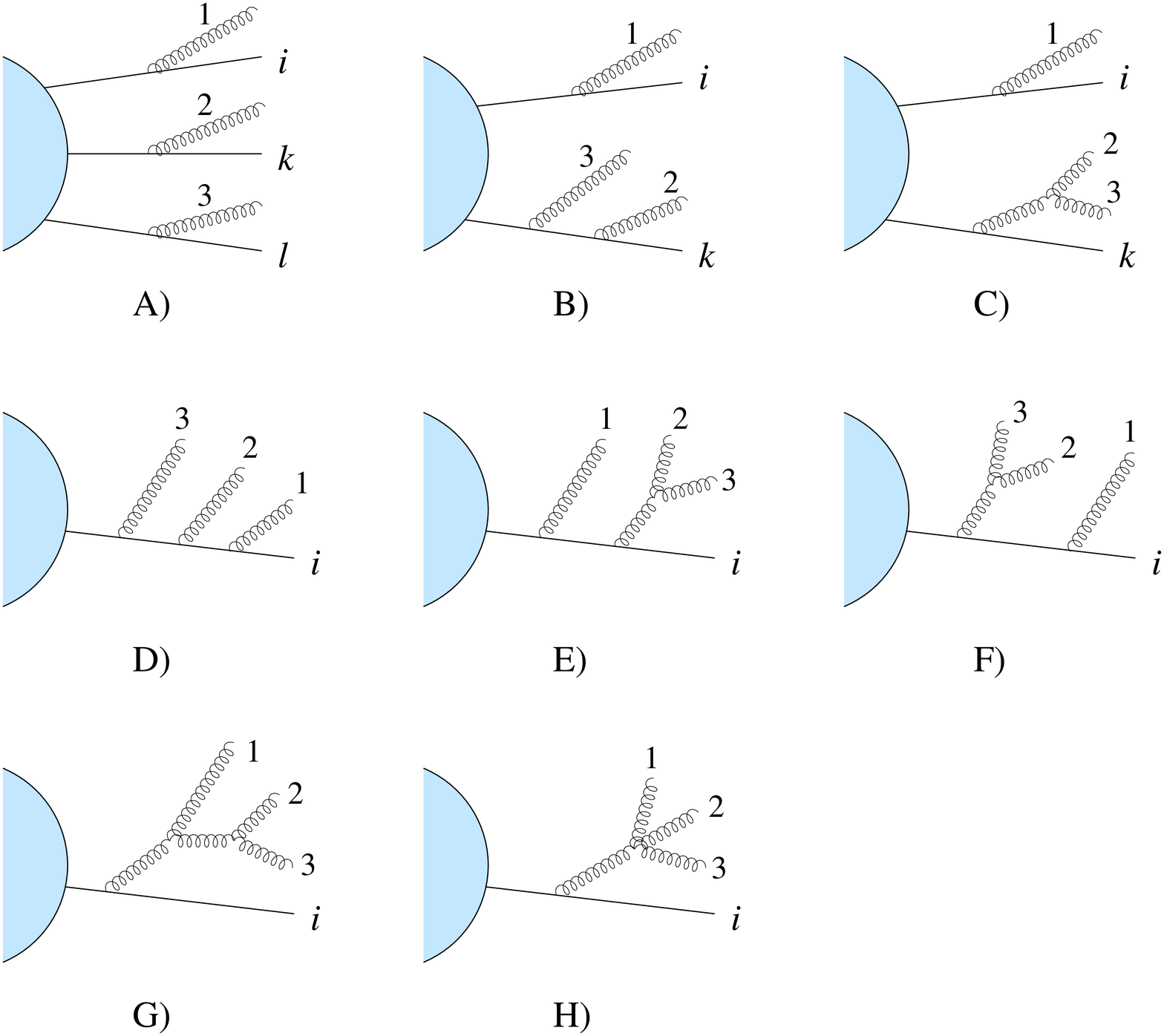}
  \Caption{Diagrams contributing to the three-gluon soft current.}
  \label{f:3s}
\end{figure}

We compute the tree-level current $\J(q_1,q_2,q_3)$ for triple gluon emission by using the same method as used in Ref.~\cite{CaGr99} to obtain $\J(q_1,q_2)$.
The three gluons are inserted onto the external legs of $\M(\{p_i\})$, and the relevant Feynman diagrams are shown in Fig.~\ref{f:3s}. Each of the soft gluons is coupled to hard-parton external lines ($i,k,l$) by using the eikonal approximation.
The soft-gluon propagators and the triple and quadruple soft-gluon vertices 
(Figs.~\ref{f:3s}C, E, F, G, H) have to be treated exactly, without introducing any soft approximation. The computation of each contributing diagram in Fig.~\ref{f:3s}
is straightforward, although the complete contribution of all the diagrams is algebraically cumbersome. The results of our computation (which were first illustrated in Ref.~\cite{thesis})
are presented below in Eqs.~(\ref{J123}), (\ref{G123}) and (\ref{gamma123}).
We note that the propagators of off-shell (internal-line) gluons are gauge dependent and, therefore, a gauge choice is required to evaluate the diagrams in 
Fig.~\ref{f:3s}. We have computed the current by using both axial and covariant gauges and we have explicitly checked that the final result for the current 
$\J(q_1,q_2,q_3)$ is gauge independent. More precisely, by using colour conservation as in Eq.~(\ref{colcons}) (i.e., neglecting contributions that vanish because of colour conservation) we are able (independently of the gauge choice) to express the current  $\J(q_1,q_2,q_3)$ in the explicit form of Eqs.~(\ref{J123}), 
(\ref{G123}) and (\ref{gamma123}).

\vspace*{2mm}

The tree-level current $\J(q_1,q_2,q_3)$ can be expressed in different (though equivalent) forms. We write our result as follows:
\beeq
\label{J123}
J_{\mu_1 \mu_2 \mu_3}^{a_1 a_2 a_3}(q_1,q_2,q_3)&=&
\sym{J_{\mu_1}^{a_1}(q_1)} {J_{\mu_2}^{a_2}(q_2)}{J_{\mu_3}^{a_3}(q_3)} \nn \\
&+& 
\left[ \sy{J_{\mu_1}^{a_1}(q_1)}{\Gamma^{(2)\,a_2 a_3}_{\mu_2 \mu_3}(q_2,q_3)} 
+ \;\left( 1 \leftrightarrow 2 \right) + \left( 1 \leftrightarrow 3 \right) 
\right]
\\
&+& \Gamma^{(3)\, a_1 a_2 a_3}_{\mu_1 \mu_2 \mu_3}(q_1,q_2,q_3) \nn \;\;,
\eeeq
where we have introduced the symbol $(\dots)_{sym}$ to denote symmetrized products.
The symmetrized products of two and three generic colour operators $O_I$ are defined
as
\begin{equation}\label{sym2}
  \sy{O_1}{O_2}\equiv\frac12\bigl(O_1 \,O_2 + O_2 \,O_1 \bigr) 
= \frac12 \left\{O_1 , O_2 \right\}\;\;,
\end{equation}
\begin{equation}
\label{sym3}
  \sym{O_1}{O_2}{O_3}\equiv\frac1{3!}\bigl(O_1 \,O_2 \,O_3 +  {\rm perms.}\{1,2,3\} \bigr) \;\;,
\end{equation}
where the right-hand side of Eq.~(\ref{sym3}) includes the sum over the $3!=6$
permutations of $O_1, O_2$ and $O_3$.

The right-hand side of Eq.~(\ref{J123}) has the structure of 
an expansion
in multigluon irreducible correlations. This structure generalizes Eq.~(\ref{J12}) to the case of $N=3$ soft gluons.

The term in the first line on the right-hand side of Eq.~(\ref{J123}) represents
the `independent' (though colour-correlated) emission of three soft gluons, and each gluon contributes through the single-emission current $J^a_\mu(q)$ in Eq.~(\ref{J1}).
This term contains products of three colour charges ($T_i^{a_1} T_k^{a_2} T_l^{a_3}$)
of the hard partons in $\M(\{p_i\})$ and a related kinematical dependence on the three
hard-parton momenta ($p_i,p_k,p_l$).
Since multiple soft-photon radiation is completely uncorrelated, this is the only term that contributes to the current in the abelian case (i.e., the current for triple soft-photon radiation is simply $J(q_1,q_2,q_3)= J(q_1) J(q_2) J(q_3)$).
The other terms in the right-hand side of Eq.~(\ref{J123}) are strictly non-abelian.

The contribution in the second line on the right-hand side of Eq.~(\ref{J123})
is given in terms of the `independent' emission of a single gluon (e.g., the current
$J^{a_1}_{\mu_1}(q_1)$) and the irreducible correlated emission of the other two gluons (e.g., the factor $\Gamma^{(2)\,a_2 a_3}_{\mu_2 \mu_3}(q_2,q_3)$).
This correlated emission is expressed through the correlation term $\G^{(2)}$
in Eqs.~(\ref{J12}) and (\ref{gamma2}). By inspection of the expression in 
Eq.~(\ref{gamma2}), we see that the second line of Eq.~(\ref{J123}) involves 
non-abelian correlations with two hard partons, with ensuing colour charge structures
of the type $f^{a_2 a_3 c} \,T_k^{c} \,T_i^{a_1}$
and a related kinematical dependence on the momenta $p_k$ and $p_i$
of the two hard partons.
Note that the contribution in the second line on the right-hand side of Eq.~(\ref{J123}) is completely symmetric with respect to any permutations of the three soft gluons. This follows from the symmetry of $\G^{(2)}(q_2,q_3)$ with respect to the exchange of the two gluons with momenta $q_2$ and $q_3$ and from the addition of the two permutations with the exchange $\left( 1 \leftrightarrow 2 \right)$ and  $\left( 1 \leftrightarrow 3 \right)$ in the square-bracket term.

The contribution in the third line on the right-hand side of Eq.~(\ref{J123})
defines the `irreducible' three-gluon correlation $\G^{(3)}(q_1,q_2,q_3)$.
We find the following result:
\beq
\label{G123}
\Gamma^{(3)\, a_1 a_2 a_3}_{\mu_1 \mu_2 \mu_3} = \sum_i \,T_i^b 
\; f^{a_1 a_2,a_3 b} \;\gamma_{i,\,\mu_1 \mu_2 \mu_3}(q_1,q_2;q_3)
+ \left( 3 \leftrightarrow 1 \right) + \left( 3 \leftrightarrow 2 \right) \;\;,
\eeq
where the kinematical function $\gamma_{i}(q_1,q_2;q_3)$ is given in 
Eq.~(\ref{gamma123})
and we have defined the product (contraction) of two structure constant
as
\begin{equation}\label{ff}
  f^{ab,cd} \equiv 
f^{abs} f^{scd} \;\;.
\end{equation}
The kinematical function $\gamma_{i}(q_1,q_2;q_3)$ is antisymmetric under the exchange
of the two gluons with momenta $q_1$ and $q_2$, and the colour coefficient 
$f^{a_1 a_2,a_3 b}$ is also antisymmetric under the exchange of the colour indices
$a_1$ and $a_2$ of these two gluons. Therefore, the first contribution on the 
right-hand side of Eq.~(\ref{G123}) is symmetric with respect to the exchange 
$\left( 1 \leftrightarrow 2 \right)$, and the addition of the two permutations
with $\left( 3 \leftrightarrow 1 \right)$ and $\left( 3 \leftrightarrow 2 \right)$
makes $\G^{(3)}(q_1,q_2,q_3)$ completely symmetric with respect to any permutations of the three soft gluons.
 
The product of two structure constants fulfils the customary Jacobi identity:
\beq
\label{jacf}
f^{ba_1,a_2a_3} + f^{ba_3,a_1a_2} + f^{ba_2,a_3a_1} = 0 \;\;.
\eeq
Therefore, the contributions in the three permutations on the right-hand side of 
Eq.~(\ref{G123}) are not linearly independent. This implies that 
$\G^{(3)}(q_1,q_2,q_3)$ can be rewritten in different ways. In particular, 
one can choose two structure constant coefficients (e.g., $f^{a_1 a_2,a_3 b}$ and
$f^{a_3 a_1,a_2 b}$)
and express 
the right-hand side of Eq.~(\ref{G123})
as sum of only two linearly independent terms.

We note that the expression of $\G^{(3)}(q_1,q_2,q_3)$ in Eq.~(\ref{G123}) is given as a sum ($\sum_i$) of contributions which involve colour and kinematical correlations
with only one hard parton at a time. The colour correlations are due to factors
(e.g., $f^{a_1 a_2,a_3 b} \,T_i^b$) that have a linear dependence on the colour charge $\T_i$ of the hard parton. 
Notably, the three-gluon correlation $\G^{(3)}$ is maximally non-abelian, since it is proportional to the product of two structure constants.

The kinematical function $\gamma_{i}(q_1,q_2;q_3)$ of Eq.~(\ref{G123}) is definitely more complex than the corresponding function $\gamma_{i}(q_1,q_2)$ of 
Eq.~(\ref{gammaC}) for the two-gluon correlation  $\G^{(2)}(q_1,q_2)$. We find
\begin{align} 
\label{gamma123}
& \gamma_{i}^{\mu_1 \mu_2 \mu_3}(q_1,q_2;q_3) = \frac1{p_i\cdot q_{123}}
\left\{ \frac{1}{12} 
\frac{p_i^{\mu_1} p_i^{\mu_2} p_i^{\mu_3} \,p_i\cdot (3q_3-q_{12})}{ p_i\cdot q_2
\; p_i\cdot q_3 \; p_i\cdot q_{12}} \right. \nn \\
& \quad \quad \left. + \frac{p_i^{\mu_3} \,p_i\cdot (q_3-q_{12})}{p_i\cdot q_3 \; p_i\cdot q_{12} \;q_{12}^2} 
\left( \frac{1}{2} g^{\mu_1 \mu_2} \,p_i\cdot q_1 + p_i^{\mu_2} \,q_2^{\mu_1} \right)
        \right. \nn \\
&\quad \quad + \left. \frac{1}{q_{123}^2 \;q_{12}^2} \Big[
    q_{12}^2 \,p_i^{\mu_1} \,g^{\mu_2\mu_3}
    + 2 q_2^{\mu_1} g^{\mu_2\mu_3} \,p_i\cdot (q_3-q_{12})
    + 4 \,q_3^{\mu_1} \,q_1^{\mu_2} \,p_i^{\mu_3} 
  \Big. \right. \nn \\
&\quad \quad + \left. \Big. 4 \,q_2^{\mu_1} \,p_i^{\mu_2} \,q_{12}^{\mu_3}
   + g^{\mu_1\mu_2} \big( q_{23}^2 \,p_i^{\mu_3}
    + q_1^{\mu_3} \,p_i\cdot(q_{13} -3 q_2) \big)
  \Big] \, \right\} - \left( 1 \leftrightarrow 2 \right)\;, 
\end{align}
where we have defined the soft momenta
\beq
\label{qmom}
q_{\ell \ell^\prime}^\mu \equiv q_{\ell}^\mu + q_{\ell^\prime}^\mu \;\;, 
\quad
q_{123}^\mu \equiv q_{1}^\mu + q_{2}^\mu + q_{3}^\mu \;\;.
\eeq

We note that the three-gluon 
current  $\J(q_1,q_2,q_3)$ in Eqs.~(\ref{J123}), 
(\ref{G123}) and (\ref{gamma123}) fulfils the current conservation relation of 
Eq.~(\ref{curcons}). This can be checked by performing a straightforward algebraic 
calculation \cite{thesis}.



Multiple soft-gluon radiation in tree-level scattering amplitudes was studied by Berends and Giele in Ref.~\cite{BeGi89}. The authors of 
Ref.~\cite{BeGi89} consider scattering amplitudes whose external legs are gluons and,
possibly, a quark-antiquark pair with or without an additional vector boson.
The colour structure of this class of tree-level amplitudes can be properly expressed in terms of colourless (though colour-ordered) subamplitudes. In the soft limit,
the colour-ordered subamplitudes fulfil a factorization formula that involves an $N$-gluon soft factor $s_{i \,{\underline {1 2 \cdots N}} \,k}$ (see, e.g., Eq.~(3.18)
in Ref.~\cite{BeGi89}). The subscripts in the soft factor 
$s_{i \,{\underline {1 \cdots N}} \,k}$ refer to the momenta $q_1, \cdots,q_N$ 
of the colour-ordered soft gluons $\{1 \cdots N\}$ and the momenta $p_i$ and $p_k$
of the their colour-connected hard partons $i$ and $j$.
The single-gluon ($s_{i \,{\underline {1}} \,k}$) and double-gluon 
($s_{i \,{\underline {1 2}} \,k}$) soft factors were explicitly computed in 
Ref.~\cite{BeGi89}. In the case of $N \geq 3$ soft gluons, Berends and Giele
derived \cite{BeGi89} a detailed iterative procedure to compute the soft factor
$s_{i \,{\underline {1 \cdots N}} \,k}$. We have applied the method of 
Ref.~\cite{BeGi89} to explicitly compute the triple-gluon soft factor 
$s_{i \,{\underline {1 2 3}} \,k}$, and the result of our computation is presented below in Eq.~(\ref{bg3g}).

The soft-gluon factorization formula in Eq.~(\ref{1gfact}) is valid to arbitrary loop
accuracy and it is also valid for arbitrary scattering amplitudes, since the colour space formulation does not require the `a priori' specification of the colour structure of the amplitude. In particular, Eq.~(\ref{1gfact}) can be applied to the tree-level scattering amplitudes that are considered in Ref.~\cite{BeGi89}, upon
implementation of the corresponding decomposition in colour-ordered subamplitudes.
This procedure can be used to obtain relations between the soft-gluon current
$\J(q_1,\cdots,q_N)$ and the soft factors $s_{i \,{\underline {1 \cdots N}} \,k}$
of Ref.~\cite{BeGi89}. The relation for $N=1$ and $N=2$ soft gluons was discussed in Ref.~\cite{CaGr99}. We have applied the procedure to triple soft-gluon radiation, and we obtain an expression for $s_{i \,{\underline {1 2 3}}\,k}$ that exactly agrees
with the result that we have independently derived by using the iterative procedure of Ref.~\cite{BeGi89}.

To present the explicit result for the soft factor $s_{i \,{\underline {1 2 3}}\,k}$,
we express it in terms of the kinematical functions 
$\gamma_{i}^{\mu_1 \mu_2}(q_1,q_2)$ and 
$\gamma_{i}^{\mu_1 \mu_2 \mu_3}(q_1,q_2;q_3)$
(see Eqs.~(\ref{gammaC}) and (\ref{gamma123})) and of the single-particle
eikonal factor $j^{\mu_\ell}_i(q_\ell)$,
\begin{equation}\label{j1}
  j^{\mu_\ell}_i(q_\ell) \equiv \frac{p_i^{\mu_\ell}}{p_i\cdot q_\ell} \;\;.
\end{equation}
We find
\beeq
\label{bg3g}
s_{i \,{\underline {1 2 3}} \,k}\!\!\!\!\! \!\!\!\!&&= 
\e_{(\sigma_1)}(1)\, \e_{(\sigma_2)}(2)\, \e_{(\sigma_3)}(3)\, 
\Bigl\{ \gamma_{i}(1,2;3) + \gamma_{i}(3,2;1) 
- \frac{1}{2} \left[ \,\gamma_{i}(1,2) \,j_i(3) + j_i(1)\, \gamma_{i}(2,3) \,\right]
 \Bigr. \nn \\ 
&&+ \Bigl. \,\gamma_{i}(1,2) \,j_k(3)
+ \frac{1}{2} \,j_i(1) \,j_i(2) \,j_k(3) - \frac{1}{6} \,j_i(1) \,j_i(2) \,j_i(3)
\Bigr\}
 - \binom{1 \leftrightarrow 3}{i \leftrightarrow k} \;\;,
\eeeq
where we have used a shorthand notation to denote the dependence on the soft momenta 
($q_\ell$) and contractions of the corresponding Lorentz indices ($\mu_\ell$).
For any functions $f^{\mu_1 \mu_2 \mu_3}(q_1,q_2;q_3)$ and
$g^{\mu_1 \mu_2 \mu_3}(q_1,q_2;q_3)$ we have defined
\beq
\label{fgnotation}
f^{\mu_1 \mu_2 \mu_3}(q_1,q_2,q_3) \;g_{\mu_1 \mu_2 \mu_3}(q_1,q_2,q_3) \equiv
f(1,2,3) \;g(1,2,3) \;\;.
\eeq
Thus, for instance, we have
\beq
\e(1)\, \e(2)\, \e(3)\, \gamma_i(1,2) \,j_k(3) = 
\e_{\mu_1}(q_1)\, \e_{\mu_2}(q_2)\, \e_{\mu_3}(q_3)\, 
\gamma_i^{\mu_1 \mu_2}(q_1,q_2) \,j_k^{\mu_3}(q_3) \;\;.
\eeq

For the sake of completeness, we also report the expressions of the single-gluon
and double-gluon soft factors $s_{i \,{\underline {1}} \,k}$ and
$s_{i \,{\underline {1 2}} \,k}$. They are \cite{BeGi89}
\beq
\label{bg1g}
s_{i \,{\underline {1}} \,k} = 
\e_{(\sigma_1)}(1)\, \left( \,j_k(1) - j_i(2) \right) \;\;,
\eeq
\beeq
\label{bg2g}
s_{i \,{\underline {1 2}} \,k} = 
\e_{(\sigma_1)}(1)\, \e_{(\sigma_2)}(2)\,  
\Bigl\{ \gamma_{i}(1,2) 
+ \frac{1}{2} \left[ \,j_i(1) \,j_i(2) - j_i(1) \, j_k(2)\,\right] \Bigr\}
+ \binom{1 \leftrightarrow 2}{i \leftrightarrow k} \;\;.
\eeeq
where we have used the same shorthand notation as in Eq.~(\ref{bg3g}).

We note that Ref.~\cite{BeGi89} considers scattering amplitudes with massless
hard partons in the external legs (specifically, a massless quark and antiquark).
The results for the soft function $s_{i \,{\underline {1 2 \cdots N}} \,k}$
that we have derived and reported in Eqs.~(\ref{bg3g}), (\ref{bg1g}) and (\ref{bg2g})
are nonetheless valid for both massless and massive hard momenta $p_i$ and $p_k$.

As specified in Sect.~\ref{s:sc12}, 
throughout the paper we consider scattering amplitudes that are computed within dimensional regularization in $d=4 - 2 \epsilon$ space-time dimensions, and we use the
customary CDR procedure \cite{cdr} with $d-2$ spin polarization states for on-shell gluons.
Dimensional regularization leads to the overall factor $(\g \,\mu_0^\epsilon)^N$
in the soft-gluon factorization formula (\ref{1gfact}), and it also leads to
a regularization dependence of the soft current $\J(q_1,\cdots,q_N)$ through the soft-gluon spin vectors $\e_{(\sigma_\ell)}^{\mu_\ell}(q_\ell)$
in Eq.~(\ref{colspinsoft}). We note, however, that the expressions of the {\em tree-level} currents $J^{a_1 \dots a_N}_{\mu_1 \dots \mu_N}(q_1,\cdots,q_N)$
in Sect.~\ref{s:cur12} ($N=1,2$) and Sect.~\ref{s:sc3}  ($N=3$)
do not have any explicit dependence on $\epsilon$. 
In particular, these tree-level expressions of the soft currents 
$J^{a_1 \dots a_N}_{\mu_1 \dots \mu_N}(q_1,\cdots,q_N)$ are valid within CDR, and they are equally valid in versions of dimensional regularizations (see, e.g., 
Ref.~\cite{Gnendiger:2017pys}) that use 2 spin polarization states for on-shell gluons
(and, possibly, four-dimensional hard-parton momenta). 

\setcounter{footnote}{2}

\section{Squared amplitudes and currents \label{s:square}}

The structure of this section parallels that of Sect.~\ref{s:sc}.
We first present a brief general discussion on soft-gluon factorization for squared amplitudes. Then we recall the known results \cite{CaGr99} on the squared currents
for single and double gluon emission at the tree level. Our results on the squared current for triple gluon emission are presented in Sect.~\ref{s:sq123}\,.

Using the notation in colour+spin space, the squared amplitude $|\M|^2$ (summed
over the colours and spins of its external legs) is written as follows
\beq
\label{squared}
|\M|^2 = \langle{\M} \ket{\,\M} \;\;.
\eeq
Accordingly, the square of the soft-gluon factorization formula in Eq.~(\ref{1gfact})
gives
\beq
\label{softsquared}
| \M(\{q_\ell\}, \{p_i\}) |^2 \simeq (\g \,\mu_0^\epsilon)^{2N} \,
\bra{\M (\{p_i\})} \;| \J(q_1,\cdots,q_N) |^2 \;\ket{\M (\{p_i\})} \;,
\eeq
where, analogously to Eq.~(\ref{1gfact}), the symbol $\simeq$ means that we have neglected contributions that are subdominant in the soft multigluon limit.
The squared current $| \J(q_1,\cdots,q_N) |^2 $ (summed over the colours and spins of the soft gluons) is still a colour operator that depends on the colour charges of the hard partons in $\M (\{p_i\})$. These colour charges produce colour correlations and,
therefore, the right-hand side of Eq.~(\ref{softsquared}) is not proportional to
$|\M(\{p_i\})|^2$ in the case of a generic scattering amplitude\footnote{Colour correlations can be simplified in the case of scattering amplitudes with two and three hard partons (see Sect.~\ref{sec:3hard} and \ref{sec:2hard}).}.

The squared current $| \J |^2 $ of Eq.~(\ref{softsquared}) can be explicitly expressed as sum over colour and spin (or Lorentz) indices of the soft gluons.
Using Eq.~(\ref{colspinsoft}) we have
\beeq
\label{spin}
| \J(q_1,\cdots,q_N) |^2 &=& \left[ J^{a_1 \dots a_N}_{\sigma_1 \dots \sigma_N}(q_1,\cdots,q_N) \right]^\dagger J^{a_1 \dots a_N}_{\sigma_1 \dots \sigma_N}(q_1,\cdots,q_N)\\
\label{lorentz}
&=& \left[ \prod_{\ell=1}^N d^{\,\mu_\ell \nu_\ell}(q_\ell) \right]
\left[ J^{a_1 \dots a_N}_{\mu_1 \dots \mu_N}(q_1,\cdots,q_N) \right]^\dagger
J^{a_1 \dots a_N}_{\nu_1 \dots \nu_N}(q_1,\cdots,q_N) \;\;,
\eeeq
where we have introduced the polarization tensor $d^{\mu\nu}(q_\ell)$ of the external
soft gluons:
\begin{equation}\label{dmunu}
  d^{\,\mu\nu}(q_\ell) \equiv 
\pol^{*\mu}_{(\sigma)}(q_\ell) \;\pol^{\nu}_{(\sigma)}(q_\ell)
  = -g^{\mu\nu} + \text{gauge terms} \;\;.
\end{equation}

In the right-hand side of Eq.~(\ref{dmunu}) we have not written the explicit expression of the gauge dependent terms, which follows from the specific choice of the physical polarization vectors $\pol^{\mu}_{(\sigma)}$. Independently of their explicit form, these gauge dependent terms are proportional to longitudinal polarizations, namely, proportional to either $q_\ell^\mu$ or $q_\ell^\nu$ in the 
right-hand side of Eq.~(\ref{dmunu}).
As a consequence of the current conservation relation in Eq.~(\ref{curcons}),
the action of longitudinal polarizations on the soft current leads to vanishing contributions onto colour-singlet states. Therefore, inserting Eq.~(\ref{lorentz})
in the right-hand side of Eq.~(\ref{softsquared}), the gauge dependent terms in 
Eq.~(\ref{dmunu}) do not contribute. It follows that, to the purpose of evaluating the soft-gluon factorization formula (\ref{softsquared}), we can simply consider the following form of the squared current:
\beq
\label{jsquare}
| \J(q_1,\cdots,q_N) |^2 \;\eqcs \;(-1)^N 
\left[ \prod_{\ell=1}^N g^{\,\mu_\ell \nu_\ell} \right]
\left[ J^{a_1 \dots a_N}_{\mu_1 \dots \mu_N}(q_1,\cdots,q_N) \right]^\dagger
J^{a_1 \dots a_N}_{\nu_1 \dots \nu_N}(q_1,\cdots,q_N) \;,
\eeq
which is explicitly gauge invariant.

\subsection{Tree-level squared current: single and double gluon emission\label{s:sq12}}

The square of the soft current $\J(q)$ in Eq.~(\ref{J1}) for single gluon emission is
\begin{equation}\label{J1sq}
| \J(q) |^2 \;\eqcs  -\sum_{i,k} \T_i\cdot \T_k \; {\cal S}_{ik}(q) \;\;,
\end{equation}
\beq
\label{S1}
{\cal S}_{ik}(q) = \frac{p_i\cdot p_k}{p_i\cdot q\; p_k\cdot q} \;\;.
\eeq
The colour charge dependence in Eq.~(\ref{J1sq}) is entirely given in terms of
{\em dipole} operators $\T_i\cdot \T_k = T_i^a T_k^a$. The insertion of the dipole operators in the factorization formula (\ref{softsquared}) leads to the factor
$\bra{\M (\{p_i\})} \; \T_j\cdot \T_k\; \ket{\M (\{p_i\})}$, which produces colour correlations between two hard partons ($j$ and $k$) in $\M(\{p_i\})$. Some properties of the colour algebra of these dipole correlations are discussed in Appendix~A
of Ref.~\cite{csdip}.

The square of the soft current $\J(q_1,q_2)$ in Eq.~(\ref{J12}) for double gluon emission was computed in Ref.~\cite{CaGr99}, and the result is 
\beq
\label{J2sq}
| \J(q_1,q_2) |^2 \;\eqcs \;\sy{W^{(1)}(q_1)}{W^{(1)}(q_2)} + W^{(2)}(q_1,q_2) \;\;,
\eeq
\begin{equation}
\label{corr2}
  W^{(2)}(q_1,q_2) = - \,C_A \sum_{i,k} \T_i\cdot \T_k \; \cS_{ik}(q_1,q_2) \;\;,
\end{equation}
where $\sy{W^{(1)}(q_1)}{W^{(1)}(q_2)}$ denotes the symmetrized product of two operators (see Eq.~(\ref{sym2})), and we have introduced the operator $W^{(1)}(q_\ell)$ to denote the square of the single-gluon current in Eq.~(\ref{J1sq}):
\beq
\label{w1def}
| \J(q) |^2 \;\eqcs \; W^{(1)}(q) \;\;.
\eeq

The form of Eq.~(\ref{J2sq}) has the structure of 
an expansion in multigluon irreducible correlations (analogously to the structure of Eq.~(\ref{J12})). The first term on the right-hand side corresponds to the `independent' (though colour correlated) emission of the two soft gluons. This term leads to colour correlations due to the iterated action of two dipole factors (e.g., $\T_i\cdot \T_m \;\T_k\cdot \T_l$). The term
$W^{(2)}(q_1,q_2)$ is definitely an irreducible-correlation contribution for double gluon emission. From Eq.~(\ref{corr2}) we see that $W^{(2)}(q_1,q_2)$ is non-abelian
(it is proportional to $C_A$) and it is expressed in terms of colour dipole operators
(analogously to the single-gluon emission contribution in Eq.~(\ref{J1sq})).

%

The function $\cS_{ij}(q_1,q_2)$ of the two-gluon correlation $W^{(2)}(q_1,q_2)$
in Eq.~(\ref{corr2}) can be written \cite{CaGr99,Czakon:2011ve} as follows
\begin{equation}\label{S2}
  \cS_{ij}(q_1,q_2) = \cS^{m=0}_{ij}(q_1,q_2) + 
\left[
   m_i^2 \; \cS^{m \neq 0}_{ij}(q_1,q_2) + m_j^2 \;
    \cS^{m \neq 0}_{ji}(q_1,q_2) 
\right] \;,
\end{equation}
where $m_i$ is the mass of the hard parton with momentum $p_i$ ($p_i^2=m_i^2$).
The `massless' contribution $\cS^{m=0}_{ij}$ is~\cite{CaGr99}
\begin{align}
  \cS^{m=0}_{ij}(q_1,q_2) &= \frac{(1-\epsilon)}{(q_1 \cdot q_2)^2}
  \frac{p_i \cdot q_1 \; p_j \cdot q_2 + p_i \cdot q_2 \; p_j \cdot
    q_1}{p_i \cdot (q_1+q_2) \; p_j \cdot (q_1+q_2)} \nonumber \\
  & - \frac{(p_i \cdot p_j)^2}{2 \; p_i \cdot q_1 \; p_j \cdot q_2 \; p_i
    \cdot q_2 \; p_j \cdot q_1} \left[ 2 - \frac{p_i \cdot q_1 \; p_j
      \cdot q_2 + p_i \cdot q_2 \; p_j \cdot q_1}{p_i \cdot (q_1+q_2) \;
      p_j \cdot (q_1+q_2)} \right] \nonumber \\
  & + \frac{p_i \cdot p_j}{2 \; q_1 \cdot q_2} \left[ \frac{2}{p_i
      \cdot q_1 \; p_j \cdot q_2} + \frac{2}{p_j \cdot q_1 \; p_i \cdot
      q_2} - \frac{1}{p_i \cdot (q_1+q_2) \; p_j \cdot (q_1+q_2)}
  \right. \nonumber \\
& \times \left. \left( 4 + \frac{(p_i \cdot q_1 \; p_j \cdot q_2 +
    p_i \cdot q_2 \; p_j \cdot q_1)^2}{\; p_i \cdot q_1 \; p_j \cdot
    q_2 \; p_i \cdot q_2 \; p_j \cdot q_1} \right) \right] \; ,\label{S2m0}
\end{align}
and the `mass-correction' contribution $\cS^{m \neq 0}_{ij}$ is \cite{Czakon:2011ve}
\begin{align}
  \cS^{m\neq0}_{ij}(q_1,q_2) &=  
\frac{p_i \cdot p_j \; p_j
    \cdot (q_1+q_2)}{2 \; p_i \cdot q_1 \; p_j \cdot q_2 \; p_i \cdot q_2 \;
    p_j \cdot q_1 \; p_i \cdot (q_1+q_2)} \nonumber \\
  & - \frac{1}{2 \; q_1 \cdot q_2 \; p_i \cdot (q_1+q_2) \; p_j \cdot
    (q_1+q_2) } \left( \frac{(p_j \cdot q_1)^2}{p_i \cdot q_1 \; p_j
      \cdot q_2} + \frac{(p_j \cdot q_2)^2}{p_i \cdot q_2 \; p_j \cdot
      q_1} \right) \; . \label{S2m}
\end{align}

The term in the square bracket on the right-hand side of Eq.~(\ref{S2}) vanishes in the case of massless hard partons, whereas $\cS^{m=0}_{ij}$ contributes in the cases of both massless and massive hard partons. We note however that the complete result in Eq.~(\ref{S2}) entirely comes from squaring the current $\J(q_1,q_2)$ in 
Eqs.~(\ref{J12})--(\ref{gammaC}), since that expression of the current is equally valid for both massless and massive hard partons. The result in Eq.~(\ref{S2m0})
was obtained in Ref.~\cite{CaGr99} by squaring $\J(q_1,q_2)$ in the case of massless
hard partons. The result in Eq.~(\ref{S2m}) was obtained in Ref.~\cite{Czakon:2011ve}
by squaring the {\em same} current, without neglecting terms with explicit dependence on $m_i^2$. We also note that the expression of $\cS^{m\neq0}_{ij}$ in Eq.~(B.9) of 
Ref.~\cite{Czakon:2011ve} differs from our explicit expression in Eq.~(\ref{S2m}).
The difference is due to the first term in the right-hand side of Eq.~(B.9) in 
Ref.~\cite{Czakon:2011ve}. Owing to colour conservation (see Eq.~(\ref{colcons})),
this term gives a vanishing total contribution to the action of the squared current 
$| \J(q_1,q_2) |^2$ in Eqs.~(\ref{J2sq}) and (\ref{corr2})
onto the (colour singlet) scattering amplitude $\M(\{p_i\})$. Therefore we have not included this harmless term in the right-hand side of Eq.~(\ref{S2m}).

We note that the squared current $| \J(q_1,q_2) |^2$ at the tree level has an explicit dependence on dimensional regularization through the $\epsilon$-dependent part of the first term in the right-hand side of Eq.~(\ref{S2m0}). This result is valid in the context of CDR, which we use throughout the paper.

As noticed at the end of Sect.~\ref{s:sc3}, the tree-level current 
$J^{a_1 a_2}_{\mu_1 \mu_2}(q_1,q_2)$ has no explicit dependence on $\epsilon$, whereas
the soft-gluon polarization vectors 
$\e_{(\sigma_\ell)}^{\mu_\ell}(q_\ell)$ in Eq.~(\ref{colspinsoft})
do depend on the specific dimensional regularization prescription that is actually used. It turns out \cite{CaGr99} that the $\epsilon$ dependence in Eq.~(\ref{S2m0})
derives from the fact that CDR uses $h_g=d-2=2(1-\epsilon)$ spin polarization states 
$h_g$ for the on-shell soft gluons. Other versions of dimensional regularizations
\cite{Gnendiger:2017pys}, such as 
dimensional reduction (DR) \cite{Siegel:1979wq}
and the four-dimensional helicity (4DH) scheme \cite{Bern:1991aq}, use $h_g=2$
spin polarization states. The result for $| \J(q_1,q_2) |^2$ in the DR and 4DH
schemes is obtained by simply setting $\epsilon = 0$ in the right-hand side of  
Eq.~(\ref{S2m0}).

Owing to colour conservation of the scattering amplitudes, the single-gluon and double-gluon contributions $W^{(1)}(q_1)$ and $W^{(2)}(q_1,q_2)$ to the squared current can be expressed in various equivalent forms. We find it convenient to rewrite them as follows
\begin{equation}
\label{w1new}
  W^{(1)}(q_1) \eqcs - \sum_{i,k} \T_i\cdot \T_k \;\frac{1}{2} \;\w_{ik}^{(1)}(q_1) \;\;,
\end{equation}
\begin{equation}
\label{w2new}
  W^{(2)}(q_1,q_2) \eqcs - \,C_A \sum_{i,k} \T_i\cdot \T_k \;\frac{1}{2} \; 
\w_{ik}^{(2)}(q_1,q_2) \;\;,
\end{equation}
where
\beq
\label{wikdef}
\w_{ik}^{(N)}(q_1,\cdots,q_N) \equiv 
{\cal S}_{ik}(q_1,\cdots,q_N) + {\cal S}_{ki}(q_1,\cdots,q_N)
- {\cal S}_{ii}(q_1,\cdots,q_N) - {\cal S}_{kk}(q_1,\cdots,q_N) \, ,
\eeq
with $N=1,2$. Note, in particular, that from Eq.~(\ref{S1}) we have
\beq
\label{w1}
\w_{ik}^{(1)}(q) = - j_{ik}^2(q) \; ,
\eeq
where $j_{ik}^\mu(q_\ell)$ is the conserved eikonal current 
($q_\mu j_{ik}^\mu(q_\ell) =0$) for radiation of the soft-gluon momentum $q$
from the harder partons $i$ and $k$:
\beq
\label{conseik}
j_{ik}^\mu(q_\ell) \equiv j_{i}^\mu(q_\ell) - j_{k}^\mu(q_\ell) = 
\frac{p_i^\mu}{p_i \cdot q_\ell} - \frac{p_k^\mu}{p_k \cdot q_\ell} \;\;.
\eeq
The equivalence between Eqs.~(\ref{J1sq})-(\ref{w1def}) and 
Eqs.~(\ref{w1new}) and (\ref{w2new}) simply follows from using colour conservation
(see Eq.~(\ref{csnotation})) and the definition of $\w_{ik}^{(N)}$ in 
Eq.~(\ref{wikdef}). The expressions in Eqs.~(\ref{w1new}) and (\ref{w2new})
have a more direct physical interpretation since the momentum-dependent functions
$\w_{ik}^{(N)}$ are straightforwardly related (see Eqs.~(\ref{jbc1g}) and 
(\ref{jbc2g}) in Sect.~\ref{sec:2hard}) to soft-gluon emission from two hard partons, $i$ and $k$, in a colour singlet configuration.

The most singular contributions due to double-soft gluon radiation originate from the region where the energies $E_1$ and $E_2$ of the two soft gluons are strongly ordered
($E_1 \ll E_2$ or $E_2 \ll E_1$).
The energy strong-ordering limit of $W^{(2)}(q_1,q_2)$ can be straightforwardly evaluated from the explicit expressions in Eqs.~(\ref{S2})--(\ref{S2m}), and we present the ensuing result for the function $\w_{ik}^{(2)}(q_1,q_2)$
in Eqs.~(\ref{w2new}) and (\ref{wikdef}). Considering the region where
$E_1 \ll E_2$,
we obtain the following compact form:
\beq
\label{w2som}
\w_{ik}^{(2)}(q_1,q_2) = j_{ik}^2(q_2) \;\;j_{i2}(q_1) \cdot j_{k2}(q_1) \;\;,
\quad \quad \quad \quad (E_1 \ll E_2) \;,
\eeq
where the conserved current $j_{i{\ell'}}(q_\ell)$ is obtained from $j_{ik}(q_\ell)$
in Eq.~(\ref{conseik}) through the replacement $p_k \to q_{\ell'}$
(i.e., $j_{i{\ell'}}(q_\ell)$ generalizes the hard-parton current $j_{ik}(q_\ell)$
to the emission of the soft gluon $q_\ell$ from the `harder' soft gluon $q_{\ell'}$).
Although $\w_{ik}^{(2)}(q_1,q_2)$ is symmetric with respect to the exchange 
$q_1 \leftrightarrow q_2$, we note that the right-hand side of Eq.~(\ref{w2som}) 
is not symmetric (the original symmetry is broken by the asymmetric constraint
$E_1 \ll E_2$). The expression in Eq.~(\ref{w2som}) is valid for both massive and massless hard partons. In the case of massless hard partons ($p_i^2=p_k^2=0$),
the symmetry with respect to $q_1 \leftrightarrow q_2$ is remarkably recovered
(as also noticed in Ref.~\cite{CaGr99}), and from Eq.~(\ref{w2som}) we obtain
(see also Eq.~(111) in Ref.~\cite{CaGr99})
\beq
\label{w2som0}
\w_{ik}^{(2) s.o.}\!(q_1,q_2) \!=\!
\left( \frac{2 p_i \cdot p_k}{q_1 \cdot q_2 \,p_i \cdot q_1 \,p_k \cdot q_2} + \bigl( 1 \leftrightarrow 2 \bigr)\right) - \frac{2 (p_i \cdot p_k)^2}{p_i \cdot q_1 \,p_k \cdot q_1 \,p_i \cdot q_2 \,p_k \cdot q_2} ,\;(p_i^2=p_k^2=0) ,
\eeq
where we have introduced the superscript $s.o.$ to explicitly denote the energy strong-ordering limit (either $E_1 \ll E_2$ or $E_2 \ll E_1$) of $\w_{ik}^{(2)}$ for the case of massless hard partons.

\section{Tree-level squared current for triple gluon emission\label{s:sq123}}

\subsection{General structure\label{s:cor123}}

We have computed the square of the three-gluon current $\J(q_1,q_2,q_3)$ in 
Eq.~(\ref{J123}). Performing straightforward (though quite cumbersome) algebraic manipulations, we are able to express the result in the following form:
\begin{align}
\label{J3sq}
| \J(q_1,q_2,q_3) |^2 &\;\eqcs \;\sym{W^{(1)}(q_1)}{W^{(1)}(q_2)}{W^{(1)}(q_3)} \nn \\
&\;+\;\left[ \sy{W^{(1)}(q_1)}{W^{(2)}(q_2,q_3)} 
+ \;\left( 1 \leftrightarrow 2 \right) + \left( 1 \leftrightarrow 3 \right) 
\right] \nn\\
&\;+ \;W^{(3)}(q_1,q_2,q_3) \;\;,
\end{align}
where we have used the symmetrized products of two and three colour operators
(see Eqs.~(\ref{sym2}) and (\ref{sym3})).
The factor $W^{(1)}(q_\ell)$ is the square of the single-gluon current in 
Eq.~(\ref{w1def}), and the factor $W^{(2)}(q_\ell,q_{\ell^\prime})$ is the two-gluon irreducible correlation that contributes to the square of the double-gluon current in
Eq.~(\ref{J2sq}). The form of Eq.~(\ref{J3sq}) generalizes the correlation structure of Eq.~(\ref{J2sq}) to the case of triple soft-gluon emission. The symmetrized products in the first two lines in the right-hand side of Eq.~(\ref{J3sq}) correspond to the iteration of single-emission and double-correlation contributions, and the term
$W^{(3)}(q_1,q_2,q_3)$ in the third line is an irreducible-correlation contribution for triple gluon emission.

The squared current $| \J(q_1,q_2,q_3) |^2$ and the terms $W^{(1)}(q_\ell)$ and
$W^{(2)}(q_\ell,q_{\ell^\prime})$ are separately gauge invariant when acting onto colour-singlet states (scattering amplitudes) and, therefore, the correlation
$W^{(3)}(q_1,q_2,q_3)$ in Eq.~(\ref{J3sq}) is also gauge invariant.

The single-emission and double-correlation terms $W^{(1)}$ and $W^{(2)}$ embody colour correlations with the hard partons in the scattering amplitude $\M(\{p_i\})$.
These colour correlations are given in terms of colour dipole operators
(see Eqs.~(\ref{J1sq}) and (\ref{corr2})).
The three-gluon correlation in Eq.~(\ref{J3sq}) also involves colour correlations. A new distinctive feature of triple-gluon radiation is that $W^{(3)}(q_1,q_2,q_3)$
cannot 
simply be
expressed in terms of colour dipole operators. We find that $W^{(3)}$ receives contributions from both dipole and {\em quadrupole} 
operators \footnote{Colour quadrupole correlations explicitly appear also in the IR divergent virtual contribution to multiparton scattering amplitudes at three-loop order
\cite{Almelid:2015jia}.}

Non-abelian colour quadrupole operators have the following typical structure:
\begin{equation}\label{Qdiff}
  \qu{imkl}^\prime = 
    2 f^{ab,cd} \, T_i^c\,  T_m^d \,T_k^a \,T_l^b  
  = 2 f^{ab,cd} \, T_k^a \, T_l^b \,T_i^c \,T_m^d  \;\;.
\end{equation}
They are obtained by the colour contraction (with respect to the colour indices of the soft gluons) of two structure constants $f$
(as defined in Eq.~\eqref{ff}) with four colour charges $\T_i$ of the hard partons.
In the notation of Eq.~(\ref{Qdiff}) the superscript `prime' in 
$\qu{imkl}^\prime$ means that we are considering the case with four distinct hard partons $\{ i, m, k, l \}$ (we are dealing with `true' quadrupoles).
In the case of four distinct hard partons the colour charges 
$\{ \T_i \,\T_m \,\T_k \,\T_l \}$ commute among themselves and their ordering in the 
right-hand side of Eq.~(\ref{Qdiff}) can be varied. If the four hard partons are not distinct (we are dealing with `pseudo' quadrupoles) the ordering of the four colour charges does matter, and the difference between various orderings can be related (by using colour algebra commutation relations) through colour dipole operators. Moreover, due to colour charge conservation (see Eq.~(\ref{csnotation})), `true' and `pseudo' quadrupoles are linearly-dependent related through their action onto the hard-parton scattering amplitude $\M(\{p_i\})$ and, therefore, any a priori distinctions between
`true' and `pseudo' quadrupoles require to break the manifest symmetry between
the hard partons.

In view of these features of colour quadrupole operators, a proper separation of dipole and quadrupole contributions to the irreducible three-gluon correlation 
$W^{(3)}(q_1,q_2,q_3)$ in Eq.~(\ref{J3sq}) requires a careful definition of quadrupole operators. We introduce the following quadrupole operator:
\begin{equation}\label{Qbase}
  \qu{imkl} \equiv 
  \frac12 f^{ab,cd} \,\left( T_k^a \{ T_i^c, T_m^d \} T_l^b + \text{h.c.} \right) \;,
\end{equation}
where the text `h.c.' means hermitian conjugate. Note that the definition in 
Eq.~(\ref{Qbase}) applies to an arbitrary set $\{i, m, k, l\}$ of hard-parton indices, with no distinction between `true' and `pseudo' quadrupoles. If the four indices are distinct, the two operators in Eqs.~(\ref{Qdiff}) and (\ref{Qbase}) are equal. As discussed in Appendix~\ref{a:quad}, the hermitian conjugate
operators in Eq.~(\ref{Qbase})
are `irreducible' quadrupole operators, in the sense that their action onto 
colour-singlet states (scattering amplitudes) do not produce colour dipole 
contributions of the type $C_A^2 \T_i\cdot \T_k$
if two or more of the indices $\{i, m, k, l\}$ are equal.

Using the quadrupole operator of Eq.~(\ref{Qbase}), the three-gluon correlation
term 
$W^{(3)}$
in Eq.~(\ref{J3sq}) can be written as 
\beq
\label{3cordq}
W^{(3)}(q_1,q_2,q_3) = W^{(3) {\rm dip.}}(q_1,q_2,q_3) + 
                       W^{(3) {\rm quad.}}(q_1,q_2,q_3) \;\;,
\eeq
where the dipole ($W^{(3) {\rm dip.}}$) and quadrupole ($W^{(3) {\rm quad.}}$)
components are
\beq
\label{3cordip}
W^{(3) {\rm dip.}}(q_1,q_2,q_3) = - \;C_A^2 \,\sum_{i,k} \T_i\cdot \T_k \; 
\cS_{ik}(q_1,q_2,q_3) \;\;,
\eeq
\beq
\label{3corquad}
W^{(3) {\rm quad.}}(q_1,q_2,q_3) = \sum_{i,m,k,l}  \;\qu{imkl} \;
\cS_{imkl}(q_1,q_2,q_3) \;\;.
\eeq
The kinematical functions $\cS_{ik}(q_1,q_2,q_3)$ and $\cS_{imkl}(q_1,q_2,q_3)$
depend on the momenta of the soft and hard partons, and their 
expressions are presented below (see Eqs.~(\ref{osik3g}) and 
(\ref{simkl})).
Note that the dipole contribution $W^{(3) {\rm dip.}}$ in Eq.~(\ref{3cordip})
is maximally non-abelian (at this perturbative order), since it is proportional to 
$C_A^2$.
The three-gluon correlation $W^{(3)}$ is gauge invariant, and we have explicitly checked that its components
$W^{(3) {\rm dip.}}$ and $W^{(3) {\rm quad.}}$ are {\em separately} gauge invariant.
This property is expected since the quadrupole operators in Eq.~(\ref{Qbase})
are `irreducible' to dipole operators of the type $C_A^2 \,\T_i\cdot \T_k$.


\setcounter{footnote}{2}

\subsection{Dipole correlation \label{s:cor123dip}}

The dependence of the dipole correlation $W^{(3) {\rm dip.}}$ on the momenta of the soft and hard partons is given by the function $\cS_{ik}(q_1,q_2,q_3)$ through
Eq.~(\ref{3cordip}). This function has a very cumbersome algebraic form.
We can present a shortened form by using the current $j_i(q_\ell)$ 
in Eq.~(\ref{j1}) 
and the momentum functions $\gamma_i(q_1,q_2)$ and $\gamma_i(q_1,q_2,q_3)$ 
in Eqs.~(\ref{gammaC}) and~(\ref{gamma123}). 
We have
\beeq
\label{osik3g}
\cS_{ik}(q_1,q_2,q_3) &=& \!\! \Bigl\{ \,\frac{1}{2} \,\gamma_k(1,2;3) \bigl[ 
\gamma_i(1,2;3) + \gamma_i(1,3;2)
+ \gamma_i(1,2) j_k(3) - \gamma_k(1,2) j_i(3) \bigr. \Bigr. \nn \\
\bigl. &+& \gamma_i(1,3) j_k(2) - \gamma_k(1,3) j_i(2) 
+ \frac{1}{2} \,j_k(1) j_i(2) \left(j_i(3) + j_k(3)\right) \bigr] \nn \\
&+& \frac{1}{2} \,\gamma_i(1,2) \bigl\{ 
\gamma_k(1,2) \bigl[ \;\frac{3}{4} j_i(3) j_k(3) - \frac{1}{2} j_i(3) j_i(3) \;\bigr]
- \frac{1}{2}\,\gamma_i(1,2) j_i(3) j_k(3) \bigr. \nn \\
&+& \frac{1}{4} \,\gamma_k(1,3) 
\bigl[ \;  j_k(2) j_i(3) + 2 j_i(2) j_k(3) - 2 j_i(2) j_i(3)\;\bigr]
- \frac{1}{2}\,\gamma_i(1,3) j_k(2) j_i(3) \bigr. \nn \\
&+& j_i(1) j_k(2)
\bigl[ \; 
\frac{7}{4} j_i(3) j_k(3)- \frac{3}{4} j_i(3) j_i(3) 
- \frac{1}{2} j_k(3) j_k(3) \;\bigr] \bigr\} \nn \\
&+& \frac{1}{2}\,j_i(1) j_i(2) j_i(3) j_k(3)
\bigl[ \; \frac{1}{3} j_i(1) j_i(2) -\frac{5}{6} j_i(1) j_k(2)+ 
\frac{31}{72} j_k(1) j_k(2) \;\bigr] \nn \\
&+& \Bigl. \frac{1}{16} \,j_i^2(1) j_i(2) j_k(2) j_k^2(3) \, \Bigr\} 
+ {\rm perms.}\{1,2,3\}\;\;,
\eeeq
where in the right-hand side the products of $j_i$ and $\gamma_i$ functions
are denoted by using the shorthand notation of Eq.~(\ref{fgnotation}).

The explicit form of the dipole correlation $W^{(3) {\rm dip.}}$ in terms of scalar products between the momenta of the hard partons and soft gluons can be obtained by inserting Eqs.~(\ref{gammaC}), (\ref{gamma123}) and (\ref{j1}) in Eq.~(\ref{osik3g}).
The ensuing algebraic expression is very long, and it is reported in 
Appendix~\ref{a:ee}.

As noticed at the end of Sect.~\ref{s:sc3}, the current $\J(q_1,q_2,q_3)$
has no explicit dependence on the dimensional regularization procedure and on
$d=4 - 2\epsilon$. Nonetheless, analogously to $| \J(q_1,q_2) |^2$
(see the comments in the paragraph above Eq.~(\ref{w1new}))), the squared
current $| \J(q_1,q_2,q_3) |^2$
does depend on the regularization procedure and, in particular, 
the correlation $W^{(3) {\rm dip.}}$ (and the right-hand side of Eq.~(\ref{osik3g}))
turns out to depend on $\epsilon$. Such dependence is explicitly shown in 
Appendix~\ref{a:ee},
where we also comment on the result of $| \J(q_1,q_2,q_3) |^2$
in various dimensional regularization schemes.

As in the case of the single and double gluon correlations $W^{(1)}(q_1$ and
$W^{(2)}(q_1,q_2)$ (see Eqs.~(\ref{w1new}) and (\ref{w2new})), we can exploit colour conservation to rewrite the dipole contribution
$W^{(3) {\rm dip.}}$
of Eq.~(\ref{3cordip}) in the following equivalent form:
\beq
\label{w3new}
W^{(3) {\rm dip.}}(q_1,q_2,q_3) \eqcs - \;C_A^2 \,\sum_{i,k} \T_i\cdot \T_k 
\;\frac{1}{2} \;\w_{ik}^{(3)}(q_1,q_2,q_3) \;\;,
\eeq
where the momentum dependent function $\w_{ik}^{(3)}$ is defined in terms of 
${\cal S}_{ik}(q_1,q_2,q_3)$ through Eq.~(\ref{wikdef}) with $N=3\,$:
\beq
\label{wikdef3}
\w_{ik}^{(3)}(q_1,q_2,q_3) \equiv 
{\cal S}_{ik}(q_1,q_2,q_3) + {\cal S}_{ki}(q_1,q_2,q_3)
- {\cal S}_{ii}(q_1,q_2,q_3) - {\cal S}_{kk}(q_1,q_2,q_3) \; .
\eeq

It is of interest to examine the momentum dependence of 
$W^{(3) {\rm dip.}}(q_1,q_2,q_3)$ in various energy ordered regions, since these regions contribute to enhanced singular contributions 
from triple soft-gluon emission.

In the energy ordered region where both $q_1$ and $q_2$ are much softer than 
$q_3$, the explicit expression of $\w_{ik}^{(3)}$ (or, equivalently,
${\cal S}_{ik}(q_1,q_2,q_3)$) is much simplified, since we obtain
\beeq
\label{w312}
\w_{ik}^{(3)}(q_1,q_2,q_3) &=& - j_{ik}^2(q_3) 
\Bigl[ \; \frac{1}{2} \left( \w_{3i}^{(2)}(q_1,q_2) + \w_{3k}^{(2)}(q_1,q_2)
- \w_{ik}^{(2)}(q_1,q_2) \right) \Bigr. \nn \\
&+& \Bigl. j_{i3}(q_1) \cdot j_{k3}(q_1) \;\;j_{i3}(q_2) \cdot j_{k3}(q_2)
\Bigr] \;,
\quad (E_1 \ll E_3, \, E_2 \ll E_3) \,,
\eeeq
where the conserved currents $j_{ik}(q_l)$ and $j_{i3}(q_l)$ are defined in 
Eq.~(\ref{conseik}) (see also Eq.~(\ref{w2som})). The function 
$\w_{ik}^{(2)}(q_1,q_2)$ on the right-hand side is the exact (i.e., without any energy ordering approximation) double-gluon correlation in Eq.~(\ref{wikdef}),
while $\w_{3k}^{(2)}$ is simply obtained from $\w_{ik}^{(2)}$ through the momentum replacement $p_i \rightarrow q_3$. Note that the right-hand side of Eq.~(\ref{w312})
is obviously symmetric with respect to $q_1 \leftrightarrow q_2$, whereas the symmetry $q_1 \leftrightarrow q_3$ (or $q_2 \leftrightarrow q_3$) is broken by the energy ordering constraint.

The function $\w_{ik}^{(3)}(q_1,q_2,q_3)$ in the region of energy strong ordering
($E_1 \ll E_2 \ll E_3$) can be directly obtained by using Eq.~(\ref{w312})
and the known behaviour of $\w^{(2)}(q_1,q_2)$ for $E_1 \ll E_2$ 
(see Eq.~(\ref{w2som})). For comparison with the expressions derived long ago 
in Ref.~\cite{CaCi84} (see Eq.~(\ref{jbc3g}) and related comments 
in Sect.~\ref{sec:3from2}),
we write the energy strong-ordering limit of $\w_{ik}^{(3)}$ in the following form:
\beeq
\label{w3som}
&& \!\!\!\!\!\!\!\! \!\!\!\!\w_{ik}^{(3)}(q_1,q_2,q_3) = - j_{ik}^2(q_3) 
\Bigl[ \; j_{i3}(q_2) \cdot j_{k3}(q_2) 
\Bigl( j_{i3}(q_1) \cdot j_{k3}(q_1) + j_{i2}(q_1) \cdot j_{k2}(q_1)\Bigr)
\Bigr. \nn \\
&&\!\!\!\!  - \Bigl. \frac{1}{2} \,j_{k3}^2(q_2) \;j_{i3}(q_1) \cdot j_{k2}(q_1)
-\frac{1}{2} \,j_{i3}^2(q_2) \;j_{k3}(q_1) \cdot j_{i2}(q_1)
\Bigr] \,,
\quad  (E_1 \ll E_2 \ll E_3) \,.
\eeeq
This expression, which is not symmetric under the exchange of the soft-gluon momenta,
is valid for both massive and massless hard partons. 
In the case of massless hard partons ($p_i^2=p_k^2=0$) the right-hand side of 
Eq.~(\ref{w3som})
remarkably turns out to be symmetric, and we obtain
\beeq
\label{w3som0}
&&\!\!\!\!\!\!\!\! \!\!\!\!
\w_{ik}^{(3) s.o.}\!(q_1,q_2,q_3) = \Bigl[ \;
\frac{2 (p_i \cdot p_k)^3}{3 (p_i \cdot q_1) (p_k \cdot q_1) 
(p_i \cdot q_2) (p_k \cdot q_2) (p_i \cdot q_3) (p_k \cdot q_3)} \Bigr. \nn \\
&& 
- \Bigl. \frac{2 (p_i \cdot p_k)^2}{(q_1 \cdot q_2) (p_i \cdot q_1) (p_k \cdot q_2) 
(p_i \cdot q_3) (p_k \cdot q_3)} \Bigr. \nn \\
&& + \Bigl. \frac{2 p_i \cdot p_k}{(q_1 \cdot q_3) (q_2 \cdot q_3) (p_i \cdot q_1) (p_k \cdot q_2)} \;
\Bigr] + {\rm perms.} \,\{1,2,3\} \;\;,
\quad \;\;(p_i^2=p_k^2=0) \;,
\eeeq
where, analogously to Eq.~(\ref{w2som0}), the superscript $s.o.$ explicitly denotes
the energy strong-ordering limit (i.e., $E_1 \ll E_2 \ll E_3$ or any other permutations of the three soft-gluon energies) of $\w_{ik}^{(3)}$ for massless hard partons.

We note that Eqs.~(\ref{w2som0}) and (\ref{w3som0}) can be rewritten in the following form:
\beeq
\label{w2bcm}
&& \!\!\!\! \!\!\!\!\!\!
\w_{ik}^{(2) s.o.}\!(q_1,q_2) + \w_{ik}^{(1)}(q_1) \,\w_{ik}^{(1)}(q_2)  \nn \\
&& \!\!\!\!\!\! 
= 2 (p_i \cdot p_k)^2 \Bigl[ \;
 \frac{1}{(p_i \cdot q_1) (q_1 \cdot q_2) (q_2 \cdot p_k) (p_k \cdot p_i)} 
+ {\rm ineq. \;perms.\;} 
\{p_i,p_k,q_1,q_2\}
\Bigr] \,,
\eeeq
\beeq
\label{w3bcm}
&& \!\!\!\! \!\!\!\!\!
\w_{ik}^{(3) s.o.}\!(q_1,q_2,q_3) + \!\bigl[\w_{ik}^{(1)}(q_1) 
\;\w_{ik}^{(2) s.o.}\!(q_2,q_3) + (1 \leftrightarrow 2) + (1 \leftrightarrow 3)
\bigr]\! + \w_{ik}^{(1)}(q_1) \,\w_{ik}^{(1)}(q_2) \,\w_{ik}^{(1)}(q_3) \nn \\
&& \!\!\!\! \!\!\!\!\! 
= 2 (p_i \cdot p_k)^2 \Bigl[ \;
 \frac{1}{(p_i \cdot q_1) (q_1 \cdot q_2) (q_2 \cdot q_3) (q_3 \cdot p_k) 
(p_k \cdot p_i)} + {\rm ineq. \;perms.\;} 
\{p_i,p_k,q_1,q_2,q_3\}
\Bigr] \,. \nn \\
&&{}
\eeeq
The relations in Eq.~(\ref{w2bcm}) and (\ref{w3bcm}) have an analogous structure.
In both equations the left-hand side has the structure of an expansion in irreducible soft-gluon correlations. In both equations the right-hand side includes a closed chain
of eikonal propagators 
$\left[ (p_i \cdot q_1) (q_1 \cdot q_2) \dots (q_N \cdot p_k) 
(p_k \cdot p_i) \right]^{-1}$
and a sum over all the permutations of the hard and soft momenta that lead to inequivalent chains (the total number of inequivalent permutations is 3 and 12 in
Eq.~(\ref{w2bcm}) and (\ref{w3bcm}), respectively).
The structure of Eqs.~(\ref{w2bcm}) and (\ref{w3bcm}) can eventually be related 
(see Eq.~(\ref{bcm}) and accompanying comments in Sect.~\ref{sec:3from2})
to a result obtained by Bassetto, Ciafaloni and Marchesini (BCM) \cite{Bassetto:1984ik}
for multiple soft-gluon radiation in scattering amplitudes with two hard gluons.
Since the soft functions $\w_{ik}^{(1)},  \w_{ik}^{(2)}$ and $\w_{ik}^{(3)}$
control the colour dipole correlations of the squared soft current
for arbitrary scattering amplitudes, the relations in Eq.~(\ref{w2bcm}) and 
(\ref{w3bcm}) can be regarded as generalized BCM formulae.

\setcounter{footnote}{2}

\subsection{Quadrupole correlation \label{s:cor123quad}}

The quadrupole component, $W^{(3) {\rm quad.}}$, of the three-gluon irreducible correlation $W^{(3)}$ (see Eqs.~(\ref{3cordq}) and (\ref{3corquad}))
is expressed in terms of the quadrupole operators $\qu{imkl}$ in Eq.~(\ref{Qbase}).
These colour operators fulfil some relevant properties (see Appendix~\ref{a:quad})
with respect to the dependence on the hard-parton indices $\{i,m,k,l\}$.
Some properties regards symmetries. The operator $\qu{imkl}$ is
\begin{itemize}
\item antisymmetric with respect to the exchange of the two adjacent indices in the first $(i \leftrightarrow m)$ or second $(k \leftrightarrow l)$ pair of indices:
\beq
\label{qant}
\qu{imkl} = - \,\qu{mikl} \;\;, \quad \quad \qu{imkl} = - \,\qu{imlk} \;\;,
\eeq
\item symmetric with respect to the exchange of the first and second pair
$(im \leftrightarrow kl)$ of indices:
\beq
\label{qsym}
\qu{imkl} =  \qu{klim} \;\;,
\eeq
\end{itemize} 
and, moreover, it fulfils the following Jacobi identity:
\beq
\label{qjac}
\qu{i k_1k_2k_3} +  \qu{i k_3k_1k_2} + \qu{i k_2k_3k_1} = 0 \;\;.
\eeq
Note that the symmetry properties in Eqs.~(\ref{qant})--(\ref{qjac}) are exactly analogous to those of the structure constant term $f^{ab,cd}$ in Eq.~(\ref{ff})
(through the replacement $abcd  \leftrightarrow imkl$).
Owing to Eq.~(\ref{csnotation}), the quadrupole operators also fulfil the following colour charge conservation relation:
\beq
\label{qcons}
\sum_i \qu{imkl} \; \eqcs \; 0 \;\;.
\eeq
As a consequence of Eqs.~(\ref{qant}) and (\ref{qsym}), Eq.~(\ref{qcons}) is valid also by replacing the sum over the first quadrupole index $i$ with the sum over any other of the indices $m,k$ and $l$ of the quadrupole.

A specific, though relevant, feature of the quadrupole operators in Eq.~(\ref{Qbase})
is that they vanish in the cases of colour singlet states of only two hard partons:
\beq
\label{q2hard}
\qu{imkl} \; \eqcs \; 0 \;\;\quad \quad ({\rm 2 \;hard \;partons}) \;\;.
\eeq
This implies that the quadrupole correlation $W^{(3) {\rm quad.}}$ does not contribute
to the soft limit of scattering amplitudes $\M(\{q_\ell\},\{p_i\})$ with three soft gluons and two hard partons. The result in Eq.~(\ref{q2hard}) is a straightforward
consequence of Eq.~(\ref{qant})--(\ref{qcons}). In particular, if $i$ and $m$ are the two hard partons $(i\neq m)$, Eq.~(\ref{qant}) implies that all non-vanishing quadrupole operators are simply proportional (through an overall sign) to 
$\qu{imim}$. Then, Eq.~(\ref{qcons}) gives $\qu{imim} \eqcs - \qu{mmim}$ and,
therefore, $\qu{imim} \eqcs 0$ since $\qu{mmim}=0$ (because of the antisymmetry in
Eq.~(\ref{qant})).

We note (see also Appendix~\ref{a:quad}) that the symmetry properties in 
Eqs.~(\ref{qant}) and (\ref{qsym}) imply that the non-vanishing quadrupole operators 
$\qu{imkl}$ can only have two pairs of distinct indices, three distinct indices or four distinct indices. Moreover, the colour conservation relation in Eq.~(\ref{qcons})
implies that quadrupoles with two pairs of distinct indices can be replaced by
a linear combination of quadrupoles with three distinct indices when acting onto a colour singlet state (see Eq.~(\ref{2qvs3q}) in Appendix~\ref{a:quad}). As a consequence, the action of the correlation $W^{(3) {\rm quad.}}$ in 
Eq.~(\ref{3corquad}) onto a scattering amplitude $\ket{\M(\{p_i\})}$ can also be expressed in terms of an equivalent sum\footnote{Note that this equivalent sum is not equal to the restriction of the sum $\sum_{i,m,k,l}$ in Eq.~(\ref{3corquad}) to the case of three and four distinct indices. Indeed, the correct restricted sum involves kinematical functions that are linear combinations of the functions $\cS_{imkl}$ in
Eq.~(\ref{3corquad}).} of colour quadrupole contributions with three or four distinct indices.

The dependence of the quadrupole correlation $W^{(3) {\rm quad.}}$ on the momenta of the soft and hard partons is given by the function $\cS_{imkl}(q_1,q_2,q_3)$ through
Eq.~(\ref{3corquad}). This function is slightly less cumbersome than the three-gluon dipole correlation function $\cS_{ik}(q_1,q_2,q_3)$. Expressing the result in terms of the current $j_i(q_\ell)$ 
(see Eq.~(\ref{j1})) 
and the functions $\gamma_i(q_1,q_2)$ and $\gamma_i(q_1,q_2,q_3)$ 
(see Eqs.~(\ref{gammaC}) and (\ref{gamma123})), we find
\beeq
\label{simkl}
\cS_{imkl}(q_1,q_2,q_3) &=& \Bigl\{
\big[ \;\frac{1}{2} \,\gamma_i(1,2;3) j_k(1) 
       - \frac{7}{24} \,j_i(1) j_k(2) j_m(1) j_l(3)  \;\big] j_l(2) j_m(3) 
\Bigr. \nn \\
&-& \gamma_i(1,2) \bigl[\; \frac{1}{2} \,j_m(1) j_k(2) j_k(3) j_l(3)
   + \frac{1}{4} \,j_l(1) j_k(2) j_i(3) j_m(3) \bigr. \nn \\
\Bigl. \bigl.  &+&  \frac{1}{2} \,\gamma_k(1,3) j_m(2) j_l(3)
\bigr] \Bigr\} + {\rm perms.}\{1,2,3\} \;\;,
\eeeq
where in the right-hand side the products of $j_i$ and $\gamma_i$ functions
are denoted by using the shorthand notation of Eq.~(\ref{fgnotation}).

We note that the function $\cS_{imkl}(q_1,q_2,q_3)$ is fully symmetric with respect to the exchange of soft-gluon momenta. This symmetry directly follows from the sum
over the six permutations in the right-hand side of Eq.~(\ref{simkl}).
We also note that the expression of  $\cS_{imkl}$ in Eq.~(\ref{simkl}) does not have the same symmetry properties as those of the quadrupole $\qu{imkl}$ with respect
to the hard-parton indices $\{i,m,k,l\}$. However, the correct hard-parton symmetries
of $W^{(3) {\rm quad.}}$ are automatically obtained by simply inserting 
Eq.~(\ref{simkl}) in the right-hand side of Eq.~(\ref{3corquad}). For instance, the symmetric part of $\cS_{imkl}$ with respect to the exchange $i \leftrightarrow m$
gives a vanishing contribution to $W^{(3) {\rm quad.}}$ because of Eq.~(\ref{qant}).
We have not `properly' symmetrized the expression of $\cS_{imkl}$
with respect to the indices $\{i,m,k,l\}$ since this produces many more terms that are actually harmless to the purpose of evaluating the relevant correlation term
$W^{(3) {\rm quad.}}$.

Using the expressions of $\gamma_i(q_1,q_2)$ and $\gamma_i(q_1,q_2,q_3)$ 
in Eqs.~(\ref{gammaC}) and (\ref{gamma123}), the function $\cS_{imkl}(q_1,q_2,q_3)$
in Eq.~(\ref{simkl}) can be explicitly written in terms of scalar products of hard and soft momenta. Such explicit expression is given in 
Appendix~\ref{a:ee}.
Here, we limit ourselves to presenting a much simpler approximated form that is valid in the energy strong-ordering region where $E_1 \ll E_2 \ll E_3$. We find
\beeq
\label{simklso}
&&\!\!\!\!\!\!\!\!\!
\!\!\!\!\!\!\!\!\!\!\!\!
\cS_{imkl}^{(1 < 2 < 3)}(q_1,q_2,q_3) = 
\frac{p_i \cdot p_k}{(p_i \cdot q_3) (p_k \cdot q_3)(p_l \cdot q_1) (p_m \cdot q_2)}\Bigl\{ 
\frac{(p_m \cdot p_l)(p_l \cdot p_i)}{3(p_l \cdot q_2)(p_i \cdot q_1)}
\Bigr. \nn \\
&& \!\!\!\!\!\!\!\!\!\!\!
\!\!\!\!\!\!\!\!
- \,\frac{p_m \cdot p_i}{3 p_i \cdot q_2}
\left[ \frac{p_l \cdot p_i}{p_i \cdot q_1} + \frac{p_l \cdot p_m}{p_m \cdot q_1}
+ \frac{2 p_l \cdot p_k}{p_k \cdot q_1}\right] 
+
 \Bigl.   \frac{p_l \cdot q_3}{q_3 \cdot q_1}
\left[ \frac{2 p_m \cdot p_i}{p_i \cdot q_2} 
- \frac{p_m \cdot q_3}{q_3 \cdot q_2} \right]
 \Bigr\} + ( 1 \leftrightarrow 2 ) \;, 
\eeeq
and we observe that such expression is remarkably symmetric with respect to the exchange $q_1 \leftrightarrow q_2$. We also specify that Eq.~(\ref{simklso})
is not algebraically identical to the energy strong-ordering limit of 
Eq.~(\ref{simkl}), but the difference is physically harmless since it does not contribute to $W^{(3) {\rm quad.}}$ in Eq.~(\ref{3corquad}).
In other words, we have simplified the right-hand side of Eq.~(\ref{simklso})
by removing some terms that give a vanishing contribution to Eq.~(\ref{3corquad})
because of the symmetry properties in Eqs.~(\ref{qant})--(\ref{qcons}) of the quadrupole operators $\qu{imkl}$.

\subsection{Collinear singularities 
\label{s:coll}}

The squared amplitude $| \M(\{q_\ell\}, \{p_i\}) |^2$ has a singular behaviour in kinematical regions where the momenta of two or more of its external {\em massless} legs become collinear. Such singular behaviour is not integrable over the angular region in four space-time dimensions, and it leads to $\epsilon$ poles (collinear divergences) by integrating in $d=4-2\epsilon$ space-time dimensions.
Independently of the angular integration, the size of the squared amplitude
is strongly enhanced in the vicinity of the singular collinear regions.

In our subsequent discussion about presence or absence of singular collinear behaviour we always refer to singular terms that lead to divergences upon angular integrations in four dimensions (for a precise formal definition of the multiparton collinear limit see Refs.~\cite{CaGr99,Campbell:1997hg}). Subdominant, though possibly singular, collinear terms (e.g., integrable singularities in four dimensions) are not considered.

As a consequence of the soft-gluon factorization formula (\ref{softsquared}),
the collinear behaviour of $| \M(\{q_\ell\}, \{p_i\}) |^2$ leads to collinear singularities in the squared current $| \J(q_1,\cdots,q_N) |^2$ for multiple 
soft-gluon emission. In the following, as the simplest example, we discuss the singular collinear behaviour for single gluon emission with some details.
Then, we present a brief discussion of our results on the collinear behaviour for double and triple soft-gluon radiation at the tree level.

The single-gluon squared current $| \J(q) |^2$ in Eq.~(\ref{J1sq}) can lead to collinear singularities if the soft-gluon momentum $q$ is parallel to the momentum of a {\em massless} hard parton, which we denote as parton $C$ (with colour charge $\T_C$
and momentum $p_C$). The possible singularities originate from the soft function
${\cal S}_{ik}(q)$ in Eq.~(\ref{S1}) with $i \neq k$ (the function 
${\cal S}_{ii}(q)$ identically vanishes if the parton $i$ is massless).
We have
\beq
\label{scoll}
{\cal S}_{ik}(q) \simeq \frac{1}{z \; p_C \cdot q} 
(\delta_{iC} + \delta_{kC}) \;\;,  
\quad  \quad i \neq k \quad \quad (q \simeq z p_C) \;\;,
\eeq
where we have extracted the singular collinear factor $(p_C \cdot q)^{-1}$, used the collinear approximation $q \simeq z p_C$ in the remaining contribution, and neglected non-singular terms. Inserting Eq.~(\ref{scoll}) in Eq.~(\ref{J1sq}), we obtain
\beq
\label{w1coll}
| \J(q) |^2 \eqcs W^{(1)}(q) \simeq - \sum_{i\neq k} \T_i \cdot \T_k \;
\frac{1}{z \; p_C \cdot q} (\delta_{iC} + \delta_{kC}) = \frac{1}{p_C \cdot q} \;
\frac{2 C_C}{z} \;\;, \quad (q \simeq z p_C) \;,
\eeq
where we have used the colour conservation (see Eq.~(\ref{csnotation}))
in the form $\sum_{k\neq C } \T_k \eqcs - \T_C$.

The singular collinear behaviour of $W^{(1)}(q)$ is fully consistent with expectations. In particular, the explicit result in Eq.~(\ref{w1coll}) exactly agrees 
with the result that can be obtained by first evaluating 
$| \M(\{q_\ell\}, \{p_i\}) |^2$ in the collinear limit $q\simeq z p_C$ and then performing the soft limit $z \to 0$ (see, e.g., Eq.~(7) and Eqs.~(9)--(13) in 
Ref.~\cite{CaGr99}). This agreement is a consequence of the commutativity between the soft and collinear limits.

The squared current $| \J(q) |^2$ embodies colour correlations between the hard partons. Nonetheless, its singular collinear behaviour in Eq.~(\ref{w1coll}) 
is proportional to the Casimir coefficient $C_C$ of the hard collinear parton, with no accompanying colour correlation effects. This absence of colour correlations is a manifestation of colour coherence. 

One way to notice the colour coherence effect
is to observe that the Casimir coefficient $C_C$ in Eq.~(\ref{w1coll}) is obtained by using $\T_C \cdot \sum_{k\neq C } \T_k \eqcs - \T_C \cdot \T_C = - C_C$.
Although the soft-gluon emission from the collinear hard parton $C$ leads to dipole correlations of the type $\T_c \cdot \T_k$ with each of the non-collinear partons,
the collinear soft gluon feels the coherent action of the non-collinear partons.
The coherent action is proportional to their total colour charge 
$\sum_{k\neq C } \T_k$ or, equivalently (because of colour conservation) to the colour charge $\T_C \eqcs - \sum_{k\neq C } \T_k$ of the collinear hard parton.
This coherent action leads to the colour coefficient $\T_C \cdot \T_C= C_C$ in 
Eq.~(\ref{w1coll}).

The tree-level squared currents $| \J(q_1,q_2) |^2$ and $| \J(q_1,q_2,q_3) |^2$
are expressed (see \, Eqs.~(\ref{J2sq}) and (\ref{J3sq})) in terms of the 
single-emission factor $W^{(1)}$ and the irreducible correlation terms
$W^{(2)}$ and $W^{(3)}$.
Therefore, the knowledge of $W^{(1)}, W^{(2)}$ and $W^{(3)}$
in the various collinear regions gives the full information on the singular collinear behaviour for double and triple soft-gluon emission. The singular collinear behaviour of $W^{(1)}$ is given in Eq.~(\ref{w1coll}). The same steps and approximations that we have used to obtain Eq.~(\ref{w1coll}) can be applied to analyze the singular collinear behaviour of $W^{(2)}$ and $W^{(3)}$ from the results in 
Eqs.~(\ref{corr2}), (\ref{S2})--(\ref{S2m}), 
(\ref{3cordq})--(\ref{3corquad}), (\ref{osik3g}) and (\ref{simkl}).
We have explicitly carried out this analysis by considering all possible collinear configurations of soft and hard partons. The results of this analysis are summarized below.

We find that $W^{(2)}(q_1,q_2)$ has singular collinear behaviour in the following configurations:
\begin{itemize}
\item[] {($c_1$)\;} 
double-collinear limit of the two soft gluons,
\item[] {($c_2$)\;}  
triple-collinear limit of the two soft gluons and a massless hard parton.
\end{itemize}
The dipole component $W^{(3) {\rm dip.}}$ of $W^{(3)}(q_1,q_2,q_3)$
has singular collinear behaviour in the following configurations:
\begin{itemize}
\item[] {($c_3$)\;} 
double-collinear limit of two soft gluons,
\item[] {($c_4$)\;} 
triple-collinear limit of the three soft gluons,
\item[] {($c_5$)\;} quadruple-collinear limit of the three soft gluons and a massless hard parton.
\end{itemize}
The quadrupole component $W^{(3) {\rm quad.}}$ is `collinear safe': it has {\em no}
singular behaviour in any collinear limits. In particular,  $W^{(3) {\rm quad.}}$
can be integrated over the soft-gluon angles in $d=4$ space-time dimensions without
encountering any collinear divergences.

We note that the list of collinear configurations at the points $(c_1)$--$(c_5)$
does not include all possible collinear configurations. The kinematical configurations that are not explicitly listed do not produce a singular collinear behaviour in 
$W^{(2)}$ and $W^{(3)}$. For instance, in the double-collinear limit of a 
hard parton and a single soft gluon both $W^{(2)}$ and $W^{(3)}$ have no collinear singularity. The absence of such collinear singularity is basically a consequence of
colour coherence (and colour conservation). The momentum dependent soft functions 
$\cS_{ik}(q_1,q_2)$, $\cS_{ik}(q_1,q_2,q_3)$ and $\cS_{imkl}(q_1,q_2,q_3)$
in Eqs.~(\ref{corr2}), (\ref{3cordip}) and (\ref{3corquad}) are separately singular in very many collinear configurations, but there are cancellations (due to colour coherence) between the separately-singular terms and, eventually, their total contribution to $W^{(2)}$ and $W^{(3)}$ produces collinear singularities only in the configurations that are listed at the points $(c_1)$--$(c_5)$. In particular, there are remarkable cancellations between the quadrupole contributions 
$\qu{imkl} \cS_{imkl}$ in Eq.~(\ref{3corquad}) that make $W^{(3) {\rm quad.}}$
collinear safe (at the algebraic level, these cancellations take place by using the colour conservation relation (\ref{qcons})
and also the Jacobi identity in Eq.~(\ref{qjac})).

We also note two key points in our discussion on the collinear behaviour, which are 
related to the expansion in irreducible correlations and the use of irreducible quadrupole operators. The squared current $| \J(q_1,\cdots,q_N) |^2$ has collinear singularities in all possible multiparton collinear configurations. Its expansion in irreducible correlations 
$W^{(N)}$ is essential to single out the restricted and definite pattern of collinear configurations at the points $(c_1)$--$(c_5)$. In the case of the three-gluon correlation $W^{(3)}$, the introduction of quadrupole operators $\qu{imkl}$
that are irreducible to colour dipole operators is essential to make the quadrupole component $W^{(3) {\rm quad.}}$ collinear safe.

We remark on the fact that our results for the singular collinear behaviour of 
$W^{(2)}$ and $W^{(3)}$ are fully consistent with the expectations and results from collinear factorization formulae \cite{CaGr99, Campbell:1997hg}
for the squared amplitude $| \M(\{q_\ell\}, \{p_i\}) |^2$.
In particular, in the singular collinear configurations at the points 
$(c_1)$--$(c_5)$ we explicitly obtain the singular collinear factors denoted as 
${\hat P}^{s s'}_{a_1 \dots a_m}$ in the collinear factorization formula (25) of 
Ref.~\cite{CaGr99}. We also note that, in some collinear configurations, we obtain the exact result (rather than a corresponding soft approximation)
for ${\hat P}^{s s'}_{a_1 \dots a_m}$ from the collinear limit of 
$| \J(q_1,\cdots,q_N) |^2$. For instance, in the triple-collinear limit at point
$(c_4)$ we find that $| \J(q_1,q_2,q_3) |^2$ is proportional to the {\em exact}
expression of the factor ${\hat P}^{\mu \nu}_{g_1 g_2 g_3}$ (see Eq.~(66) in
Ref.~\cite{CaGr99}) for the collinear splitting process of three gluons. This is a
consequence of the fact that our result for $| \J(q_1,q_2,q_3) |^2$ is valid for arbitrary relative energies of the three (soft) gluons. In contrast,
in other collinear configurations that involve soft and hard partons
(e.g., those at the points $(c_2)$ and $(c_5)$), we can only obtain the soft limit of the corresponding collinear factor ${\hat P}^{s s'}_{a_1 \dots a_m}$
(analogously to what happens in Eq.~(\ref{w1coll})).

In the case of the collinear limit of {\em four} partons, no explicit expressions
for the corresponding collinear factor ${\hat P}^{s s'}_{a_1 a_2 a_3 a_4}$
at the squared amplitude level are available in the literature
(the collinear factors at the amplitude level are explicitly known in $d=4$ space-time dimensions \cite{DelDuca:1999iql, Birthwright:2005ak}).
Using our result for $| \J(q_1,q_2,q_3) |^2$ we can compute the {\em quadruple}
collinear factor ${\hat P}^{s s'}_{g_1 g_2 g_3 C}$ in the soft limit, and we obtain
the following result (we recall that ${\hat P}^{s s'}_{g_1 g_2 g_3 C}$ is defined as in Eq.~(25) of Ref.~\cite{CaGr99}):
\begin{align}
\label{4coll}
{\hat P}^{s s'}_{g_1 g_2 g_3 C} &\simeq \delta^{s s'} 
\left[ \,p_C\cdot (q_1+q_2+q_3)\,\right]^3 C_C \left\{
C_C^2 \;\w_{nC}^{(1)}(q_1) \;\w_{nC}^{(1)}(q_2) \;\w_{nC}^{(1)}(q_3) \right. \nn \\
&\left. +\;C_C C_A \left[ \w_{nC}^{(1)}(q_1) \;\w_{nC}^{(2)}(q_2,q_3)
+ \;\left( 1 \leftrightarrow 2 \right) + \left( 1 \leftrightarrow 3 \right) 
\right] + \; C_A^2 \;\w_{nC}^{(3)}(q_1,q_2,q_3)
\right\} \;\;.
\end{align}
Here the subscript $C$ denotes the hard collinear parton with momentum
$p_C$ ($C_C=C_F$ if the hard parton is a quark or antiquark, and $C_C=C_A$
if the hard parton is a gluon) and $\{s, s'\}$ are the spins of the parent collinear parton. The momentum dependent functions $\w_{nC}^{(N)}$ ($N=1,2,3$) are obtained
by the soft functions $\w_{ik}^{(N)}$ in Eqs.~(\ref{w1new}), (\ref{w2new}) and
(\ref{w3new}) through the momentum assignement $p_k^\mu \to p_C^\mu$ and
$p_i^\mu \to n^\mu$, where $n^\mu$ is an arbitrary light-like ($n^2=0$) vector such that $n \cdot p_C \neq 0$. The auxiliary vector $n^\mu$ \cite{CaGr99} is needed simply to specify the longitudinal-momentum fraction 
$z_\ell \equiv (n\cdot q_\ell)/n \cdot (p_C+q_1+q_2+q_3)$ of the collinear partons
(i.e., $q_\ell \simeq z_\ell (p_C+q_1+q_2+q_3)$ in the collinear limit).
The collinear splitting function in Eq.~(\ref{4coll}) depends on the subenergies
$q_\ell \cdot p_C$ and $q_\ell \cdot q_{\ell^\prime}$ and on the momentum fractions
$z_\ell$, and it is valid in the soft approximation $z_\ell \ll 1$  with $\ell=1,2,3$
(the symbol `$\simeq$' in Eq.~(\ref{4coll}) refers to the soft approximation).
Note that the spin of the parent collinear parton is unchanged within the soft approximation (see the factor $\delta^{s s'}$ in the right-hand side of 
Eq.~(\ref{4coll})).

We conclude our comments on the singular collinear behaviour of the squared currents
$| \J(q_1,\cdots,q_N) |^2$ by observing that such behaviour is consistent with the known angular-ordered (colour coherence) features \cite{Ermolaev:1981cm, Mueller:1981ex}
of multiple soft-gluon radiation. The hierarchical structure of the singular collinear configurations at the points $(c_1)$--$(c_5)$ directly corresponds
to the angular-ordered pattern of soft-gluon cascades
\cite{Ermolaev:1981cm, Mueller:1981ex, Dokshitzer:1982xr, Bassetto:1984ik, Catani:1984dp}. 
Such angular pattern was
examined in detail in Sect.~6.3 of Ref.~\cite{Dokshitzer:1991wu} by considering the dominant soft contributions for gluon radiation, namely, by computing the squared current $| \J(q_1,\cdots,q_N) |^2$ in the energy strong-ordering region where
$E_1 \ll E_2 \dots \ll E_N$. Our result for $| \J(q_1,q_2,q_3) |^2$ is valid for arbitrary soft-gluon energies, and it can be used to get an
improved understanding of angular-ordering features beyond the dominant soft contributions.

\setcounter{footnote}{2}

\section{
Processes with soft gluons and three hard partons}
\label{sec:3hard}

The soft-gluon factorization formula (\ref{softsquared}) leads to colour correlations between the squared current $| \J(q_1,\cdots,q_N) |^2$ and the hard-parton scattering amplitude $\M(\{p_i\})$. In the cases of scattering amplitudes with two or three hard partons, the colour correlation structure can be completely (for two hard partons) or partly (for three hard partons) worked out in factorized $c$-number form. 
These features are discussed in this section for the three hard-parton case and in Sect.~\ref{sec:2hard} for the two hard-parton case.

\subsection{All-order features
}
\label{sec:3hardgeneral}

We consider a generic scattering amplitude $\M_{ABC}(\{q_\ell\}, \{p_i\})$ whose external legs are three hard partons (denoted as $A, B, C$), soft gluons and additional colourless particles. Owing to flavour conservation, the three hard partons can be either a gluon and a $q{\bar q}$ pair ($\{ ABC \} = \{ gq{\bar q} \}$)
or three gluons ($\{ ABC \} = \{ ggg \}$).
The corresponding scattering amplitude $\M_{ABC}(\{p_i\})$ without soft gluons
is a colour singlet state formed by the three hard partons $A, B$ and $C$.

We first consider the case $\{ ABC \} = \{ gq{\bar q} \}$. Here there is only {\em one} possible colour singlet configuration of the three hard partons, which is generated by the colour vector $\ket{A B C}$. Modulo overall normalization, we can set
$\bra{\,a\beta {\bar \gamma}\,} \,ABC\,\rangle = t^a_{\beta {\bar \gamma}}$, where
$t^a_{\beta {\bar \gamma}}$ is the colour matrix in the fundamental representation and
$a,\beta$ and ${\bar \gamma}$ are the colour indices of the gluon ($A$), quark
($B$) and antiquark ($C$). Therefore, we straightforwardly have
\beq
\label{Jgqq}
| \J(q_1,\cdots,q_N) |^2 \;\,\ket{A B C} = \ket{A B C} 
\;| \J(q_1,\cdots,q_N) |^{2}_{\; ABC} \;, 
\;\;\; \quad (\{ ABC \} = \{ gq{\bar q} \}) \;,
\eeq
and the soft-gluon factorization formula in Eq.~(\ref{softsquared}) becomes
\beq
\label{Mgqq}
| \M_{ABC}(\{q_\ell\}, \{p_i\}) |^2 \simeq \!(\g \,\mu_0^\epsilon)^{2N} \,
| \M_{ABC}(\{p_i\}) |^2 \;| \J(q_1,\cdots,q_N) |^{2}_{\; ABC} \;,
\; (\{ ABC \} = \{ gq{\bar q} \}) \,.
\eeq
The result in Eq.~(\ref{Jgqq}) follows from the fact that the colour operator
$| \J(q_1,\cdots,q_N) |^2$ is colour conserving and it acts onto a one-dimensional colour space generated by $\ket{A B C}$. Therefore, $\ket{A B C}$ is necessarily
an eigenvector of $| \J(q_1,\cdots,q_N) |^2$, with a corresponding eigenvalue that is denoted by the $c$-number $| \J(q_1,\cdots,q_N) |^{2}_{\; ABC}$. 
Since $\ket{\M_{ABC}(\{p_i\})} \propto \ket{A B C}$, Eq.~(\ref{Mgqq}) directly follows
from Eqs.~(\ref{softsquared}) and (\ref{Jgqq}).

The right-hand side of Eq.~(\ref{Mgqq}) is directly proportional to the square 
$| \M_{ABC}(\{p_i\}) |^2$ of the hard-parton scattering amplitude, so that colour correlations are `effectively' removed (though $| \J(q_1,\cdots,q_N) |^{2}_{\; ABC}$
does depend on $SU(N_c)$ colour coefficients) in $c$-number form, analogously to soft-photon factorization formulae in QED.
We remark on the fact that the soft-gluon factorization formula (\ref{Mgqq})
is valid at {\em arbitrary} loop orders in the perturbative expansion of both the soft current and the scattering amplitude (i.e., the validity of Eq.~(\ref{Mgqq})
is not limited to tree-level currents and scattering amplitudes).

We now consider the case $\{ ABC \} = \{ ggg \}$. Here the colour singlet space spanned by the three hard partons is two-dimensional. It is convenient to choose the basis formed by the orthogonal colour state vectors $\ket{(ABC)_f \,}$ and
$\ket{(ABC)_d \,}$ that are defined as follows
\beq
\label{gggstates}
\bra{\,abc\,} \left(ABC\right)_f \,\rangle \equiv i f^{abc}
\,,
\;\;\;
\bra{\,abc\,} \left(ABC\right)_d \,\rangle \equiv  d^{abc}
\,,
\;\;\; \quad (\{ ABC \} = \{ ggg \}) \;\;,
\eeq
where $a, b, c$ are the colour indices of the three gluons, $f^{abc}$ is the structure constant and the fully-symmetric tensor $d^{abc}$ is 
$d^{abc}= 2 {\rm Tr}(\{t^a, t^b\} t^c)$. The generic scattering amplitude
$\ket{\M_{ABC} (\{p_i\})}$ is, in general, a linear combination\footnote{For example, in the case of the Higgs boson ($H$) decay amplitude $H \to ggg$ the three gluons are in a colour antisymmetric state. At variance, the $Z$ boson amplitude $Z \to ggg$
has both the symmetric and antisymmetric colour components.} of the colour antisymmetric state
$\ket{(ABC)_f \,}$ and the colour symmetric state
$\ket{(ABC)_d \,}$. 

The squared current $| \J(q_1,\cdots,q_N) |^2$ is colour conserving and, therefore, it can produce only colour correlations between the two states $\ket{(ABC)_f \,}$
and $\ket{(ABC)_d \,}$ of the three hard-gluon amplitude 
$\ket{\M_{ABC} (\{p_i\})}$.
However, we note that the two states in Eq.~(\ref{gggstates}) have a different charge conjugation. Therefore, since the squared current $| \J(q_1,\cdots,q_N) |^2$
for multiple {\em soft-gluon} radiation is invariant under charge 
conjugation\footnote{Such charge conjugation invariance would not apply to radiation
of soft $q{\bar q}$ pairs.}, it follows that it has a diagonal action onto the colour state vectors of Eq.~(\ref{gggstates}). We have
\beeq
\label{Jgggf}
\!\!\!\!\!\!\!\!\!\!\!~| \J(q_1,\cdots,q_N) |^2 \;\,\ket{(A B C)_f\,} 
\!\!&=&\!\! \ket{(A B C)_f\,}
\;| \J(q_1,\cdots,q_N) |^{2}_{\; ABC} \;, 
\;(\{ ABC \} = \{ ggg \}) \,,\\
\label{Jgggd}
\!\!\!\!\!\!\!\!\!\!\!~| \J(q_1,\cdots,q_N) |^2 \;\,\ket{(A B C)_d\,} 
\!\!&=&\!\! \ket{(A B C)_d\,}
\;| \J(q_1,\cdots,q_N) |^{2}_{\;(A B C)_d } \;,
\eeeq
where $| \J(q_1,\cdots,q_N) |^{2}_{\; ABC}$ and $| \J(q_1,\cdots,q_N) |^{2}_{\;(A B C)_d }$ are the corresponding $c$-number eigenvalues (note that the eigenvalue for the symmetric state $\ket{(A B C)_d\,}$ is explicitly denoted by the subscript 
$(A B C)_d$ in $| \J(q_1,\cdots,q_N) |^{2}_{\;(A B C)_d }$, whereas for the antisymmetric state $\ket{(A B C)_f\,}$ we simply use the notation 
$| \J(q_1,\cdots,q_N) |^{2}_{\; ABC}\,$).
Owing to the diagonalization in Eqs.~(\ref{Jgggf}) and (\ref{Jgggd}),
the soft-gluon factorization formula in Eq.~(\ref{softsquared}) 
for scattering amplitudes with three hard gluons
becomes
\beeq
\label{Mggg}
&&| \M_{ABC}(\{q_\ell\}, \{p_i\}) |^2 \simeq (\g \,\mu_0^\epsilon)^{2N} \,
\left\{ \frac{| \bra{\,(ABC)_f\,} \, \M_{ABC} (\{p_i\}) \rangle |^2}{\bra{\,(ABC)_f\,} (ABC)_f \rangle} \;| \J(q_1,\cdots,q_N) |^{2}_{\; ABC}
\right. \nn \\
&& \left. \;\;\; \;\!\!\!\!
 + \;
\frac{| \bra{\,(ABC)_d\,} \, \M_{ABC} (\{p_i\}) \rangle |^2}{\bra{\,(ABC)_d\,} (ABC)_d \rangle} \;| \J(q_1,\cdots,q_N) |^{2}_{\; (A B C)_d}
\right\}\;, \;\;\;(\{ ABC \} = \{ ggg \}) \,.
\eeeq
In other words, using the colour state basis in Eq.~(\ref{gggstates}),
the colour correlations produced by the colour operator $| \J(q_1,\cdots,q_N) |^2$
are effectively taken into account by a {\em diagonal} $2 \times 2$ colour matrix
(the use of a different colour basis would instead lead to a non-diagonal $2\times 2$ matrix).

As in the case of Eq.~(\ref{Mgqq}), we remark that also the soft-gluon factorization formula in Eq.~(\ref{Mggg}) is valid at {\em arbitrary} loop orders in the perturbative expansion of both the soft current and the scattering amplitude. To our knowledge, this all-order feature of multiple soft-gluon radiation from three hard gluons has not been noticed before in the literature on soft-gluon factorization.

In the case of single and double soft-gluon radiation at the tree level, the two eigenvalues of $| \J(q_1,\cdots,q_N) |^2$ in Eqs.~(\ref{Jgggf}) and (\ref{Jgggd})
are equal and, therefore, the right-hand side of Eq.~(\ref{Mggg}) is exactly factorized \cite{CaGr99}
in the same form as in Eq.~(\ref{Mgqq}). As explicitly shown in the next subsection,
in the case of triple soft-gluon radiation the two eigenvalues are no longer degenerate. The quadrupole colour correlations of Eqs.~(\ref{3cordq}) 
and (\ref{3corquad})
are responsible for removing the degeneracy of the two eigenvalues.

\subsection{Soft-gluon radiation at the tree level}
\label{sec:3hardtree}

We consider soft-gluon radiation at the tree level, and we present the explicit expressions for emission of $N=1,2$ and 3 soft gluons from the scattering amplitude
$\M_{ABC}$ with three hard partons. In the final part of this subsection
we also present some results for emission of $N \geq 4$ soft gluons with energy strong ordering.

We fix our notation by always denoting $A=g$ (so that the squared colour charge
$\T^2_A=C_A$ of the parton $A$ coincides with the customary notation for the Casimir coefficient $C_A=N_c$) and, thus, $\{ BC \} = \{ q{\bar q} \}$ or
$\{ BC \} = \{ gg \}$ (the corresponding squared colour charges are $\T^2_B=C_B$
and $\T^2_C=C_C$ with $C_B=C_C$, since $\{ BC \}$ is a particle--antiparticle pair).

The eigenvalues of the squared current $| \J(q_1,\cdots,q_N) |^2$ in 
Eqs.~(\ref{Jgqq}), (\ref{Jgggf}) and (\ref{Jgggd})
are computed by applying the general results in Sect.~\ref{s:sq12} and 
Sect.~\ref{s:sq123}.
In the cases of single and double soft-gluon emission the colour dependence of the squared current is entirely given in terms of products of dipole factors 
$\T_i \cdot \T_k$. 
Using colour conservation, the action of dipole factors onto colour singlet
states of three hard partons can be evaluated in terms of quadratic Casimir coefficients (see the Appendix~A of Ref.~\cite{csdip}). In particular,
we have $2\, \T_A \cdot \T_B \,\ket{ABC} = (C_C - C_A -C_B) \,\ket{ABC}$ and related permutations of $A,B,C$. Note that these colour algebra relations are valid for a generic colour singlet state $\ket{ABC}$. Therefore, the two eigenvalues in the 
right-hand side of Eqs.~(\ref{Jgggf}) and (\ref{Jgggd}) are equal. Combining colour algebra with the momentum dependence of the expressions in Eqs.~(\ref{w1def}),
(\ref{w1new}), (\ref{J2sq}) and (\ref{w2new}),
we straightforwardly obtain the following results:
\beq
\label{1gabc}
| \J(q) |^{2}_{\; ABC} \;= \; C_B \;\w_{BC}^{(1)}(q) 
+ C_A \;\w_{ABC}^{(1)}(q)\;\;,
\eeq
\beeq
\label{2gabc}
| \J(q_1,q_2) |^{2}_{\; ABC} \;&=& \;C_B^2 \;\w_{BC}^{(1)}(q_1) \;\w_{BC}^{(1)}(q_2) 
\nn \\
&+& C_B C_A \left[ \;\w_{BC}^{(2)}(q_1,q_2) 
+ \w_{BC}^{(1)}(q_1) \;\w_{ABC}^{(1)}(q_2) + \w_{BC}^{(1)}(q_2) \;\w_{ABC}^{(1)}(q_1)
\right] \nn \\
&+& C_A^2 \left[ \;\w_{ABC}^{(2)}(q_1,q_2)
+ \w_{ABC}^{(1)}(q_1) \;\w_{ABC}^{(1)}(q_2) \right] \;\;.
\eeeq
In the right-hand side of Eqs.~(\ref{1gabc}) and (\ref{2gabc}),
the functions $\w_{BC}^{(N)}$  and $\w_{ABC}^{(N)}$ depend on the momenta of the soft gluons and the momenta $p_A, p_B$ and $p_C$ of the hard partons, but they do not depend on colour coefficients. The two hard-parton functions $\w_{BC}^{(N)}$ ($N=1,2$)
are those in Eqs.~(\ref{w1new})--(\ref{wikdef}). Moreover, we have defined the following three hard-parton functions $\w_{ABC}^{(N)}$:
\beq
\label{wabc}
\w_{ABC}^{(N)}(q_1,\cdots,q_N) \equiv 
\frac{1}{2} \left[ \w_{AB}^{(N)}(q_1,\cdots,q_N) + 
\w_{AC}^{(N)}(q_1,\cdots,q_N) - \w_{BC}^{(N)}(q_1,\cdots,q_N) 
\right] \;\;.
\eeq
Note that $\w_{ABC}^{(N)}$ is not symmetric with respect to the dependence on the three hard-parton momenta (it is symmetric only under the exchange 
$B \leftrightarrow C$).

The results in Eqs.~(\ref{1gabc}) and (\ref{2gabc}) are valid for both massless and massive hard partons. In particular, they generalize the tree-level massless results in Eq.~(58) of Ref.~\cite{Catani:2000pi}
and Eq.~(A.4) of Ref.~\cite{CaGr99} to the case of massive hard partons.

The dependence of Eqs.~(\ref{1gabc}) and (\ref{2gabc}) on the colour state of the hard-parton scattering amplitude $\M_{ABC}(\{p_i\})$ is entirely expressed through
the Casimir coefficient $C_B$ ($C_B=C_C$) of the hard parton $B$. 
We have
$C_B=C_F$ if $\{ABC \}= \{ gq{\bar q}\}$ (see Eqs.~(\ref{Jgqq}) and (\ref{Mgqq})),
while we have $C_B=C_A$ if $\{ABC \}= \{ ggg\}$ 
(see Eqs.~(\ref{Jgggf})--(\ref{Mggg}))). In particular, knowing 
$| \J(q_1,\cdots,q_N) |^{2}_{\; ABC}$ for the case $\{ABC \}= \{ gq{\bar q}\}$,
we obtain $| \J(q_1,\cdots,q_N) |^{2}_{\; ABC}$ for the pure gluon case
$\{ABC \}= \{ ggg \}$ through the replacement $C_F \to C_A$. This replacement can be regarded as a `Casimir scaling' relation. Such a property is valid for $N=1$ and 
$N=2$, but we anticipate that it is violated for triple soft-gluon radiation
(see Eqs.~(\ref{3ggggd}) and (\ref{3gabc})).

The squared current $| \J(q_1,q_2,q_3) |^2$ for triple soft-gluon radiation is given in Eqs.~(\ref{J3sq}), (\ref{3cordq}), (\ref{3corquad}) and (\ref{w3new}),
and it includes both colour dipole and colour quadrupole contributions.
The colour dipole contribution for emission from the three hard partons $\{ABC \}$
is denoted by $| \J(q_1,q_2,q_3) |^{2 \,({\rm dip.})}_{\; ABC}$, and it can be computed analogously to Eqs.~(\ref{1gabc}) and~(\ref{2gabc}). 
We obtain the following result:
\begin{align}
\label{3gdipabc}
&| \J(q_1,q_2,q_3) |^{2 \,({\rm dip.})}_{\; ABC} \;= \;
C_B^3 \;\w_{BC}^{(1)}(q_1) \;\w_{BC}^{(1)}(q_2) \;\w_{BC}^{(1)}(q_3)
 \nn \\
&\;+\;C_B^2 C_A \left[ \w_{BC}^{(1)}(q_1) \;\w_{BC}^{(2)}(q_2,q_3)
+ \w_{ABC}^{(1)}(q_1) \;\w_{BC}^{(1)}(q_2) \;\w_{BC}^{(1)}(q_3)
+ \;\left( 1 \leftrightarrow 2 \right) + \left( 1 \leftrightarrow 3 \right) 
\right] \nn\\
&\;+ \;C_B C_A^2 \left\{ \;\w_{BC}^{(3)}(q_1,q_2,q_3)
+ \left[ \w_{BC}^{(1)}(q_1) \;\w_{ABC}^{(1)}(q_2) \;\w_{ABC}^{(1)}(q_3)
\right. \right. \nn \\
&\left. \left. \;+ \w_{BC}^{(1)}(q_1) \;\w_{ABC}^{(2)}(q_2,q_3)
+ \w_{ABC}^{(1)}(q_1) \;\w_{BC}^{(2)}(q_2,q_3)
+ \;\left( 1 \leftrightarrow 2 \right) + \left( 1 \leftrightarrow 3 \right)
\right]
\right\} \nn \\
&\;+\;C_A^3 \left\{ \;\w_{ABC}^{(3)}(q_1,q_2,q_3)
+ \w_{ABC}^{(1)}(q_1) \;\w_{ABC}^{(1)}(q_2) \;\w_{ABC}^{(1)}(q_3)
\right. \nn \\
&\;+\;\left. \left[ \w_{ABC}^{(1)}(q_1) \;\w_{ABC}^{(2)}(q_2,q_3)
+ \;\left( 1 \leftrightarrow 2 \right) + \left( 1 \leftrightarrow 3 \right)
\right]
\right\}
\;\;,
\end{align}
where the two hard-parton function $\w_{BC}^{(3)}$ is given in Eq.~(\ref{wikdef3})
and the three hard-parton function $\w_{ABC}^{(3)}$ is obtained from Eq.~(\ref{wabc}).
We recall that the expression in Eq.~(\ref{3gdipabc}) is valid for a generic colour singlet state $\ket{ABC}$, and we also note that such expression fulfils Casimir scaling (analogously to Eqs.~(\ref{1gabc}) and (\ref{2gabc})).

The colour dipole contribution in Eq.~(\ref{3gdipabc}) has to be supplemented by the contribution from colour quadrupole correlations. We have explicitly evaluated the action of the quadrupole operators $\qu{imkl}$ of Eq.~(\ref{3corquad})
onto colour singlet states $\ket{ABC}$ (see Eqs.~(\ref{qqqg})--(\ref{qgggd}) in 
Appendix~\ref{a:quad}).
In agreement with the general reasoning that leads to Eqs.~(\ref{Jgqq}), 
(\ref{Jgggf}) and (\ref{Jgggd}), we find that the three colour singlet states
$\ket{gq{\bar q}\,}$, $\ket{(ggg)_f}$  and $\ket{(ggg)_d}$ are eigenstates of the quadrupole operators. In particular, the gluon symmetric state $\ket{(ggg)_d}$
is annihilated by the quadrupole operators, whereas the state $\ket{gq{\bar q}\,}$
and the gluon antisymmetric state $\ket{(ggg)_f}$ correspond to different 
eigenvalues of the quadrupole operators. Combining the algebra of the colour quadrupole with the momentum dependence as given in Eq.~(\ref{3corquad}),
we obtain the quadrupole contributions for triple soft-gluon radiation from three hard partons.
  
The complete (dipole and quadrupole) results for the eigenvalues in 
Eqs.~(\ref{Jgqq}), (\ref{Jgggf}) and (\ref{Jgggd}), are summarized as follows.
In the case of the gluon colour symmetric state $\ket{(ggg)_d}$ 
(see Eq.~(\ref{Jgqq})) we find
\beq
\label{3ggggd}
| \J(q_1,q_2,q_3) |^{2}_{\; (ABC)_d} =
| \J(q_1,q_2,q_3) |^{2 \,({\rm dip.})}_{\; ABC} \;\;,
\eeq
where the dipole contribution is given in Eq.~(\ref{3gdipabc}).
In the cases of $\{ABC \} = \{ gq{\bar q} \}$ (see Eq.~(\ref{Jgqq})) and of the gluon colour antisymmetric state $\ket{(ggg)_f}$ (see Eq.~(\ref{Jgggf}))  we find
\beq
\label{3gabc}
| \J(q_1,q_2,q_3) |^{2}_{\; ABC} =
| \J(q_1,q_2,q_3) |^{2 \,({\rm dip.})}_{\; ABC}
+ \lambda_B N_c \;\w_{ABC}^{(3) {\rm quad.}}(q_1,q_2,q_3) \;\;,
\eeq
where 
$\lambda_B=\lambda_F$ if $B$ is a quark (or antiquark) and 
$\lambda_B=\lambda_A$ if $B$ is a gluon, with
\beq
\label{lquad}
\lambda_F=\frac{1}{2} \;\;,\quad  \quad \lambda_A=3 \;\;.
\eeq
The quadrupole soft function $\w_{ABC}^{(3) {\rm quad.}}(q_1,q_2,q_3)$ in 
Eq.~(\ref{3gabc}) does not depend on colour coefficients, and it depends on the momenta $q_1,q_2,q_3$ of the soft gluons and the momenta $p_A,p_B,p_C$ of the three hard partons. Its expression is
\beeq
\label{w3abcquad}
\w_{ABC}^{(3) {\rm quad.}}(q_1,q_2,q_3) &=& 
\left[ \, {\cal S}_{ABAB} - {\cal S}_{ABBA}
+{\cal S}_{ABCA} - {\cal S}_{ABAC} + {\cal S}_{BAAC} - {\cal S}_{BACA}
\, \right] \nn \\
&+& {\rm perms.} \;\{A,B,C\} \;\;,
\eeeq
where ${\cal S}_{imkl}= {\cal S}_{imkl}(q_1,q_2,q_3)$ is the momentum dependent function\footnote{The function ${\cal S}_{imkl}$ can also be expressed as in 
Eq.~(\ref{sbar}).} in Eqs.~(\ref{3corquad}) and (\ref{simkl}).

Owing to the quadrupole contribution in the right-hand side of 
Eqs.~(\ref{3ggggd}) and (\ref{3gabc}), we note that triple soft-gluon radiation in
$\{ABC \} = \{ gq{\bar q} \}$ and $\{ABC \} = \{ ggg \}$ hard-parton processes cannot be related by using the Casimir scaling
replacement $C_F \to C_A$.

Note that the quadrupole coefficient $\lambda_B$ depends on the colour state of the three hard partons not only at the strictly-formal level but also at a sizeable quantitative level. For instance, in the case of Eq.~(\ref{3gabc}) we have
$\lambda_A/\lambda_F= 6$.
The violation of Casimir scaling for tree-level radiation of $N=3$ soft gluons persists for larger soft-gluon multiplicities $(N \geq 4)$.

The expressions in Eqs.~(\ref{1gabc}), (\ref{2gabc}), (\ref{3gdipabc})--(\ref{3gabc})
are valid for generic energies of the soft gluons. Using the 
approximations in Eqs.~(\ref{w2som}), (\ref{w312})--(\ref{w3bcm}) and (\ref{simklso}),
these expressions can be simplified in various kinematical subregions with energy ordering of the soft gluons.

In particular, considering the energy strong-ordering region where $E_1 \ll E_2 \ll
E_3$ and using Eq.~(\ref{simklso}), we are able to express the quadrupole function in Eq.~(\ref{w3abcquad}) in a compact form, which highlights some of the main features of
$\w_{ABC}^{(3) {\rm quad.}}$. We find
\beeq
\label{wquadso}
\w_{ABC}^{(3) {\rm quad.}}(q_1,q_2,q_3) &=& 
\left[ \; \frac{1}{2}\; \w_{AB}^{(1)}(q_3) \; G_{3BAC}(q_2) \;G_{3ABC}(q_1) 
\,+ \left( 1 \leftrightarrow 2\right)\right] \nn \\
&+& {\rm perms.} \;\{A,B,C\} \;\;,
\quad\quad\quad~~~~~~~~~ (E_1 \ll E_2 \ll E_3)\;\;,
\eeeq
where we have introduced the `quadrupole eikonal function' $G_{imkl}(q)$,
\beq
\label{gquad}
G_{imkl}(q) \equiv g_{\mu \nu} j_{im}^\mu(q)  j_{kl}^\nu(q)
= \left[ \frac{k_i \cdot k_k}{(k_i \cdot q) (k_k \cdot q)} 
       - \frac{k_i \cdot k_l}{(k_i \cdot q) (k_l \cdot q)}
\right] + \binom{i \leftrightarrow m}{k \leftrightarrow l} \;\;.
\eeq
Here, $j_{ik}^\mu(q)$ is the conserved eikonal current in Eq.~(\ref{conseik}),
and the momenta $k_i$ generically denote momenta of hard partons or momenta of soft gluons that are harder than $q$ (e.g., the subscript `3' of $G_{3BAC}(q_2)$
in Eq.~(\ref{wquadso}) refers to the soft-gluon momentum $q_3$).

We note that the symmetry properties of $G_{imkl}(q)$ with respect to its momentum indices $\{i,m,k,l\}$ are exactly the same as those of the quadrupole operators 
$\qu{imkl}$ (see Eqs.~(\ref{qant})--(\ref{qjac})). Indeed, we have
\beeq
\label{gant}
&& G_{imkl}(q) = - \,G_{mikl}(q) \;\;, 
\quad \quad G_{imkl}(q) = - G_{imlk}(q) \;\;, \\
\label{gsym}
&& G_{imkl}(q) = \;G_{klim}(q) \;\;,\\
\label{gjac}
&& G_{i k_1k_2k_3}(q) +  G_{i k_3k_1k_2}(q) + G_{i k_2k_3k_1}(q) = 0 \;\;.
\eeeq

We also remark on some distinctive features of the momentum function $G_{imkl}(q)$
with respect to its angular dependence. The various contributing terms in the 
right-hand side of Eq.~(\ref{gquad}) can separately lead to collinear singularities. However, if the four hard momenta $\{ k_i, k_m, k_k, k_l \}$ are {\em distinct}
(more precisely, if they point towards different directions), the collinear singularities cancel and $G_{imkl}(q)$ is {\em collinear safe} (i.e., the angular integration over the direction of the soft-gluon momentum $q$ does not lead to collinear divergences). At variance, if two of the hard momenta are both massless and collinear to $q$, $G_{imkl}(q)$ can have a singular collinear behaviour: this happens if the two collinear momenta refer to indices in the first and second pair of indices
(e.g., $k_i$ collinear to $k_l$). In contrast, if the two momenta in the first
or second pair of indices are proportional, $G_{imkl}(q)$ identically vanishes.

As discussed in Sect.~\ref{s:coll}, the quadrupole correlation 
$W^{(3) {\rm quad.}}(q_1,q_2,q_3)$ is collinear safe with respect to the angular integration over the soft-gluon momenta. This general property can be easily checked 
by considering the energy strong-ordering approximation in Eq.~(\ref{wquadso})
and 
the collinear features of the quadrupole eikonal function 
$G_{imkl}(q)$. For instance, we can examine the behaviour of the term
\beq
\w_{AB}^{(1)}(q_3) \; G_{3BAC}(q_2) \;G_{3ABC}(q_1)
\eeq
of Eq.~(\ref{wquadso}) in the collinear configurations where one of the soft momenta $q_1, q_2$ or $q_3$ is collinear to the hard-parton momentum $p_B$:
if either $q_1$ or $q_2$ is collinear to $p_B$, the functions $G_{3ABC}(q_1)$ and 
$G_{3BAC}(q_2)$ are collinear safe; is $q_3$ is collinear to $p_B$, the singular collinear behaviour of $\w_{AB}^{(1)}(q_3)$ is cancelled by the vanishing behaviour of the factor $G_{3BAC}(q_2)$ (which follows from the antisymmetry property in
Eq.~(\ref{gant})). Exploiting the properties of $G_{imkl}(q)$, a similar reasoning can be applied to check that the expression of 
$\w_{ABC}^{(3) {\rm quad.}}(q_1,q_2,q_3)$ in Eq.~(\ref{wquadso}) is collinear safe in all possible multiparton collinear configurations of the soft-gluon momenta.

In the context of the energy strong-ordering approximation, we present a 
more general result that is valid for the emission of 
an arbitrary number of soft gluons.
We consider the general factorization formula in Eq.~(\ref{Mggg})
for the emission of $N$ soft gluons from three hard gluons. We limit ourselves to considering {\em tree-level} soft-gluon radiation in the strong ordering region where
$E_1 \ll E_2 \ll \dots \ll E_N$. In this energy region, the factorization formula 
(\ref{Mggg}) becomes
\beeq
\label{mgggso}
| \M_{ABC}(\{q_\ell\}, \{p_i\}) |^2 \!\!&\simeq& \!\!\!(\g \,\mu_0^\epsilon)^{2N} \,
| \M_{ABC}(\{p_i\}) |^2  \\
\!\! &\times& \!\!\!\!
\Bigl\{ | \J(q_1,\cdots,q_N) |^{2 \,(s.o.)}_{\; g(A)g(B)g(C)} 
+ {\cal O}((N_c)^{N-2})
\Bigr\} \;,
\; (\{ ABC \} = \{ ggg \}) \,, \nn
\eeeq
where the squared current is
\beq
\label{jgggso}
| \J(q_1,\cdots,q_N) |^{2 \,(s.o.)}_{\; g(A)g(B)g(C)} = C_A^N 
(p_A \cdot p_B) (p_B\cdot p_C) (p_C \cdot p_A) \;
F_{eik}^{(N+3)}(p_A,p_B,p_C,q_1,\cdots,q_N) \;.
\eeq

In Eq.~(\ref{jgggso}), $p_A, p_B$ and $p_C$ are the momenta of the three hard gluons
and $F_{eik}^{(N)}$ (with $N \geq 3$) is the following `multi-eikonal function' of the generic momenta
$k_i$:
\beq
\label{feik}
F_{eik}^{(N)}(k_1,\cdots,k_N) \equiv \!\bigl[ 
(k_1 \cdot k_2) (k_2\cdot k_3) \dots (k_{N-1} \cdot k_N) (k_{N} \cdot k_1)
\bigr]^{-1}\! + {\rm ineq. \;perms.} \{k_1,\cdots,k_N\} \,.
\eeq
The terms on the right-hand side are closed chains of product of eikonal propagators
$(k_i\cdot k_j)^{-1}$, and $F_{eik}^{(N)}$ is obtained by summing over all the permutations that lead to inequivalent chains. There are $(N-1)!/2$
inequivalent permutations in Eq.~(\ref{feik}).

We recall that the multi-eikonal function also arises by computing the square of maximal helicity violating (MHV) amplitudes \cite{Parke:1986gb}
for pure multigluon scattering. However, the soft limit in Eq.~(\ref{mgggso})
is not directly related to MHV amplitudes. The amplitude 
$\M_{ABC}(\{q_\ell\}, \{p_i\})$ in Eq.~(\ref{mgggso}) has $N+3$ gluons and necessarily (because of momentum conservation in the soft limit) additional colourless particles in its external legs. Moreover, the squared amplitude 
$| \M_{ABC}(\{q_\ell\}, \{p_i\}) |^2$ is obtained by summing over all the helicity configurations of his external gluons.

We note that the squared current in Eq.~(\ref{jgggso}) is proportional to $C_A^N= (N_c)^N$ and, therefore, the result in Eq.~(\ref{mgggso}) neglects corrections (denoted by the term ${\cal O}((N_c)^{N-2}$) that are formally subdominant to leading order in the large-$N_c$ approximation. One can explicitly check that the 
squared current in Eq.~(\ref{jgggso}) exactly coincides with the energy strong ordering limit of Eqs.~(\ref{1gabc}) and (\ref{2gabc}). Therefore the factorization formula (\ref{mgggso}) is exact (i.e., with no ${\cal O}((N_c)^{N-2})$ corrections)
for emission of $N=1$ and 2 soft gluons. In the case of triple soft-gluon emission,
the squared current in Eq.~(\ref{jgggso}) exactly coincides with the 
energy strong ordering limit of the dipole current in Eq.~(\ref{3gdipabc}).
Therefore, if $N=3$ the terms of ${\cal O}((N_c)^{N-2})$ in Eq.~(\ref{mgggso})
are those  due to the quadrupole contributions that are proportional to the function
$\w_{ABC}^{(3) {\rm quad.}}$
(which does not vanish in the case of energy strong ordering) in Eqs.~(\ref{3gabc}). 
The comparison between Eqs.~(\ref{Mggg}) and (\ref{mgggso}) also implies that, in the case of energy strong ordering, the two eigenvalues of the squared current in 
Eqs.~(\ref{Jgggf}) and (\ref{Jgggd})
becomes degenerate, modulo corrections of ${\cal O}((N_c)^{N-2})$.

Our derivation (which also uses the BCM formula of Ref.~\cite{Bassetto:1984ik})
of the results in Eqs.~(\ref{mgggso}) and (\ref{jgggso}) is presented in 
Sect.~\ref{sec:2hard} (see Eq.~(\ref{bcmabc}) and accompanying comments).
To our knowledge, these results are new and they have not appeared in the previous literature.

\setcounter{footnote}{2}

\section{Processes with soft gluons and two hard partons}
\label{sec:2hard}

In this section we discuss
soft-gluon radiation in scattering processes with two hard partons. 
In Sect.~\ref{sec:2hardgeneral} we 
consider soft-gluon factorization to arbitrary orders in the loop expansion. We comment on the colour structure and then, considering certain kinematical regions with energy ordering of the soft gluons, we directly relate the squared currents for emission of $N$ soft gluons from two and three hard partons.
In Sect.~\ref{sec:3from2}
we explicitly consider radiation of $N=1,2$ and 3 soft gluons at the tree level.
Then, in Sect.~\ref{sec:quadruple}
we present some results on the tree-level emission of $N=4$ soft gluons.

\subsection{All-order features
}
\label{sec:2hardgeneral}

We consider a generic scattering amplitude $\M_{BC}(\{q_\ell\}, \{p_i\})$ whose external legs are two hard partons (denoted as $B$ and $C$), soft gluons and additional colourless particles. Owing to flavour conservation, the two hard partons can be either a $q{\bar q}$ pair ($\{ BC \} = \{ q{\bar q} \}$)
or two gluons ($\{ BC \} = \{ gg \}$).

The corresponding scattering amplitude $\M_{BC}(\{p_i\})$ without soft gluons
is a colour singlet state. There is only {\em one} colour singlet configuration of the two hard partons, and the corresponding one-dimensional colour space is generated by the colour state vector denoted as $\ket{B C}$.
Modulo overall normalization,  in the case of a $\{ BC \} = \{ q{\bar q} \}$ state we can set $\bra{\,\beta {\bar \gamma}\,} \,BC\,\rangle = \delta_{\beta {\bar \gamma}}$
where $\beta$ and ${\bar \gamma}$ are the colour indices of the quark
($B$) and antiquark ($C$), while  in the case of a $\{ BC \} = \{ gg \}$ state 
we can set $\bra{\,b c\,} \,BC\,\rangle = \delta_{b c}$ where $b$ and $c$ are the colour indices of the two gluons.

The squared current $| \J(q_1,\cdots,q_N) |^2$ in Eq.~(\ref{softsquared}) conserves the colour charge of the hard partons and, consequently, the state 
$| \J |^2 \,\ket{B C}$ is also proportional to $\ket{B C}$. We write
\beq
\label{jbc}
| \J(q_1,\cdots,q_N) |^2 \;\,\ket{B C} = \ket{B C} 
\;| \J(q_1,\cdots,q_N) |^{2}_{\; BC} \;, 
\eeq
and it follows that the soft-gluon factorization formula (\ref{softsquared}) 
can be written as
\beq
\label{sbc}
| \M_{BC}(\{q_\ell\}, \{p_i\}) |^2 \simeq \!(\g \,\mu_0^\epsilon)^{2N} \,
| \M_{BC}(\{p_i\}) |^2 \;| \J(q_1,\cdots,q_N) |^{2}_{\; BC} \;.
\eeq
Note that the squared current factor $| \J(q_1,\cdots,q_N) |^{2}_{\; BC}$
in Eq.~(\ref{jbc}) is the eigenvalue of the colour operator $| \J |^2$ onto the colour state $\ket{B C}$. Therefore $| \J(q_1,\cdots,q_N) |^{2}_{\; BC}$ is a 
$c$-number, and the soft-gluon formula (\ref{sbc}) for the squared amplitude has 
a factorized $c$-number form, with no residual correlation effects in colour space.
In this respect, the structure of Eq.~(\ref{sbc}) is similar to that of soft-photon factorization formulae in QED. The non-abelian (colour correlation) effects are effectively embodied in the dependence of the $c$-number factor 
$| \J(q_1,\cdots,q_N) |^{2}_{\; BC}$ on $SU(N_c)$ colour coefficients
(see, e.g., Eqs.~(\ref{jbc1g})--(\ref{jbc3g}) and (\ref{j4bc})).

We note that the factorized structure of Eqs.~(\ref{jbc}) and (\ref{sbc}) simply and directly follows from the fact that two hard partons in a colour singlet configuration generate a one-dimensional colour space. Therefore, 
Eqs.~(\ref{jbc}) and (\ref{sbc}) are valid at {\em arbitrary} loop orders in the perturbative expansion of both the soft-gluon current and the scattering amplitude.

Combining Eqs.~(\ref{jbc}) and (\ref{sbc}) with the discussion in 
Sect.~\ref{sec:3hardgeneral}, 
we can derive a general result that relates multiple soft-gluon radiation from scattering amplitudes with two and three hard partons. Specifically, we examine the radiation of $N$ soft gluons from two hard partons $B$ and $C$ in the energy region where one soft gluon (say, the gluon with momentum $q_N$) is much harder than the other $N-1$ gluons. Therefore, we are considering the energy region where 
$E_N \gg E_\ell$ without additional constraints on the energies $E_\ell$ $(\ell=1,\dots,N-1)$
of the softer gluons. In this energy region the squared current 
$| \J(q_1,\cdots,q_N) |^{2}_{\; BC}$ of Eq.~(\ref{sbc}) fulfils the following relation:
\beq
\label{gsorel}
| \J(q_1,\cdots,q_N) |^{2}_{\; BC} =\! | \J(q_1,\cdots,q_{N-1}) |^{2}_{\; ABC}
\;| \J(q_N) |^{2}_{\; BC} \;, \,(p_A=q_N, E_N \gg E_\ell \;{\rm with\;} \ell < N) \,,
\eeq
where $| \J(q_1,\cdots,q_{N-1}) |^{2}_{\; ABC}$ is the squared current for emission of the softer gluons from three hard partons, namely, the two hard partons ($B$ and $C$) and the hardest soft gluon $A$ with momentum $p_A=q_N$. More precisely, 
$| \J(q_1,\cdots,q_{N-1}) |^{2}_{\; ABC}$ is the squared current in Eq.~(\ref{Jgqq})
if $\{ BC \} = \{ q{\bar q}\}$, while it is the squared current eigenvalue
in Eq.~(\ref{Jgggf}) if $\{ BC \} = \{ gg\}$. The proof of Eq.~(\ref{gsorel}), which is given below, is based on the relations in Eqs.~(\ref{Mgqq}), (\ref{Mggg}) and
(\ref{sbc}) (and also on the structure of Eq.~(\ref{j1gg})).
Since these relations are valid to all orders, 
Eq.~(\ref{gsorel}) is equally valid at {\em arbitrary} perturbative orders in the loop expansion of the squared currents.

We note that the factor $| \J(q_1,\cdots,q_{N-1}) |^{2}_{\; ABC}$ in 
Eq.~(\ref{gsorel}) is evaluated in the energy region where the momentum $p_A=q_N$
is softer than the momenta $p_B$ and $p_C$. However, this energy constraint has no effect on $| \J(q_1,\cdots,q_{N-1}) |^{2}_{\; ABC}$, which is exactly equal to the squared currents in Eqs.~(\ref{Jgqq}) and (\ref{Jgggf}) with no energy constraints.
This statement follows from the fact that the soft limit $q_\ell \to 0 \;
(\ell=1,\dots,N-1)$ is insensitive to the actual size of the energies of the hard partons. More formally, the statement is a consequence of the invariance of the soft current $\J$ under the rescaling $p_i \to \xi_i p_i$ ($\xi_i$ are arbitrary positive definite parameters) of each hard-parton momentum $p_i$ (such invariance is evident
in the computation of the current by using the eikonal approximation for soft emission from the hard partons). We also note that the energy ordering requirement
$E_N \gg E_\ell \;{\rm with\;} \ell \leq N-1$ in Eq.~(\ref{gsorel}) only affects
the expression of $| \J(q_1,\cdots,q_N) |^{2}_{\; BC}$ in the left-hand side, without affecting  $| \J(q_1,\cdots,q_{N-1}) |^{2}_{\; ABC}$.

The relation (\ref{gsorel}) can be exploited in different ways. For instance, it can be used to extract information on $| \J |^{2}_{\; ABC}$ from the knowledge
of $| \J |^{2}_{\; BC}$ (see, e.g.,  Eqs.~(\ref{jgggso}) and 
(\ref{bcm})),
or viceversa (see, e.g., Sect.~\ref{sec:quadruple}). Obviously, it can also be used to check
explicit computations of the currents.

The proof of Eq.~(\ref{gsorel}) proceeds as follows. To compute the squared current
$| \J |^{2}_{\; BC}$ for emission of $\{q_1,\cdots,q_N\}$ from B and C, we first consider the emission of $\{q_1,\cdots,q_{N-1}\}$ from three harder partons as given by the partons $B$ and $C$ and the hardest gluon $A$ with momentum
$p_A=q_N$. Then we consider the emission of the soft gluon with momentum $p_A=q_N$
from the hard partons $B$ and $C$. If the hard partons are $\{B C\}=\{q {\bar q}\}$,
this procedure simply amounts to apply first the factorization formula (\ref{Mgqq})
for $N-1$ soft gluons and then the factorization formula (\ref{sbc}) for the emission of a single soft gluon with momentum $q_N$: the combination of Eqs.~(\ref{Mgqq})
and (\ref{sbc}) straightforwardly gives Eq.~(\ref{gsorel}).
If the hard partons $B$ and $C$ are two gluons, the procedure requires the use of the factorization formulae in Eqs.~(\ref{Mggg}) and (\ref{sbc}), and this leads to a subtle point since (at variance with Eq.~(\ref{Mgqq})) Eq.~(\ref{Mggg}) involves the emission of the soft gluon with momentum $p_A=q_N$ from two possible colour state configurations of the amplitude $\ket{\M_{ABC}(\{p_i\}}$. Nonetheless,
in the soft limit $p_A=q_N \to 0$ we find (see Eqs.~(\ref{mbcq}) and (\ref{j1gg})
and accompanying comments) that the amplitude
$\ket{\M_{ABC}(\{p_i\}}$ is necessarily proportional to the antisymmetric colour state $\ket{(ABC)_f \,}$ in Eq.~(\ref{gggstates})
(i.e., $\langle (ABC)_d \,\ket{\M_{ABC}(\{p_i\}}$ vanishes in the soft limit 
$p_A=q_N \to 0$). Therefore, the use of (\ref{Mggg}) leads to 
Eq.~(\ref{gsorel}), where $| \J(q_1,\cdots,q_{N-1}) |^{2}_{\; ABC}$
is the squared current eigenvalue of the antisymmetric colour state.

The all-order structure of Eqs.~(\ref{jbc}), (\ref{sbc}) and (\ref{gsorel})
refers to the squared current $| \J |^{2}$ for multiple soft-gluon emission from two hard partons. We also 
comment 
on the all-order structure of the soft current $\J$, although we limit our discussion to single soft-gluon radiation.

We consider the scattering amplitude $\M_{BC}(q, \{p_i\})$ with two hard partons $B$ and $C$ (with momenta $p_B$ and $p_C$) and a gluon with momentum $q$ and colour index $a$, and we perform the soft limit $q \to 0$. The dominant singular behaviour in the soft limit is given by the factorization formula (\ref{1gfact}),
which we explicitly rewrite:
\begin{equation}
\label{mbcq}
  \ket{\M_{BC}(q, \{p_i\})} \simeq
  \g \,\mu_0^\epsilon \;\J(q) \; \ket{\M_{BC}(\{p_i\})} \;.
\end{equation}
Here $\ket{\M_{BC}(\{p_i\})}$ is the all-loop scattering amplitude with the two hard partons $B$ and $C$ in a colour singlet configuration, and $\J(q)$ is the all-loop
current for single soft-gluon emission from the two hard partons in the colour
singlet state $\ket{BC\,}$.

To present the all-order structure of $\J(q)$ and to highlight its colour structure,
we explicitly denote the dependence on the colour indices.
The amplitude $\M_{BC}(q, \{p_i\})$ depends on the soft-gluon index $a$ and on the colour indices of the hard partons $B$ and $C$.
If $\{ BC \} = \{ gg\}$ the hard-gluon colour indices are denoted by $b$ and $c$,
while if the two hard partons are a quark ($B$) and an antiquark ($C$) their colour indices are denoted as $\beta$ and ${\bar \gamma}$, respectively.
The all-order structure of the current in Eq.~(\ref{mbcq}) is as follows
\beq
\label{j1qq}
\bra{a\,\beta\, {\bar \gamma}\,} \; \e^{(\sigma)}_{\mu}(q) \, \J^\mu(q) \;\ket{BC\,}
= t^{a}_{\beta{\bar \gamma}} \,\e^{(\sigma)}_{\mu}(q)
\left( \frac{p_B^\mu}{p_B\cdot q} - \frac{p_C^\mu}{p_C\cdot q}\right) 
\Bigl[ 1 +  f_{h.o.}^{(q{\bar q})}
\Bigr] \;,
\;\;(\{ BC \} = \{ q{\bar q}\}) \;,
\eeq
\beq
\label{j1gg}
\!
\bra{a\,b\, c\,} \; \e^{(\sigma)}_{\mu}(q) \, \J^\mu(q) \;\ket{BC\,} \!=\!
i\,f^{bac} \,\e^{(\sigma)}_{\mu}(q)
\left( \frac{p_B^\mu}{p_B\cdot q} - \frac{p_C^\mu}{p_C\cdot q}\right) 
\Bigl[ 1 +  f_{h.o.}^{(gg)}
\Bigr] \;,
\;\;(\{ BC \} = \{ gg\}) \;,
\eeq
where $\e^{(\sigma)}_{\mu}(q)$ is the physical spin polarization vector of the soft gluon.
As briefly discussed below, Eqs.~(\ref{j1qq}) and (\ref{j1gg}) directly derive from extending and refining the reasoning in Refs.~\cite{Catani:2000pi, Li:2013lsa, Duhr:2013msa}.

We first comment on the case  $\{ BC \} = \{ q{\bar q}\}$. There is a sole colour singlet state $\ket{ABC\,}$ that can be formed by the soft gluon and the 
$q{\bar q}$ pair and, consequently 
(since $\langle a\, \beta\, {\bar \gamma}\, \ket{ABC\,}  
\propto t^a_{\beta {\bar \gamma}}$),
the right-hand side of Eq.~(\ref{j1qq})
is necessarily proportional to the colour matrix $t^a_{\beta {\bar \gamma}}$
in the fundamental representation of $SU(N_c)$.
The soft-gluon current $\J^\mu(q)$ 
depends on the momenta $q,p_B$ and $p_C$. It follows \cite{Catani:2000pi, Li:2013lsa, Duhr:2013msa} that it is necessarily proportional
to the colourless current $j_{BC}^\mu(q)=
p_B^\mu/p_B\cdot q - p_C^\mu/p_C\cdot q$ because of gauge invariance, namely,
because of the current conservation relation $q_\mu \J^\mu(q)=0$
(any terms proportional to $q^\mu$ in $\J^\mu(q)$ do not contribute to 
Eq.~(\ref{j1gg}) since $q^\mu \e^{(\sigma)}_{\mu}(q)=0$ for physical spin polarizations). This simple discussion leaves undetermined only the dimensionless scalar function $f_{h.o.}^{(q{\bar q})}$ in Eq.~(\ref{j1qq}). Setting 
$f_{h.o.}^{(q{\bar q})}=0$, Eq.~(\ref{j1qq}) coincides with
the tree-level current in Eq.~(\ref{J1})
(one can simply use colour conservation, namely, 
$\T_C \,\ket{BC\,} = - \T_B \ket{BC\,}$ in Eq.~(\ref{J1})).
Therefore, the function $f_{h.o.}^{(q{\bar q})}$ is due to loop corrections at higher perturbative orders. The momentum dependence of $f_{h.o.}^{(q{\bar q})}$ is 
constrained by symmetry properties \cite{Catani:2000pi, Li:2013lsa, Duhr:2013msa}
such as, for instance, the invariance of the soft current with respect to
the rescaling $p_B \to \xi_B p_B$ and $p_C \to \xi_C p_C$ of the hard-parton momenta.
In the case of massless quark and antiquark, the functional dependence of 
$f_{h.o.}^{(q{\bar q})}$ is 
\beq
\label{arg}
f_{h.o.}^{(q{\bar q})} =
f_{h.o.}^{(q{\bar q})}\!\left(\g^2 
\Bigl(
\frac{\mu_0^2 \;(- 2 p_B\cdot p_C -i0)}{(- 2 p_B\cdot q -i0) (- 2 p_C\cdot q -i0)}
\Bigr)^{\! \epsilon}
\right) \,, \;\;\;\;(p_B^2=p_C^2=0 ) \;,
\eeq
since the argument of $f_{h.o.}^{(q{\bar q})}$ in the right-hand side of 
Eq.~(\ref{arg}) is the sole Lorentz invariant and dimensionless function that is invariant under the rescaling of the {\em massless} hard momenta
(the infinitesimal part `$-i 0$' arises from the Feynman prescription for analytic continuation in different physical kinematical regions of the momenta $q,p_B,p_C$).
In the case of massive hard partons, additional forms of momentum dependence can 
be present in $f_{h.o.}^{(q{\bar q})}$ (see Ref.~\cite{Bierenbaum:2011gg} at one-loop order). 
For instance, if $p_B^2 \neq 0$ the higher-order function
$f_{h.o.}^{(q{\bar q})}$ can also depend on the argument
$\g^2 [ \mu_0^2 p_B^2/ (- 2 p_B\cdot q -i0)^2 ]^\epsilon$.

The all-order form of the soft-gluon current in Eq.~(\ref{j1gg}) for radiation from two hard gluons follows from a discussion that is analogous to the discussion that leads to Eq.~(\ref{j1qq}) for the $q{\bar q}$ case. The only difference regards a subtle point related to the colour structure, whereas the discussion about the dependence on the momenta $q, p_B$ and $p_C$ is unchanged. The soft gluon can be combined with the two hard gluons $B$ and $C$ in two possible colour singlet states, as given by the antisymmetric and symmetric states in Eq.~(\ref{gggstates}). Therefore, the soft-gluon current 
in Eq.~(\ref{j1gg}) could have two colour components proportional to 
$f^{abc}$ and $d^{abc}$, respectively. The right-hand side of Eq.~(\ref{j1gg})
instead includes only the $f^{abc}$ component, while the $d^{abc}$ component is absent.
At the tree level, the explicit result in Eq.~(\ref{J1}) for the soft-gluon current implies that the soft gluon is produced in the antisymmetric state. We recall
(see the accompanying comments to Eqs.~(\ref{Jgggf}) and (\ref{Jgggd}))
that the antisymmetric and symmetric states have a different charge conjugation.
Since QCD radiative corrections at loop level are invariant under charge conjugation,
high-order loop corrections to the tree-level soft-gluon current cannot produce transitions from the antisymmetric to the symmetric state. This explains the absence of the $d^{abc}$ component in the right-hand side of Eq.~(\ref{j1gg}).
Although the scattering amplitude $\ket{\M_{BC}(q, \{p_i\})}$ in the left-hand side of Eq.~(\ref{mbcq}) can have both antisymmetric and symmetric components, the result in Eq.~(\ref{j1gg}) implies that the symmetric component is dynamically suppressed in the soft limit $q \to 0$.

The single soft-gluon current in Eqs.~(\ref{mbcq})--(\ref{j1gg}) has been explicitly computed at one-loop order (for both massless \cite{Bern:1999ry, Catani:2000pi}
and massive \cite{Bierenbaum:2011gg} hard partons) and two-loop order (for massless hard partons \cite{Badger:2004uk, Li:2013lsa, Duhr:2013msa}). The explicit one-loop
and two-loop results agree with the general structure in 
Eqs.~(\ref{j1qq}) and (\ref{j1gg}). In the case of {\em massless} hard partons,
the functions $f_{h.o.}^{(q{\bar q})}$ and $f_{h.o.}^{(gg)}$ turns out to be equal up to two-loop order. Such an equality is not expected to be valid starting from the three-loop order \cite{Li:2013lsa}. This expectation follows from observing that the colour coefficient dependence of $f_{h.o.}^{(BC)}$ at {\em three-loop} order
is in close correspondence (through Feynman diagrams with similar topologies)
with the colour coefficient dependence of the squared current 
$| \J(q_1,q_2,q_3,q_4) |^{2}_{\; BC}$ for {\em quadruple} soft-gluon radiation at the {\em tree level}. As shown and discussed in Sect.~\ref{sec:quadruple},
such squared current exhibits correlations with colour coefficients that violate
Casimir scaling. The same colour coefficients contribute (modulo accidental cancellations of their overall factor) to $f_{h.o.}^{(BC)}$ at three-loop order,
leading to an explicit dependence of $f_{h.o.}^{(BC)}$ on the colour representation of the hard partons $B$ and $C$.

\subsection{Soft-gluon radiation at the tree level
}
\label{sec:3from2}

The squared current factor $| \J(q_1,\cdots,q_N) |^{2}_{\; BC}$ of Eq.~(\ref{sbc})
for emission of $N=1,2$ and 3 soft gluons at the tree level can be explicitly computed by applying the general results of Sects.~\ref{s:sq12} and \ref{s:sq123} to the case of two hard partons. We recall that colour quadrupole operators do not contribute to soft-gluon radiation from two hard partons (see Eq.~(\ref{q2hard})).
Therefore, only colour dipole operators $\T_i \cdot \T_k$ (and their products)
are involved in the colour structure of the squared current $| \J(q_1,\cdots,q_N) |^2$ for emission of $N=1$
(see Eqs.~(\ref{w1def}) and (\ref{w1new})),
$N=2$ 
(see Eqs.~(\ref{J2sq}) and (\ref{w2new}))
and $N=3$ (see Eqs.~(\ref{J3sq}), (\ref{3cordq}) and (\ref{w3new}))
soft gluons from two hard partons.
The action of colour dipole operators onto a colour singlet state $\ket{BC}$ of two hard partons can be straightforwardly evaluated by using colour charge conservation
($\T_B \;\ket{BC} = - \T_C \;\ket{BC}$), and it leads to $\T_B \cdot \T_C \;\ket{BC}
= -C_B\;\ket{BC})$ (note that $C_C=C_B$, since the two hard partons $\{B, C\}$ are a particle-antiparticle pair). It follows that $| \J(q_1,\cdots,q_N) |^{2}_{\; BC}$ 
with $N=1,2$ and 3 can be directly expressed in terms of the Casimir coefficient 
$C_B$ of the hard partons and of the factors $C_A^{N-1} \w_{BC}^{(N)}$ in 
Eqs.~(\ref{w1new}), (\ref{w2new}) and (\ref{w3new}) for soft-gluon correlated emission. We obtain the following results:
\beq
\label{jbc1g}
| \J(q) |^{2}_{\; BC} \;= \; C_B \;\w_{BC}^{(1)}(q) \;\;.
\eeq
\beq
\label{jbc2g}
| \J(q_1,q_2) |^{2}_{\; BC} \;= \;C_B^2 \;\w_{BC}^{(1)}(q_1) \;\w_{BC}^{(1)}(q_2)  
+ C_B C_A \;\w_{BC}^{(2)}(q_1,q_2) \;\;,
\eeq
\begin{align}
\label{jbc3g}
| \J(q_1,q_2,q_3) |^{2}_{\; BC} &\;= \;
C_B^3 \;\w_{BC}^{(1)}(q_1) \;\w_{BC}^{(1)}(q_2) \;\w_{BC}^{(1)}(q_3)
 \nn \\
&\;+\;C_B^2 C_A \left[ \w_{BC}^{(1)}(q_1) \;\w_{BC}^{(2)}(q_2,q_3)
+ \;\left( 1 \leftrightarrow 2 \right) + \left( 1 \leftrightarrow 3 \right) 
\right] \nn\\
&\;+ \;C_B C_A^2 \;\w_{BC}^{(3)}(q_1,q_2,q_3)
\;\;,
\end{align}
where $\w_{BC}^{(N)}(q_1,\dots,q_N)$ depends on the momenta of the soft gluons and the momenta $p_B$ and $p_C$ of the two hard partons. 
Some brief comments on Eqs.~(\ref{jbc1g})--(\ref{jbc3g}) are presented below.

From the viewpoint of the dependence on the colour coefficients, the expansion 
of Eqs.~(\ref{jbc1g})--(\ref{jbc3g}) in irreducible correlations for multiple 
soft-gluon emission corresponds to an expansion in maximally non-abelian colour factors $C_B^k C_A^{N-k} (k=1,\dots,N)$. We also note that the dependence of 
$| \J(q_1,\cdots,q_N) |^{2}_{\; BC}$ on the colour of the two hard partons $B$ and $C$ is entirely determined by the Casimir invariant $C_B=C_C$ of the two hard partons.
In particular, knowing $| \J(q_1,\cdots,q_N) |^{2}_{\; BC}$ for the case
$\{B,C\}=\{q, {\bar q}\}$ (i.e., $C_B=C_F$), the simple replacement $C_F \to C_A$ is sufficient to get
$| \J(q_1,\cdots,q_N) |^{2}_{\; BC}$ for the case $\{B,C\}=\{g,g\}$ (i.e., $C_B=C_A$).
The correspondence between the $\{q {\bar q}\}$ and $\{gg\}$ cases through this 
simple replacement of colour coefficients can be regarded as a `Casimir scaling' relation.

As shown in Sect.~\ref{sec:3hardtree}, 
in the case of tree-level soft-gluon radiation from three hard partons,
Casimir scaling is fulfilled for emission of $N=1$ and 2 soft gluons,
but it is violated for emission of $N=3$ soft gluons. 
In the case of tree-level soft-gluon radiation from two hard partons,
Eqs.~(\ref{jbc1g})--(\ref{jbc3g}) show the validity of Casimir scaling for
emission of $N=1,2,3$ soft gluons, and we anticipate (see Sect.~\ref{sec:quadruple})
that Casimir scaling is violated for emission of $N=4$ soft gluons. 
In the case of three hard partons, the violation of Casimir scaling is directly due to
colour quadrupole correlations. These quadrupole correlations are also responsible
(though indirectly) for the violation of Casimir scaling in the case of two hard partons (Sect.~\ref{sec:quadruple}).

The results in Eqs.~(\ref{jbc1g})--(\ref{jbc3g}) are valid for both massless and massive hard partons. In the massless case, Eqs.~(\ref{jbc1g}) and (\ref{jbc2g})
coincide with the results in Ref.~\cite{CaGr99}.
In the massive case, Eq.~(\ref{jbc2g}) can straightforwardly be obtained from the soft-current expression in Ref.~\cite{Czakon:2011ve}.

The expression in Eq.~(\ref{jbc3g}) for triple soft-gluon radiation from two hard partons derives from the general results in Sect.~\ref{s:sq123}, and it is valid for arbitrary energies of the soft gluons. Previous results in the literature 
\cite{Bassetto:1984ik, CaCi84, Dokshitzer:1991wu}
are limited to the region of energy strong ordering ($E_1 \ll E_2 \ll E_3$)
of the three soft gluons, and to the case of massless hard partons.
Using the energy strong ordering approximations of $\w_{ik}^{(3)}$ and
$\w_{ik}^{(3)}$ in Eqs.~(\ref{w2som}) and (\ref{w3som}),
we have compared the result in Eq.~(\ref{jbc3g}) with those in 
Refs.~\cite{Bassetto:1984ik, CaCi84, Dokshitzer:1991wu} and we find full agreement.
We add some comments on the comparison.

Triple soft-gluon radiation with energy strong ordering from two generic (either $q{\bar q}$ or $gg$) massless hard partons was computed in Refs.~\cite{CaCi84} and
\cite{Dokshitzer:1991wu}.
The results in Eqs.~(31), (47), (51) and (52) of Ref.~\cite{CaCi84}
and those in our Eq.~(\ref{jbc3g}) have exactly the same structure and, in particular,
the irreducible gluon correlation in Eq.~(52) of Ref.~\cite{CaCi84} identically agrees
with the expression in our Eq.~(\ref{w3som}). Since Eq.~(\ref{w3som}) is valid for both massless and massive hard partons, the formal expressions in Ref.~\cite{CaCi84}
turn out to be valid also in the massive case, although they were derived for massless hard partons.

In Ref.~\cite{Bassetto:1984ik} BCM considered the tree-level emission of $N$ soft gluons from two hard gluons, and they derived an explicit expression for the corresponding squared current $| \J(q_1,\cdots,q_N) |^{2}_{\; g(B)g(C)}$
that is valid in the kinematical region where the soft-gluon energies $E_\ell$ are strongly ordered ($E_1 \ll E_2 \ll \dots \ll E_N)$. The BCM formula is
\beq
\label{bcm}
| \J(q_1,\cdots,q_N) |^{2 \,(s.o.)}_{\; g(B)g(C)} = 2 \,C_A^N \;(p_B\cdot p_C)^2 
\; F_{eik}^{(N+2)}(p_B,p_C,q_1,\cdots,q_N) +{\cal O}((N_c)^{N-2}) \;\;,
\eeq
where $F_{eik}^{(N+2)}$ is the multi-eikonal function in Eq.~(\ref{feik}). The term denoted by ${\cal O}((N_c)^{N-2})$ in the right-hand side of Eq.~(\ref{bcm})
represents contributions that are formally subdominant in the large-$N_c$ approximation \cite{Fiorani:1988by}. These contributions actually vanish for 
$N \leq 3$ \cite{Bassetto:1984ik}, and they first explicitly appear for emission of $N=4$ soft gluons 
(see Eq.~(\ref{bcm4}) in Sect.~\ref{sec:quadruple}).

We have explicitly checked that Eqs.~(\ref{jbc2g}) and (\ref{jbc3g}) agree with the BCM formula (\ref{bcm}). Actually, the check is immediate by observing that setting
$C_B= C_A$ in the expressions of Eqs.~(\ref{jbc2g}) and (\ref{jbc3g}), these expressions are proportional to the energy strong ordering results in 
Eqs.~(\ref{w2bcm}) and (\ref{w3bcm}), whose right-hand side directly involves the 
multi-eikonal function $F_{eik}^{(N+2)}$ with $N=2$ and $N=3$.

The result that we have presented in Eqs.~(\ref{mgggso}) and (\ref{jgggso}) is
a straightforward consequence of Eqs.~(\ref{gsorel}) and (\ref{bcm}). To see this, 
we first specify Eq.~(\ref{gsorel}) to the case of hard gluons 
($\{ B, C \}= \{ g, g\}$),
and we rewrite it as
\beq
\label{bcmabc}
| \J(q_1,\cdots,q_{N-1}) |^{2}_{\; g(A) g(B) g(C)} =
\frac{| \J(q_1,\cdots,q_N) |^{2}_{\; g(B) g(C)}} 
{| \J(q_N) |^{2}_{\; g(B) g(C)}} \;, \;\;( E_N \gg E_\ell \;{\rm with\;} \ell < N) \,,
\eeq
where $q_N=p_A$ (as remarked in the accompanying comments to Eq.~(\ref{gsorel}), the fact that $p_A$ is softer than $p_B$ and $p_C$ has no effect on the left-hand side of 
Eq.~(\ref{bcmabc})). Then we restrict Eq.~(\ref{bcmabc}) to the case of tree-level soft-gluon radiation in the region of strongly ordered energies 
$(E_1 \ll \dots \ll E_N)$. In the right-hand side, we use 
$| \J(q_N) |^{2}_{\; g(B) g(C)}= 2C_A (p_B\cdot p_C)/[(p_C\cdot p_A) (p_A\cdot p_B)]$
(see Eq.~(\ref{jbc1g})) and the BCM formula (\ref{bcm}) for 
$| \J(q_1,\cdots,q_N) |^{2}_{\; g(B) g(C)}$. This directly leads to the expression
of $| \J |^{2 \,(s.o.)}_{\; g(A)g(B)g(C)}$ 
in Eq.~(\ref{jgggso}).


\subsection{Quadruple soft-gluon radiation at the tree level}
\label{sec:quadruple}


Multiple soft-gluon radiation ($N \geq 4$) at the tree level can be examined by extending the analysis of Sects.~\ref{s:sc3} and \ref{s:sq123} to higher soft-gluon multiplicities.

The general expression of the tree-level squared current
$| \J(q_1,q_2,q_3,q_4) |^2$ for {\em quadruple} soft-gluon radiation in a generic process with an arbitrary number of hard-parton external legs can be presented in the following form:
\begin{align}
\label{j24g}
| \J(q_1,q_2,q_3,q_4) |^2 &\;\eqcs 
\;\symm{W^{(1)}(q_1)}{W^{(1)}(q_2)}{W^{(1)}(q_3)}{W^{(1)}(q_4)} \nn \\
&\;+\;\left[ \sym{W^{(2)}(q_1,q_2)}{W^{(1)}(q_3)}{W^{(1)}(q_4)} 
+ \;{\rm ineq.~perms.} \;\{1,2,3,4\} 
\right] \nn\\
&\;+\;\left[ \sy{W^{(2)}(q_1,q_2)}{W^{(2)}(q_3,q_4)}
+ \;{\rm ineq.~perms.} \;\{1,2,3,4\} 
\right] \nn\\
&\;+\;\left[ \sy{W^{(1)}(q_1)}{W^{(3)}(q_2,q_3,q_4)}
+ \;\left( 1 \leftrightarrow 2 \right) + \left( 1 \leftrightarrow 3 \right) 
+ \left( 1 \leftrightarrow 4 \right)
\right] \nn\\
&\;+ \;W^{(4)}(q_1,q_2,q_3,q_4) \;\;.
\end{align}

The expansions in irreducible gluon correlations, $W^{(N)}(q_1,\cdots,q_N)$,
of the squared currents for single, double and triple soft-gluon radiation are given 
by the expressions in Eqs.~(\ref{w1def}), (\ref{J2sq}) and (\ref{J3sq}).
The expression in Eq.~(\ref{j24g}) generalizes these expansions to the case of quadruple soft-gluon radiation. In particular, since $W^{(N)}$ with $N \leq 3$ is 
determined by the corresponding squared currents with $N \leq 3$,
the computation of $| \J(q_1,q_2,q_3,q_4) |^2$ is equivalent to that of the irreducible four-gluon correlations $W^{(4)}$ in the fifth line of Eq.~(\ref{j24g}).

We recall that the correlations $W^{(N)}$ are colour operators, and they appear in the form of symmetrized products in the right-hand side of Eq.~(\ref{j24g}).
The first line involves the symmetrized product of four operators $W^{(1)}(q_\ell)$.
The symmetrized product of four generic colour operators $O_I$ is defined as
\begin{equation}
\label{sym4}
  \symm{O_1}{O_2}{O_3}{O_4}\equiv\frac1{4!}\bigl(O_1 \,O_2 \,O_3 \,O_4+  {\rm perms.}\{1,2,3,4\} \bigr) \;\;,
\end{equation}
where the right-hand side includes the sum over the $4!=24$
permutations of $O_1, O_2, O_3$ and $O_4$. The other lines in the right-hand side of 
Eq.~(\ref{j24g}) involve symmetrized products of two and three colour operators
(these symmetrized products are defined in Eqs.~(\ref{sym2}) and (\ref{sym3}))
and sums over permutations that lead to inequivalent contributions. 
The square-bracket term in the second line involves a sum over 6 inequivalent permutations. 
The square-bracket term in the third line involves a sum over 3 inequivalent permutations. The 4 inequivalent permutations that contribute to the 
square-bracket term in the fourth line are explicitly denoted therein.

The general expression of the irreducible four-gluon correlation 
$W^{(4)}(q_1,q_2,q_3,q_4)$ can be computed by applying the techniques of 
Sects.~\ref{s:sc3} and \ref{s:sq123} to quadruple soft-gluon radiation.
In this subsection we limit ourselves to presenting explicit results for 
quadruple soft-gluon radiation in scattering processes with two hard partons.

In the case of emission of four soft gluons from two hard partons, $B$ and $C$,
we have to evaluate the action of the squared current 
$| \J(q_1,q_2,q_3,q_4) |^2$ or, equivalently, of the correlation operator
$W^{(4)}(q_1,q_2,q_3,q_4)$  onto the colour singlet state $\ket{ B C\,}$,
as in Eq.~(\ref{jbc}). We find the results that are reported below.

The irreducible correlation $W^{(4)}$ has the following form:
\beq
\label{w4bc}
W^{(4)}(q_1,q_2,q_3,q_4) \ket{BC} = \ket{BC} \,
C_B \left[ C_A^3 \,\w^{(4)(L)}_{BC}(q_1,q_2,q_3,q_4) 
+ \lambda_B N_c \,\w^{(4)(S)}_{BC}(q_1,q_2,q_3,q_4)\right] ,
\eeq
where the dependence on the colour state is due to the Casimir coefficient $C_B$
and the colour dependent coefficient $\lambda_B$ of Eq.~(\ref{lquad}).
The functions $\w^{(4)(L)}_{BC}$ and $\w^{(4)(S)}_{BC}$ are colour independent, and they depend on the momenta $\{p_B, p_C, q_1,q_2,q_3,q_4 \}$ of the hard and soft partons. We note that the two contributions in the right-hand side of 
Eq.~(\ref{w4bc}) behave differently in the large-$N_c$ limit: the dominant term
is proportional to $\w^{(4)(L)}_{BC}$, while $\w^{(4)(S)}_{BC}$ produces
a subdominant term that is formally suppressed by a relative factor of order
$1/N_c^2$.
The expression of the squared current is
\begin{align}
\label{j4bc}
& | \J(q_1,q_2,q_3,q_4) |^{2}_{\; BC} \;= \;
C_B^4 \;\w_{BC}^{(1)}(q_1) \;\w_{BC}^{(1)}(q_2) \;\w_{BC}^{(1)}(q_3) 
\;\w_{BC}^{(1)}(q_4)
 \nn \\
&\;+\;C_B^3 C_A \left[ 
\frac{1}{4} \w_{BC}^{(1)}(q_1) \;\w_{BC}^{(1)}(q_2) \;\w_{BC}^{(2)}(q_3,q_4)
+ \;{\rm perms.} \;\{1,2,3,4\} 
\right] \nn\\
&\;+\;C_B^2 C_A^2 \left[ \frac{1}{6} \w_{BC}^{(1)}(q_1) \;\w_{BC}^{(3)}(q_2,q_3,q_4)
+ \frac{1}{8} \w_{BC}^{(2)}(q_1,q_2) \;\w_{BC}^{(2)}(q_3,q_4)
+ \;{\rm perms.} \;\{1,2,3,4\} 
\right] \nn\\
&\;+ C_B C_A^3 \,\w^{(4)(L)}_{BC}(q_1,q_2,q_3,q_4) 
+ C_B \lambda_B N_c \,\w^{(4)(S)}_{BC}(q_1,q_2,q_3,q_4)
\;\;,
\end{align}
where the functions $\w^{(4)(L)}_{BC}$ and $\w^{(4)(S)}_{BC}$ in the fourth line are those in Eq.~(\ref{w4bc}), and the other momentum-dependent correlations $\w^{(N)}$
with $N \leq 3$ are explicitly known from lower-multiplicity results
(see Eqs.~(\ref{w1new}), (\ref{w2new}) and (\ref{w3new})).
The results in Eqs.~(\ref{w4bc}) and (\ref{j4bc}) are valid for both massless and massive hard partons, and for arbitrary energies of the soft gluons. 

We have not computed the functions $\w^{(4)(L)}_{BC}$ and $\w^{(4)(S)}_{BC}$ in explicit form for arbitrary soft-gluon energies. We can present the results in a wide energy ordered region where one gluon is much harder than the others. In the region where
$E_4 \gg E_\ell$ with $\ell=1,2,3$ (note that no other restriction is applied on the energies $E_1$, $E_2$ and $E_3$ of the softer gluons), we find
\begin{align}
\label{j4bcl}
\w^{(4)(L)}_{BC}(q_1,q_2,q_3,q_4) &= \w_{BC}^{(1)}(q_4)
\left\{ \w_{4BC}^{(3)}(q_1,q_2,q_3) + 
\w_{4BC}^{(1)}(q_1) \;\w_{4BC}^{(1)}(q_2) \;\w_{4BC}^{(1)}(q_3) \right. \nn\\
&+ \!\left. \left[ \w_{4BC}^{(1)}(q_1) \;\w_{4BC}^{(2)}(q_2,q_3) + \;\left( 1 \leftrightarrow 2 \right) + \left( 1 \leftrightarrow 3 \right) \right]
\right\} , \; (E_\ell \ll E_4, \;\ell \leq 3) ,
\end{align}
\beq
\label{j4bcs}
\w^{(4)(S)}_{BC}(q_1,q_2,q_3,q_4) = \w_{4BC}^{(3) {\rm quad.}}(q_1,q_2,q_3) 
\;\w_{BC}^{(1)}(q_4) \;, \quad (E_\ell \ll E_4, \;\ell \leq 3) \;\;,
\eeq
where $\w_{4BC}^{(N)}$ with $N \leq 3$ is the lower-multiplicity dipole correlation in Eq.~(\ref{wabc}), and $\w_{4BC}^{(3) {\rm quad.}}$ is the quadrupole correlation function in Eq.~(\ref{w3abcquad}) for triple soft-gluon radiation from three hard partons (we have to set $p_A=q_4$ 
in both Eqs.~(\ref{wabc}) and (\ref{w3abcquad})). Obviously, simple permutations of the four gluon momenta in Eqs.~(\ref{j4bcl}) and (\ref{j4bcs}) are sufficient to give the results in the other three regions where $E_3$, or $E_2$, or $E_1$ is the larger soft-gluon energy.
 
We illustrate the derivation of the results in Eqs.~(\ref{w4bc})--(\ref{j4bcs}).
To compute the squared current
$| \J(q_1,q_2,q_3,q_4) |^{2}_{\; BC}$ we start from the general expression
in Eq.~(\ref{j24g}), and we evaluate its action onto the colour singlet state 
$\ket{ B C\,}$. The right-hand side of Eq.~(\ref{j24g}) involves the four-gluon
correlation $W^{(4)}$ and the correlations $W^{(N)}$ for lower gluon multiplicities
$(N \leq 3)$. The contribution of $W^{(N)}$ with $N \leq 3$ to 
$| \J(q_1,q_2,q_3,q_4) |^{2}_{\; BC}$ is given in the first three lines in the 
right-hand side of Eq.~(\ref{j4bc}). This contribution can be obtained in a straightforward way since the correlations $W^{(N)}$ with $N \leq 3$ are known
(see Eqs.~(\ref{w1new}), (\ref{w2new}) and (\ref{w3new})),
and their action onto $\ket{ B C\,}$ is completely given in terms of colour dipole operators (analogously to the computation of Eqs.~(\ref{jbc1g})--(\ref{jbc3g})).
The last line in the right-hand side of Eq.~(\ref{j4bc}) is the contribution
of $W^{(4)}$ (see Eq.~(\ref{w4bc})) to $| \J(q_1,q_2,q_3,q_4) |^{2}_{\; BC}$.

To obtain the results in Eqs.~(\ref{w4bc}), (\ref{j4bcl}) and (\ref{j4bcs})
we exploit Eqs.~(\ref{gsorel}) and (\ref{3gabc}). More explicitly, we first use
Eq.~(\ref{gsorel}) at the tree-level with $N=4$ in the corresponding energy ordering region, and we have
\beq
\label{sorel4g}
| \J(q_1,q_2,q_3,q_4) |^{2}_{\; BC} = | \J(q_4) |^{2}_{\; BC} \;\;
| \J(q_1,q_2,q_3) |^{2}_{\; ABC}
\;, \;\;\;\;\;\; (E_\ell \ll E_4 ,\; \ell \leq 3) \,,
\eeq
where $p_A=q_4$. Then we use $| \J(q_4) |^{2}_{\; BC}$ as given in Eq.~(\ref{jbc1g})
and the result in Eq.~(\ref{3gabc}) for $| \J(q_1,q_2,q_3) |^{2}_{\; ABC}$.
We obtain
\beq
\label{sorelvs3}
| \J(q_1,q_2,q_3,q_4) |^{2}_{\; BC} = C_B \;\w_{BC}^{(1)}(q_4)
\left[ \;| \J(q_1,q_2,q_3) |^{2 \,({\rm dip.})}_{\; 4BC}
+ \lambda_B N_c \;\w_{4BC}^{(3) {\rm quad.}}(q_1,q_2,q_3) \;\right] \;\;,
\eeq
where we have set $p_A=q_4$. Finally, the expression (\ref{3gdipabc})
of $| \J(q_1,q_2,q_3) |^{2 \,({\rm dip.})}_{\; ABC}$ (with $p_A=q_4$) can be inserted in Eq.~(\ref{sorelvs3}), and the result can be compared with the right-hand side of 
Eq.~(\ref{j4bc}). The comparison has to be performed in the energy ordering region
where $E_\ell \ll E_4$ (with $\ell \leq 3$) by exploiting
the corresponding energy ordering approximations in Eqs.~(\ref{w2som}) 
and (\ref{w312}). The comparison shows that Eq.~(\ref{sorelvs3}) correctly reproduces all the terms with colour factors $C_B^4, C_B^3 C_A$ and $C_B^2 C_A^2$ in 
Eq.~(\ref{j4bc}). Moreover, the comparison is used to determine the remaining terms in the right-hand side of Eq.~(\ref{j4bc}), and we obtain the results in 
Eqs.~(\ref{w4bc}), (\ref{j4bcl}) and (\ref{j4bcs}).

The derivation of Eq.~(\ref{w4bc}) that we have just described is valid in the energy region where $E_\ell \ll E_4$ with $\ell \leq 3$. However, the energy restriction is only relevant to determine the corresponding expressions of $\w^{(4)(L)}_{BC}$ and
$\w^{(4)(S)}_{BC}$ in Eqs.~(\ref{j4bcl}) and (\ref{j4bcs}). Indeed, the colour coefficient dependence in the right-hand side of Eq.~(\ref{w4bc}) is the correct and general dependence for arbitrary energies of the soft gluons. This conclusion about the colour structure of Eq.~(\ref{w4bc}) is a consequence of the following observation. The computation of $| \J(q_1,q_2,q_3,q_4) |^{2}_{\; BC}$ in the energy ordered region requires the evaluation of {\em all} the topologically-distinct Feynman diagrams that contribute to the squared current for arbitrary energies of the four soft gluons. The explicit evaluation of the corresponding colour coefficients
directly leads to the colour structure of Eq.~(\ref{w4bc}). The energy ordering approximation only affects the actual computation of the momentum dependence of the 
Feynman diagrams and the ensuing momentum dependence of the squared soft current.

The comparison between Eqs.~(\ref{j4bc}) and (\ref{sorelvs3}) also leads to a direct interpretation of the colour structure of the irreducible four-gluon correlation 
$W^{(4)}$ in Eq.~(\ref{w4bc}). In the energy ordered region, the term 
$| \J(q_4) |^{2}_{\; BC}| \,\J(q_1,q_2,q_3) |^{2 \,({\rm dip.})}_{\; 4BC}$
in Eq.~(\ref{sorelvs3}) is due to the iteration of colour dipole correlations for subsequent radiation of soft gluons. This term produces the contributions to 
Eq.~(\ref{j4bc}) that are proportional to the colour factors $C_B^{4-k} C_A^k$
($k=0,1,2,3$). In particular, the contribution $C_B C_A^3 \,\w^{(4)(L)}_{BC}$
to Eq.~(\ref{w4bc}) can be regarded as the maximally non-abelian (irreducible)
colour dipole correlation for quadruple soft-gluon radiation.
The term 
$| \J(q_4) |^{2}_{\; BC} \, \lambda_B N_c \;\w_{4BC}^{(3) {\rm quad.}}$
in Eq.~(\ref{sorelvs3}) originates from colour quadrupole correlations for triple-soft gluon radiation. This term produces the contribution
$C_B \lambda_B N_c \,\w^{(4)(S)}_{BC}$ to Eq.~(\ref{w4bc}), which can be regarded 
as a `quadrupole-induced' irreducible correlation for radiation of four soft gluons from two hard partons.

The squared current for quadruple soft-gluon radiation form two massless hard partons $\{B,C\}$ was examined in Chapter~6.3 of Ref.~\cite{Dokshitzer:1991wu}.
The results of Ref.~\cite{Dokshitzer:1991wu} refer to the kinematical region
where the 
soft-gluon energies are strongly ordered (e.g., $E_1 \ll E_2 \ll E_3 \ll E_4$).
In particular, the authors of Ref.~\cite{Dokshitzer:1991wu} pointed out the presence
of an irreducible-correlation contribution with colour factor $C_B N_c$,
which was named `colour monster' contribution, and they explicitly computed it for
the case of soft-gluon radiation from a quark--antiquark pair 
($\{B,C\}= \{q,{\bar q}\}$). These findings are fully consistent with the colour structure of Eq.~(\ref{w4bc}), where the term $C_B \lambda_B N_c \,\w^{(4)(S)}_{BC}$,
with $C_B \lambda_B= C_F/2$,
corresponds to the colour monster contribution.

The results in Eqs.~(\ref{w4bc})--(\ref{j4bcs}) extend the analysis of 
Ref.~\cite{Dokshitzer:1991wu} in many respects. Equations~(\ref{w4bc}) 
and (\ref{j4bc})
are fully general: they are valid for both massless and massive hard partons and for arbitrary soft-gluon energies. 
In the following, we generically refer to the term 
$C_B \lambda_B N_c \,\w^{(4)(S)}_{BC}$ in Eqs.~~(\ref{w4bc}) and (\ref{j4bc})
as the colour monster contribution, independently of its actual dependence on the momenta of the hard and soft partons.
In particular,
Eq.~(\ref{w4bc}) shows that the colour monster contributions for $\{B,C\}= \{q,{\bar q}\}$ and $\{B,C\}= \{g,g\}$
hard processes are directly proportional through the colour factor
$(C_F \lambda_F)/(C_A \lambda_A)= C_F/(6 C_A)$. Energy ordering approximations only affect the expressions in Eqs.~(\ref{j4bcl}) and (\ref{j4bcs}), which are nonetheless valid in an
energy region that is wider than the strongly-ordered energy region examined in 
Ref.~\cite{Dokshitzer:1991wu}. Using the energy strong-ordering approximations in Eqs.~(\ref{w2som0}), (\ref{w3som0}) and (\ref{wquadso}),
we have explicitly checked that Eqs.~(\ref{j4bc})--(\ref{j4bcs}) agree with the results of Ref.~\cite{Dokshitzer:1991wu}. 
In particular, in the case of radiation from a 
massless quark and antiquark 
($\{B,C\}= \{q,{\bar q}\}$), the result in Eqs.~(\ref{j4bc}) and (\ref{j4bcs}) agrees
with the colour monster contribution in Eqs.~(6.51a) and (6.51b) of 
Ref.~\cite{Dokshitzer:1991wu}.

The colour monster contribution for radiation from two hard gluons 
($\{B,C\}= \{g,g\}$) was not explicitly evaluated in Ref.~\cite{Dokshitzer:1991wu}.
Using Eqs.~(\ref{j4bc})--(\ref{j4bcs}) we can compute the squared current for quadruple soft-gluon radiation from two hard gluons in the energy strong-ordering region where $E_1 \ll E_2 \ll E_3 \ll E_4$. We find
\beeq
\label{bcm4}
| \J(q_1,q_2,q_3,q_4) |^{2 \,(s.o.)}_{\; g(B)g(C)} &=& 2 \,C_A^4 \;(p_B\cdot p_C)^2 
\; F_{eik}^{(6)}(p_B,p_C,q_1,q_2,q_3,q_4) \nn \\
&+& 3 \,N_c^{2} \;\w_{BC}^{(4) (S) \,(s.o.)}(q_1,q_2,q_3,q_4)
\;\;,
\eeeq
where $\w_{BC}^{(4) (S) \,(s.o.)}$ is the energy 
strong-ordering
approximation of $\w_{BC}^{(4) (S)}$ and it is presented in Eq.~(\ref{cm0}).
The expression in Eq.~(\ref{bcm4}) is fully consistent with the 
BCM formula (\ref{bcm}). In particular, the second term in the right-hand side of 
Eq.~(\ref{bcm4}) gives the first explicit correction to the multi-eikonal
BCM result of Eq.~(\ref{bcm}). Owing to the angular-ordering features of multiple soft gluon radiation, such correction is expected \cite{Bassetto:1984ik} to be dynamically suppressed in collinear regions. Our explicit result actually shows a very strong suppression. The contribution of the multi-eikonal function $F_{eik}^{(6)}$ to 
Eq.~(\ref{bcm4})
is singular in all the multiple collinear limits that involve either soft gluons
or soft gluons and one hard parton. The contribution of $\w_{BC}^{(4) (S) \,(s.o.)}$
is instead singular only in the five-parton collinear limit of the four soft gluons and one hard parton (see Eq.~(\ref{cm0}) and accompanying comments).
Therefore, performing the $d$-dimensional integration over the angles of the four soft gluons,  $F_{eik}^{(6)}$ can produce collinear divergences of 
${\cal O}(1/\epsilon^4)$, while $\w_{BC}^{(4) (S) \,(s.o.)}$ produces a collinear divergence of ${\cal O}(1/\epsilon)$.

In the energy strong-ordering region where $E_1 \ll E_2 \ll E_3 \ll E_4$, inserting Eq.~(\ref{wquadso}) in Eq.~(\ref{j4bcs}) we can straightforwardly obtain a compact expression for the colour monster function $\w_{BC}^{(4) (S)}$. In the case of massive hard partons $(p_B^2\neq 0, p_C^2\neq 0)$ we find the expression
\beeq
\label{cmm}
\w_{BC}^{(4) (S)}(q_1,q_2,q_3,q_4) = \w_{BC}^{(1)}(q_4)
\Bigl\{ \!\!\!\!\!\!\!\!\!\!&&  
\Bigl[ \,\w_{4C}^{(1)}(q_3) \;G_{34CB}(q_1) 
             + \frac{1}{2} \, \w_{BC}^{(1)}(q_3) \;G_{3B4C}(q_1) \Bigr] 
       G_{3C4B}(q_2) \Bigr. \nn \\
\!&&\Bigl. + \;( B \leftrightarrow C) \Bigr\} + ( 1 \leftrightarrow 2) \;,
~~~~~(E_1 \ll E_2 \ll E_3 \ll E_4) \,, \nn \\
&&{}
\eeeq
which, in the massless case, can be rewritten as
\beeq
\label{cm0}
&& \!\!\!\!\!\!\!\!\!\!\!\!
\w_{BC}^{(4) (S) \,(s.o.)}(q_1,q_2,q_3,q_4) =
2 \,(p_B \cdot p_C)^2
\Bigl\{  
\Bigl[ \,\frac{2 \,G_{34CB}(q_1)}{(q_3 \cdot q_4) (p_C \cdot p_B)} 
             + \frac{G_{3B4C}(q_1)}{(q_3 \cdot p_B) (q_4 \cdot p_C)} \Bigr] 
\Bigr. \nn \\
&& \Bigl.  \times \, \frac{G_{3C4B}(q_2)}{(q_3 \cdot p_C) (q_4 \cdot p_B)} 
+ \;( 3 \leftrightarrow 4) \Bigr\} + ( 1 \leftrightarrow 2) \;\;,
~~~~~~(E_1 \ll E_2 \ll E_3 \ll E_4) \,, 
\Bigr. 
\eeeq
where $G_{imkl}(q_\ell)$ is the quadrupole eikonal function in Eq.~(\ref{gquad}).
We note that Eq.~(\ref{cmm}) is symmetric with respect to the exchange 
$q_1 \leftrightarrow q_2$, whereas its massless limit in Eq.~(\ref{cm0}) is remarkably symmetric also with respect to the exchange $q_3 \leftrightarrow q_4$.

The colour monster contribution $\w^{(4)(S)}_{BC}$ has collinear singularities, as first noticed in Ref.~\cite{Dokshitzer:1991wu}. The expression in Eq.~(\ref{cm0})
is particularly suitable to highlight the collinear behaviour of the colour monster in the energy
strong-ordering region. We see that $\w_{BC}^{(4) (S) \,(s.o.)}$ is a sum of terms that include the product of two factors with the following structure:
\beq
\label{goverk}
\frac{G_{imkl}(q_\ell)}{(k_i \cdot k_m) (k_k \cdot k_l)} \;\;, 
~~~~~~~~~~~~\ell=1,2 \;\;,
\eeq
where the four {\em distinct} momenta $\{k_i, k_m, k_k, k_l\}$ are either hard-parton momenta or soft-gluon momenta harder than $q_\ell$.
As discussed in the accompanying comments to Eq.~(\ref{gquad}), the quadrupole eikonal function $G_{imkl}(q_\ell)$ is collinear safe with respect to the direction of $q_\ell$. Moreover, $G_{imkl}(q_\ell)$ vanishes if $k_i$ and $k_m$ 
(or, $k_k$ and $k_l$) are collinear, and such vanishing behaviour cancels the collinear singularity of Eq.~(\ref{goverk}) in the limit 
$k_i \cdot k_m \to~0$ (or, $k_k \cdot k_l \to 0$). Such cancellation mechanism is effective in various multiparton collinear limits, and eventually 
$\w_{BC}^{(4) (S) \,(s.o.)}(q_1,q_2,q_3,q_4)$
is singular only in the five-parton collinear limit of the four soft gluons and one of the two massless hard partons. The same conclusion on the collinear behaviour of the colour monster function $\w^{(4)(S)}_{BC}$ applies to the energy ordering approximation in Eq.~(\ref{j4bcs}). In the expression of Eq.~(\ref{j4bcs})
the factor $\w_{BC}^{(1)}(q_4)$ for the emission of the hardest soft gluon
is singular if the momentum $q_4$ is collinear to one of the massless hard-parton momenta $p_B$ and $p_C$. However, such collinear singularity is partly screened by
the vanishing behaviour of the quadrupole
correlation factor $\w_{4BC}^{(3) {\rm quad.}}(q_1,q_2,q_3)$, which is collinear safe
with respect to the directions of $q_1,q_2$ and $q_3$.

We comment on the colour structure of quadruple soft-gluon radiation from two hard partons and, in particular, on the dependence on the colour representation of the hard partons. This dependence is encoded by the colour coefficients $C_B$ and $\lambda_B$ in Eqs.~(\ref{w4bc}) and (\ref{j4bc}). We note that the simple replacement
$C_F \to C_A$ is not sufficient to relate soft radiation from two hard quarks and two hard gluons. Therefore, quadruple soft-gluon radiation at the tree level produces {\em violation} of Casimir scaling. The violation is due to the irreducible correlation
$W^{(4)}$ in Eq.~(\ref{w4bc}) and, specifically, to its colour monster contribution, which depends on the colour coefficient $\lambda_B$. Although the colour monster contribution is formally suppressed in the large-$N_c$ limit, its relative effect in 
$\{B,C\}= \{g,g\}$ and $\{B,C\}= \{q,{\bar q}\}$ scattering amplitudes is sizeably different since $\lambda_A/\lambda_F=6$ (see Eq.~(\ref{lquad})). We also note that the momentum dependence of the colour monster contribution is the same in
$\{B,C\}= \{q,{\bar q}\}$ and $\{B,C\}= \{g,g\}$ scattering amplitudes. Therefore,
the corresponding squared currents for quadruple soft-gluon emission can be related
through a {\em generalized} form of Casimir scaling. The generalized Casimir scaling relation involves the simultaneous replacement $C_F \to C_A$  and 
$\lambda_F \to \lambda_A$.

In the case of colour singlet quantities such as the squared currents for soft-gluon radiation from two hard partons, violation of Casimir scaling at high perturbative orders is expected. The expectation is based on the fact that the colour factors of the corresponding Feynman diagrams depend not only on quadratic Casimir coefficients
(e.g., $C_F$ and $C_A$) but also on additional Casimir invariants (see, e.g., Refs.~\cite{vanRitbergen:1998pn}). In the case of quadruple soft-gluon radiation from two hard partons, the additional invariants are the `quartic' Casimir coefficients
$d_{AB}^{(4)}$, with $B=F$ if $\{B,C\}= \{q,{\bar q}\}$ and $B=A$ if
$\{B,C\}= \{g,g\}$. We use the definition and normalization of $d_{AB}^{(4)}$
as specified in Eqs.~(2.6)--(2.9) of Ref.~\cite{Moch:2018wjh}.
In particular, for $SU(N_c)$ QCD we have 
\beq
\label{cas4}
\frac{d_{AA}^{(4)}}{N_A} = \frac{N_c^2 (N_c^2 + 36)}{24} \;\;,
\quad \quad
\frac{d_{AF}^{(4)}}{N_F} = \frac{(N_c^2 - 1) (N_c^2 + 6)}{48} \;\;,
\eeq
where $N_B$ is the dimension of the colour representation of the hard parton $B$
($N_B=N_F=N_c$ if $B$ is a quark, and $N_B=N_A=N_c^2-1$ if $B$ is a gluon).

The coefficient of the colour monster contribution to Eq.~(\ref{w4bc}) can be expressed in terms of the quadratic and quartic Casimir invariants. Indeed, we find
\beq
\label{cm4}
C_B \,\lambda_B \,N_c = 2 \,\frac{d_{AB}^{(4)}}{N_B} - \frac{1}{12}\, C_B \,C_A^3 
\;\;.
\eeq
Correspondingly, the irreducible four-gluon correlation $W^{(4)}$ in 
Eq.~(\ref{w4bc}) can be rewritten as follows
\beq
\label{w4cas}
W^{(4)}(q_1,q_2,q_3,q_4) = 
C_B \,C_A^3 \,\w^{(4)[2]}_{BC}(q_1,q_2,q_3,q_4) + 
2 \,\frac{d_{AB}^{(4)}}{N_B} \,\w^{(4)[4]}_{BC}(q_1,q_2,q_3,q_4) \;\;,
\eeq
where $\w^{(4)[4]}_{BC}=\w^{(4)(S)}_{BC}$ and
\beq
\w^{(4)[2]}_{BC}(q_1,q_2,q_3,q_4) = \w^{(4)(L)}_{BC}(q_1,q_2,q_3,q_4) 
- \frac{1}{12} \,\w^{(4)(S)}_{BC}(q_1,q_2,q_3,q_4) \;\;.
\eeq
Therefore, the violation of Casimir scaling in $W^{(4)}$, as due to the 
colour monster contribution, can be equivalently view as the effect produced by the quartic Casimir invariant. The generalization of Casimir scaling that we have already noticed appears in Eq.~(\ref{w4cas}) through the dependence on the quadratic and quartic Casimir coefficients $C_B$ and $d_{AB}^{(4)}/N_B$.

Our discussion on the colour structure of the squared current
$| \J(q_1,q_2,q_3,q_4) |^{2}_{\; BC}$ eventually depends only on the colour coefficients of the corresponding Feynman diagrams. Therefore, colour monster contributions do appear (modulo accidental, though unlikely, cancellations)
in related soft factors for radiation from two hard partons at ${\cal O}(\as^4)$. 
Specifically, we refer to the soft factors for emission of three gluons at one loop,
two gluons at two loops, and a single gluon at three loops.

The squared current $| \J(q_1,q_2,q_3,q_4) |^{2}_{\; BC}$ in Eq.~(\ref{j4bc})
for quadruple soft-gluon radiation has collinear singularities, analogously to the squared currents for lower soft-gluon multiplicities (see Sect.~\ref{s:coll}).
In particular, the colour monster contribution $\w^{(4)(S)}_{BC}$ has collinear singularities, as we have previously discussed.

The collinear singularity of the colour monster term implies that, at the inclusive level, it also contributes to the soft limit of the collinear evolution kernel 
of the parton distribution functions \cite{Moch:2017uml, Moch:2018wjh}.
The soft limit of the evolution kernel is proportional to the cusp anomalous 
dimension \cite{Korchemsky:1988si}. At ${\cal O}(\as^4)$ the cusp anomalous dimension for quark and gluon evolution violates quadratic Casimir scaling, and the violation is due to terms proportional to quartic Casimir invariants \cite{Moch:2018wjh}. The relation between the
cusp anomalous dimension and soft multiparton radiation is discussed in 
Ref.~\cite{Catani:2019rvy}. The term of the cusp anomalous dimension that is proportional to the quartic Casimir invariants $d_{AA}^{(4)}$ and $d_{AF}^{(4)}$ of Eqs.~(\ref{cas4}) and (\ref{cm4}) (such term is presently known in approximated numerical form \cite{Moch:2018wjh}) is due to the colour monster contribution in 
Eq.~(\ref{j4bc}) (see also Eq.~(\ref{w4cas}))
and to analogous contributions from loop corrections to the squared currents 
$| \J(q_1,\cdots,q_N) |^{2}_{\; BC}$ with $N \leq 3$.

We have devoted the entire Sect.~\ref{sec:2hard} to examine multiple soft-gluon radiation in a generic hard-scattering amplitude $\M_{BC}(\{q_\ell\}, \{p_i\})$
with two hard partons, and we have explicitly discussed the physically most relevant cases in which the two hard partons $B$ and $C$ are either two gluons or a $q{\bar q}$ pair. Most of the discussion and, in particular, the main final results can be directly extended to arbitrary hard partons. In the following we briefly comment on such extension.

We consider soft-gluon radiation in the general case in which the hard partons $B$ and $C$ (with arbitrary massless or massive momenta $p_A$ and $p_B$) of  
$\M_{BC}(\{q_\ell\}, \{p_i\})$ belong to generic {\em irreducible} colour representations of $SU(N_c)$. The only constraint is that $B$ and $C$ belong to colour conjugate representations\footnote{If $B$ and $C$ do not belong to
conjugate representations, the hard-parton state $\ket{\M_{BC}(\{p_i\})}$ cannot be a colour singlet state.}, say, $B$ in the representation $R$ and
$C$ in the representation ${\overline R}$ that is conjugate to $R$
($R$ and ${\overline R}$ are not necessarily inequivalent, and we can also have
$R={\overline R}$ such as, for instance, if $\{B, C\} = \{g, g\}$).
Therefore, there is only {\em one} colour singlet configuration $\ket{BC\,}$ of the two hard partons, and Eqs.~(\ref{jbc}) and (\ref{sbc}) are valid. In particular, we can consider the generic $c$-number squared current 
$| \J(q_1,\cdots,q_N) |^{2}_{\; BC}$, whose colour factors depend on the colour representation $R$ of the hard partons.

The results in Eqs.~(\ref{jbc1g})--(\ref{jbc3g}), (\ref{w4bc}) and (\ref{j4bc}) 
about the squared currents
for single, double, triple and quadruple soft-gluon radiation at the tree level are valid for the generic colour representation $R$. In these equations the soft functions
$\w_{BC}^{(N)}$ (with $N=1,2,3$), $\w^{(4)(L)}_{BC}$ and $\w^{(4)(S)}_{BC}$ have a pure kinematical dependence on the parton momenta and 
no dependence
on colour coefficients. The colour dependence is embodied in the coefficients $C_B$ and 
$\lambda_B$.

The coefficient $C_B$ is the Casimir coefficient of the hard parton $B$, namely, it is
the quadratic Casimir $C_B = T^a_R \,T^a_R \equiv \T_R \cdot \T_R$ of the colour matrices $T^a_R$ of the representation $R$. The colour quadrupole coefficient 
$\lambda_B$ of the hard parton $B$ is defined by generalizing 
Eqs.~(\ref{qqqg}) and (\ref{qgggf}) to the case of a generic representation $R$.
The generalization is obtained by considering a specific colour singlet state, 
$\ket{(ABC)_R \,}$, which is formed by the gluon $A$ and the partons $B$ and $C$ in the representations $R$ and ${\overline R}$. This colour singlet state is defined as
\beq
\label{abcR}
\langle a \, \beta \, {\overline \gamma}\; \ket{\,(ABC)_R \,} = 
\left( T^a_R \right)_{\beta \, {\overline \gamma}} \;\;,
\eeq
where $a, \beta$ and ${\overline \gamma}$ are the colour indices of the gluon $A$, parton $B$ and parton $C$. The colour quadrupole coefficient $\lambda_B$ of the parton $B$ in the colour representation $R$ is then defined as follows by the expectation value of the quadrupole operator $\qu{BCBC}$ onto this colour singlet state
(see Eqs.~(\ref{qqqg}) and (\ref{qgggf}) for comparison):
\beq
\label{lambdaR}
N_c \; \lambda_B \equiv \frac{\bra{\,(ABC)_R \,} \; \qu{BCBC} \;
\ket{\,(ABC)_R \,} }{\langle \, (ABC)_R \, \ket{\,(ABC)_R \,}} \;\;.
\eeq

Using this definition of $\lambda_B$, the derivation of Eqs.~(\ref{w4bc}) and 
(\ref{j4bc}) that we have previously illustrated applies to the generic hard partons 
$B$ and $C$. Moreover, using $SU(N_c)$ colour algebra we find that the colour monster
coefficient $C_B \lambda_B N_c$ and the gluon ($A$) quartic Casimir invariant  
$d_{AB}^{(4)}$ (we still refer to the definition in Eqs.~(2.6) and (2.7) of 
Ref.~\cite{Moch:2018wjh}) fulfil the relation in Eq.~(\ref{cm4}) for an arbitrary 
irreducible colour representation $R$ of the parton $B$.

\section{Generating functional and exponentiation
\label{s:exp}}

The expansion of the squared current $| \J(q_1,\cdots,q_N) |^{2}$ in terms of irreducible correlations $W^{(N)}(q_1,\cdots,q_N)$ can be recast in a compact form by introducing the corresponding generating functional. Considering the soft-gluon squared current for a generic scattering amplitude, we define the generating functional $\Psi[u]$ as follows
\beq
\label{gf}
\Psi[u] \equiv 1 + \sum_{N=1}^{\infty} \frac{1}{N!} 
\int \left( \prod_{\ell=1}^N \,[dq_\ell] \;u(q_\ell) \right)
| \J(q_1,\cdots,q_N) |^{2} \;\;,
\eeq
where $u(q)$ is an auxiliary weight function of the soft-gluon momentum $q$ and the soft-gluon $d$-dimensional phase space is denoted as
\beq
\label{phsp}
[dq_\ell] \equiv \frac{d^d q_\ell}{\left(2\pi\right)^{d-1}} \;\delta_+(q_\ell^2) \;\;.
\eeq
Equivalently, the squared current is obtained through the functional derivative
$\delta/(\delta u(q))$ of
$\Psi[u]$ with respect to the weight function:
\beq
\label{dgf}
| \J(q_1,\cdots,q_N) |^{2} \equiv \left.
\left( \prod_{\ell=1}^N \frac{\delta}{\delta u(q_\ell)}\right) 
\,\Psi[u] \,
\right|_{u=0} \;\;.
\eeq

The irreducible correlations $W^{(N)}(q_1,\cdots,q_N)$ are therefore defined through
the logarithm of $\Psi[u]$. We write
\beq
\label{defexp}
\Psi[u] \equiv \exp \bigl\{ W[u] \bigr\} \;\;,
\eeq
and we have
\beq
\label{defwn}
 W^{(N)}(q_1,\cdots,q_N) = \left.
\left( \prod_{\ell=1}^N \frac{\delta}{\delta u(q_\ell)}\right) 
\,W[u] \,
\right|_{u=0} \;\;,
\eeq
or, equivalently,
\beq
\label{wudef}
W[u] = \sum_{N=1}^{\infty} \frac{1}{N!} 
\int \left( \prod_{\ell=1}^N \,[dq_\ell] \;u(q_\ell) \right)
W^{(N)}(q_1,\cdots,q_N)  \;\;.
\eeq
Note that, analogously to $| \J |^{2} $ and $W^{(N)}$, the generating functionals 
$\Psi[u]$ and $W[u]$ are colour operators acting on the hard-parton
scattering amplitude $\ket{\M (\{p_i\})}$. 

The exponential form of $\Psi[u]$ in Eq.~(\ref{defexp}) is just a definition of the irreducible correlations $W^{(N)}$, with no additional physics content. The physics content is introduced by considering the explicit results in 
Eqs.~(\ref{w1new}), (\ref{w2new}), (\ref{3cordq}), (\ref{3corquad}), and (\ref{w3new})
and inserting them in Eqs.~(\ref{defexp}) and (\ref{wudef}).
We obtain
\beq
\label{gfexpl}
\Psi[u] = \exp \left\{ - \frac{1}{2} \sum_{i,k} \T_i \cdot \T_k 
\;W_{(p_i,p_k)}[u;N_c]
+ \sum_{i,m,k,l} \qu{imkl} \;W_{(p_i,p_m,p_k,p_l)}[u]
+{\cal O}(u^4) 
\right\} \;,
\eeq
where the dipole ($W_{(p_i,p_k)}[u;N_c]$) and quadrupole ($W_{(p_i,p_m,p_k,p_l)}[u]$)
generating functionals are
\beeq
\label{dipw}
W_{(p_i,p_k)}[u;N_c] &=& \int [dq] \;u(q) \;\w_{ik}^{(1)}(q)
+ \frac{1}{2} \,C_A \int [dq_1] [dq_2] \;u(q_1) u(q_2) \;\w_{ik}^{(2)}(q_1,q_2)
\nn \\
&+& \frac{1}{3!} \,C_A^2 \int [dq_1] [dq_2] [dq_3] \;u(q_1) u(q_2) u(q_3)
\;\w_{ik}^{(3)}(q_1,q_2,q_3) \;\;,
\eeeq
\beq
\label{quadw}
W_{(p_i,p_m,p_k,p_l)}[u] = \frac{1}{3!} \int [dq_1] [dq_2] [dq_3] 
\;u(q_1) u(q_2) u(q_3) \;\cS_{imkl}(q_1,q_2,q_3) \;\;.
\eeq
We note that the quadrupole operators $\qu{imkl}$ are irreducible to dipole operators
(see Sect.~\ref{s:cor123} and Appendix~\ref{a:quad})
and, therefore, the distinction between the dipole and quadrupole terms in 
Eq.~(\ref{gfexpl}) is physically meaningful
(e.g., gauge invariant).
 
We recall that, according to our notation, the tree-level current $\J(q_1,\cdots,q_N)$
is defined by factorizing the overall power $\g^N$ of the QCD coupling 
(see Eq.~(\ref{1gfact})). Therefore, the expansions in powers of $u$ in 
Eqs.~(\ref{gfexpl})--(\ref{quadw}) are analogous to expansions in powers of $\as$,
with the formal correspondence ${\cal O}(u^n) \sim {\cal O}(\as^n)$.

We note that the entire dependence of $\Psi[u]$ on the colour charges of the hard partons is embodied in the exponentiated dipole ($\T_i \cdot \T_k$) and quadrupole
($\qu{imkl}$) operators of Eq.~(\ref{gfexpl}). The generating functionals 
$W_{(p_i,p_k)}[u;N_c]$ and  $W_{(p_i,p_m,p_k,p_l)}[u]$ in Eqs.~(\ref{dipw}) and 
(\ref{quadw}) depend on the momenta of the hard partons (but they do not depend on their colour charges) and of the soft gluons (through differentiation with respect to $u(q_\ell)$), and they also depend on $N_c$. In particular, the dependence on $N_c$
in Eq.~(\ref{dipw}) is maximally non-abelian, i.e. it is proportional to
$C_A^{n-1} u^n \sim {\cal O}(C_A^{n-1} \as^n)$.

If we consider the generating functional $\Psi_{(p_B,p_C)}^{(BC)}[u]$ for soft-gluon radiation from two hard partons $\{B C \}$, the results in 
Eqs.~(\ref{jbc1g})--(\ref{jbc3g}) can be expressed as
\beq
\label{qqgf}
\Psi_{(p_B,p_C)}^{(q {\bar q})}[u] = \exp \bigl\{ C_F \;W_{(p_B,p_C)}[u;N_c] 
+ {\cal O}(u^4) \bigr\} \;,
\eeq
\beq
\label{gggf}
\Psi_{(p_B,p_C)}^{(gg)}[u] = \exp \bigl\{ C_A \;W_{(p_B,p_C)}[u;N_c] 
+ {\cal O}(u^4) \bigr\} \;,
\eeq
where $W_{(p_B,p_C)}[u;N_c]$ is exactly equal to the generating functional in Eq.~(\ref{dipw}).
Equations (\ref{qqgf}) and (\ref{gggf}) clearly show the exponentiated Casimir scaling correspondence $C_F \leftrightarrow C_A$ between radiation from $q {\bar q}$
and $gg$ colour singlets. At ${\cal O}(u^4)$ Casimir scaling violation (generalization) effects occur according to the results in Eq.~(\ref{j4bc}) and the accompanying discussion in Sect.~\ref{sec:quadruple}.

Neglecting quadrupole contributions, the exponentiated result in Eq.~(\ref{gfexpl})
is obtained from Eqs.~(\ref{qqgf}) or (\ref{gggf}) by `simply' replacing the colour factor $C_F$ or $C_A$ with the colour dipole factor $\T_i \cdot \T_k$ and summing over colour dipole terms. Therefore the generic generating functional in 
Eq.~(\ref{gfexpl}) fulfils exponentiated `colour dipole scaling', which is violated by
colour quadrupole interactions between three or more hard partons 
(see Eq.~(\ref{q2hard})) starting at ${\cal O}(u^3) \sim {\cal O}(\as^3)$.
From our discussion in Sect.~\ref{s:coll}, we also recall that the colour quadrupole term in Eq.~(\ref{gfexpl}) has no collinear singularities\footnote{In the case of three hard partons we also recall from Sect.~\ref{sec:3hardtree} 
that the quadrupole terms at 
${\cal O}(\as^3)$ are suppressed with respect to colour dipole terms by a relative factor of ${\cal O}(1/N_c^2)$ in the large-$N_c$ limit.},
while the dipole scaling term has collinear singularities whose dominant contributions are effective in angular-ordered regions (see the points 
$(c_1)$--$(c_5)$ in Sect.~\ref{s:coll}).

The exponentiated form in Eq.~(\ref{gfexpl}) derives from the explicit results
on triple-soft gluon radiation at the tree level that we have computed in this paper.
At the purely technical level, this exponentiated form {\em partly} follows
from an observation on non-abelian eikonal exponentiation that was made long ago in Ref.~\cite{Gatheral:1983cz}. In particular, the structure of Eq.~(\ref{gfexpl})
is {\em partly} similar to that of the virtual (loop-level) IR divergences of scattering amplitudes 
\cite{Aybat:2006mz, Becher:2009qa, Gardi:2009qi, Gardi:2010rn, Mitov:2010rp, Almelid:2015jia, Almelid:2017qju}. However, as remarked below, there are important conceptual and physical differences between the exponentiation of the generating functional and the
exponentiation of virtual IR divergences.

In the case of virtual IR divergences, the exponentiated functionals 
$W_{(p_i,p_k)}[u;N_c]$ and  $W_{(p_i,p_m,p_k,p_l)}[u]$ of Eq.~(\ref{gfexpl})
are replaced by dipole and quadrupole functions that only depend on the hard-parton momenta and $N_c$ (see Eqs.~(2), (3), (4) and (7) in Ref.~\cite{Almelid:2015jia}).
These `virtual-radiation' functions are obtained by integration over the loop momenta.
In contrast, the exponentiated functionals in Eq.~(\ref{gfexpl}) embody the full information on the structure of multiple final-state radiation of soft-gluons at the completely {\em exclusive} level. Owing to the physical differences between exclusive
and loop-level radiation, the structure of the generating functional $\Psi[u]$ can be much complex and quite different from the structure of the factor $Z$ 
(see Eqs.~(1) and (2) in Ref.~\cite{Almelid:2015jia}) that controls the virtual IR divergences of the scattering amplitudes. Relevant differences on the IR divergences still persist after the integration over the phase space of the radiated soft gluons. Indeed, the $d$-dimensional phase space integration in
Eq.~(\ref{dipw}) over the soft-gluon momenta produces $\epsilon$ poles of the type
$(\as/\epsilon^2)^n$ in the {\em exponent} of the tree-level generating functional 
$\Psi[u]$ in Eq.~(\ref{gfexpl}), whereas the exponent of the virtual-radiation factor $Z$ has milder IR divergences of the type $\as^n/\epsilon^{n+1}$. The stronger IR divergences embodied in the tree-level generating functional arise from the singular collinear behaviour of the irreducible correlations $\w^{(N)}(q_1, \dots, q_N)$ (see Sect.~\ref{s:coll}).
Analogous IR divergences are produced by supplementing the tree-level generating functional with loop-level radiative corrections to the squared current 
$| \J(q_1,\cdots,q_N) |^{2}$ and, eventually IR divergences cancel in the computation of physical cross sections. Incidentally, we note that the inclusion of the one-loop
correction to single soft-gluon radiation produces colour correlations of the type 
$f^{a b c} T_i^a T_k^b T_m^c$ between {\em four} or more hard partons (see Sect.~3 in Ref.~\cite{Catani:2000pi}) 
in the {\em exponent} of the generating functional $\Psi[u]$
at ${\cal O}(u^2)$.

The exponentiated structure of the generating functional $\Psi[u]$ is relevant for studies of soft multiparton radiation at the exclusive level. Such structure is certainly important in the context of the calculation and resummation of soft-gluon logarithmic corrections to physical observables. The relation with soft-gluon resummation is particularly evident for physical observables whose phase space factorizes in the soft limit (see, e.g., Ref.~\cite{Bonciani:2003nt}).
In these cases the auxiliary weight function $u(q)$ in $\Psi[u]$ can be properly replaced with the phase space dependence of the corresponding physical observable, and Eq.~(\ref{gfexpl}) directly leads to the exponentiation of the large logarithmic contributions from multiple soft-gluon radiation at the tree-level. In this paper we have explicitly computed the exponentiated structure of the 
tree-level generating functional in Eq.~(\ref{gfexpl}) up to ${\cal O}(u^3)$, which contributes to logarithmic resummation up to ${\cal O}(\as^3)$ for generic hard-scattering 
processes.

\section{Summary}
\label{sec:sum}

We have considered the radiation of three soft gluons in QCD hard scattering. We have computed the tree-level current $\J(q_1,q_2,q_3)$ for triple soft-gluon emission
in a generic scattering amplitude with an arbitrary number of external hard partons.
The result is valid for arbitrary relative energies of the three soft gluons.
The current is expressed in terms of irreducible correlations for emission of one, two and three gluons. Such expression highlights the maximally non-abelian character
of the irreducible correlations. The soft currents acts in colour space, and it is written in terms of the colour charges and momenta of the generic hard partons. In the specific case of pure multigluon amplitudes, we have obtained the explicit result of the colour stripped current for the decomposition in colour ordered subamplitudes.

We have computed the tree-level squared current $| \J(q_1,q_2,q_3) |^2$ and the ensuing colour correlations for squared amplitudes with both massless and massive hard partons. We have shown the presence of non-abelian colour quadrupole correlations between three or more hard partons, in addition to the colour dipole correlations that appear in the cases of single and double soft-gluon radiation.
We have properly identified colour quadrupole operators that are irreducible to dipole operators. Such identification is essential to introduce a physically 
meaningful (e.g., gauge invariant) distinction between dipole and quadrupole correlation effects.
We have also discussed the singular collinear behaviour of $| \J(q_1,q_2,q_3) |^2$,
which turns out to be consistent with collinear factorization properties
and angular-ordering features of multiple soft-gluon radiation.
In particular, colour quadrupole correlations have no associated collinear singularities.

In the specific cases of processes with two and three hard partons, the colour structure can be simplified, and we have discussed some ensuing features of multiple soft-gluon radiation to all orders in the loop expansion.

We have evaluated the tree-level squared current $| \J(q_1,q_2,q_3) |^2$ for radiation from three hard partons by explicitly computing the colour quadrupole coefficients. We find that the quadrupole contributions for triple soft-gluon radiation breaks the Casimir scaling symmetry (due to colour dipole interactions) between hard quarks and gluons.
We have also considered multiple soft-gluon radiation with energy strong ordering,
and we have derived a generalization of the multi-eikonal BCM formula to processes with three hard gluons.

In the case of processes with two hard partons, we have extended our analysis to the emission of four soft gluons at the tree level. We have presented the general colour structure of the squared current $| \J(q_1,q_2,q_3,q_4) |^2$, and we have explicitly computed it in a wide energy region where one of the four soft gluons is harder than the others. The result includes a colour monster contribution (suppresses by 
${\cal O}(1/N_c^2)$ in the large-$N_c$ limit) that had been previously noticed by applying energy strong-ordering approximations. We have computed the colour monster contribution for radiation from both hard quarks and gluons, and we point out that it is directly related to quartic Casimir invariants for quarks and gluons. Consequently
the colour monster contribution violates quadratic Casimir scaling, and it leads to a generalized form of Casimir scaling between quarks and gluons at ${\cal O}(\as^4)$.
The colour monster term has collinear singularities and, therefore, it contributes
to the cusp anomalous dimension at ${\cal O}(\as^4)$.
Considering quadruple soft-gluon radiation from two hard gluons we have also explicitly computed the first correction of ${\cal O}(1/N_c^2)$ to the 
multi-eikonal BCM formula. 

We have also introduced the generating functional for multiple soft-gluon radiation, and we have discussed its exponentiation for tree-level radiation up to 
${\cal O}(\as^3)$ in generic hard-scattering processes. The exponentiated structure fulfils colour dipole scaling, which is
violated at ${\cal O}(\as^3)$ by colour quadrupole correlations between three or more hard partons.


\setcounter{footnote}{2}

\section*{Appendices}
\appendix

\section{Current conservation
\label{a:cons}}

We present a general (i.e., for emission of an arbitrary number of soft gluons and
to all orders in the loop expansion) and formal proof of the current conservation relation in Eq.~(\ref{curcons}). The proof exploits the fact that the current
$\J(q_1,\cdots,q_N)$ in the soft-gluon factorization formula (\ref{1gfact}) has a certain degree of arbitrariness: different forms of $\J$ are physically equivalent provided they differ by terms that give vanishing contributions when acting onto the colour singlet amplitude $\ket{\M (\{p_i\})}$.

We start our proof by considering a generic expression 
${\widetilde J}^{a_1 \dots a_N}_{\mu_1 \dots \mu_N}(q_1,\cdots,q_N)$ of the soft current that fulfils the non-abelian Ward identity in Eq.~(\ref{wardid}). Having
${\widetilde \J}$ at our disposal we define the current $\J$ as follows:
\beq
\label{jtil}
J^{a_1 \dots a_N}_{\mu_1 \dots \mu_N}(q_1,\cdots,q_N)
= \left( \prod_{\ell=1}^{N} \left[ - \;{\tilde d}_{\mu_\ell \nu_\ell}(q_\ell; n_\ell) \right] \right) \;{\widetilde J}_{a_1 \dots a_N}^{\nu_1 \dots \nu_N}(q_1,\cdots,q_N)
\;\;,
\eeq
\beeq
\label{dep}
{\tilde d}^{\mu_\ell \nu_\ell}(q_\ell; n_\ell) &=& \sum_\sigma 
{\tilde \e}_{(\sigma)}^{\mu_\ell}(q_\ell; n_\ell) \; 
\left[ {\tilde \e}_{(\sigma)}^{\nu_\ell}(q_\ell; n_\ell)\right]^* \\
\label{dgauge}
&=& - g^{\mu_\ell \nu_\ell} + {\rm gauge \; terms} \;
(\propto q_\ell^{\mu_\ell} {\rm or} \;q_\ell^{\nu_\ell} ) \;\;,
\eeeq
where ${\tilde d}^{\mu_\ell \nu_\ell}(q_\ell; n_\ell)$ is the spin polarization tensor of the soft gluon with momentum $q_\ell$, as obtained by summing over a complete set of physical polarizations ${\tilde \e}_{(\sigma)}^{\mu_\ell}(q_\ell; n_\ell)$
(see Eq.~(\ref{dep})). The auxiliary physical polarization vectors ${\tilde \e}_{(\sigma)}^{\mu_\ell}(q_\ell; n_\ell)$ of the gluon $\ell$ are specified by some gauge fixing conditions, and we choose the axial gauge condition 
$n_\ell^\mu  {\tilde \e}^{(\sigma)}_{\mu}(q_\ell; n_\ell)=0$, where $n_\ell^\mu$ is an arbitrary auxiliary vector  that can be different for each of the soft gluons.
Independently of the choice of the gauge vector $n_\ell$, the polarization tensor 
${\tilde d}^{\mu_\ell \nu_\ell}(q_\ell; n_\ell)$ has the form in Eq.~(\ref{dgauge}),
where the gauge terms are proportional to either $q_\ell^{\mu_\ell}$ or 
$q_\ell^{\nu_\ell}$. 

The two currents ${\widetilde \J}$ and $\J$ of Eq.~(\ref{jtil}) are physically equivalent onto colour singlet amplitudes since they fulfil the following identity:
\beq
\label{Jident}
\left(\prod_{\ell=1}^{N}
\e_{(\sigma_{\ell})}^{\mu_{\ell}}(q_{\ell}) \right)
J^{a_1 \dots a_N}_{\mu_1 \dots \mu_N}(q_1,\cdots,q_N)
\eqcs
\left(\prod_{\ell=1}^{N}
\e_{(\sigma_{\ell})}^{\mu_{\ell}}(q_{\ell}) \right)
{\widetilde J}^{a_1 \dots a_N}_{\mu_1 \dots \mu_N}(q_1,\cdots,q_N) \;\;,
\eeq
where  $\e_{(\sigma_{\ell})}^{\mu_{\ell}}(q_{\ell})$ are the 
polarization vectors of the scattering amplitude 
$\ket{\M(\{q_\ell\}, \{p_i\})}$ 
in the soft-gluon factorization formula (see Eqs.~(\ref{1gfact}) and 
(\ref{colspinsoft})).
To prove Eq.~(\ref{Jident}), we first notice that, due to the definition of $\J$ in Eq.~(\ref{jtil}), the difference between the left-hand and right-hand sides of Eq.~(\ref{Jident}) can only be due to the gauge terms proportional to either
$q_{\ell}^{\mu_\ell}$ or $q_{\ell}^{\nu_\ell}$
in Eq.~(\ref{dgauge}). However the terms $q_{\ell}^{\mu_\ell}$ of 
${\tilde d}^{\mu_\ell \nu_\ell}(q_\ell; n_\ell)$ gives a vanishing contribution to 
Eq.~(\ref{Jident}) because $\e_{(\sigma_{\ell})}^{\mu_{\ell}}(q_{\ell})$ is a physical polarization ($q_{\ell}^{\mu} \e^{(\sigma_{\ell})}_{\mu}(q_{\ell}) =0$).
Analogously, the term $q_{\ell}^{\nu_\ell}$ of 
${\tilde d}^{\mu_\ell \nu_\ell}(q_\ell; n_\ell)$ gives a vanishing contribution to
Eq.~(\ref{Jident}), since it contributes in the form
\beq
\label{modwi}
q_\ell^{\nu_\ell} \;\left[ \prod_{\ell^\prime \neq \ell} 
{\overline \e}_{(\sigma_{\ell^\prime})}^{\nu_{\ell^\prime}}(q_{\ell^\prime}) \right]
{\widetilde J}^{a_1 \dots a_\ell \dots a_N}_{\nu_1 \dots \nu_\ell  \dots \nu_N}(q_1,\cdots,q_N) \;\eqcs  \;0 \;\;,
\eeq
which vanishes because of Eq.~(\ref{wardid}). Note that, by iteratively removing the vanishing contributions of $q_{1}^{\nu_1}, q_{2}^{\nu_2}, \dots$ to Eq.~(\ref{Jident}),
the Lorentz indices of ${\widetilde J}^{\nu_1 \dots \nu_N}$ are multiplied by polarization vectors that can be either $\e^\nu(q_\ell)$ or 
${\tilde \e}^\nu(q_\ell;n_\ell)$, and these polarization vectors are generically denoted by ${\overline \e}^{\nu}(q_\ell)$ in Eq.~(\ref{modwi}). However,
${\overline \e}^{\nu}(q_\ell)$ is always a physical polarization vector and, therefore, the physical equivalence of ${\widetilde \J}$ and $\J$ 
(see Eq.~(\ref{Jident})) is eventually a simple consequence of the fact that ${\widetilde \J}$ fulfils the non-abelian Ward identity in Eq.~(\ref{wardid}).

The auxiliary polarization vectors ${\tilde \e}^\nu(q_\ell;n_\ell)$ in Eq.~(\ref{dep})
are physical ($q_\ell \cdot {\tilde \e}(q_\ell;n_\ell) =0$) and, consequently,
the soft-gluon current $\J$ in Eq.~(\ref{jtil}) is conserved. Therefore, we have proven that we can always find an explicit expression of the current $\J$ that fulfils current conservation as in Eq.~(\ref{curcons}). Actually, we have explicitly constructed many different expressions of physically equivalent and conserved soft-gluon currents. These expressions are directly obtained by choosing the auxiliary gauge vectors $n_\ell$ in Eq.~(\ref{jtil}).

Since the current $\J$ in Eq.~(\ref{jtil}) depends on the auxiliary gauge vectors $n_\ell$, one may wonder whether the definition of a conserved soft-gluon current necessarily introduces an explicit dependence on external (unphysical) momenta. 
This is not the case, since one can use the soft-gluon momenta to define the gauge vectors. For example, in the case of $N=2$ soft gluons one can set $n_1=q_2$ and $n_2=q_1$ in Eq.~(\ref{jtil}). Analogously, in the case of $N=3$ soft gluons
one can use both choices $\{  n_1=q_2, n_2=q_3, n_3=q_1 \}$ and
$\{  n_1=q_3, n_2=q_1, n_3=q_2 \}$ and, one can also define $\J$ by symmetrizing
with respect to these two choices of gauge vectors. Such a symmetrization procedure
can be straightforwardly extended to an arbitrary nunmber $N$ of soft gluons.
In summary, the gluon polarization tensors  
${\tilde d}(q_\ell; n_\ell)$ of Eq.~(\ref{jtil}) can be introduced in such a way that 
the difference between ${\widetilde \J}$ and the conserved current $\J$ is due to contributions that only depend on the soft-gluon momenta, and with a 
fully symmetric dependence on them.

In this Appendix we have presented a general proof of the current conservation relation in Eq.~(\ref{curcons}). The formal proof uses the construction in Eq.~(\ref{jtil}).
However, we remark on the fact that in the cases with $N=2$ and $N=3$ soft gluons
the explicit results of the conserved currents in Eqs.~(\ref{J12})--(\ref{gammaC}) \cite{CaGr99} and in Sect.~\ref{s:sc3}
(see Eqs.~(\ref{J123}), (\ref{G123}) and (\ref{gamma123})) are obtained without using
the construction in Eq.~(\ref{jtil}). These results are straightforwardly obtained by
computing $\J$ and by simply applying Eq.~(\ref{colcons}) to neglect terms
(which are possibly gauge dependent) that give a vanishing contribution onto colour singlet scattering amplitudes. We expect that a similar straightforward procedure
directly leads to conserved currents also for higher soft-gluon multiplicities.

\section{Colour quadrupole operators
\label{a:quad}}

We consider colour quadrupole operators, which are obtained by the colour contraction (with respect to the colour indices of the soft gluons) of two structure constants 
$f^{ab,cd}$ (see Eq.~\eqref{ff}) with four colour charges $\T_i$ of the hard partons.
If the four hard partons are distinct, the quadrupole operators are defined in 
Eq.~(\ref{Qdiff}), and any other definition simply differs by an overall normalization factor. If the four hard partons are not distinct, we are dealing with pseudo quadrupoles. In this case the ordering of the four colour charges does matter, and we distinguish three types of {\em hermitian conjugate} (pseudo) quadrupoles:
\beeq
\label{quadU}
\qu{klmi}^{(U)}
    &\equiv& f^{ab,cd} \;(T_k^a T_l^b T_m^c T_i^d + \text{h.c.} ) \;\;,\\
\label{quadS}
\qu{klmi}^{(S)} 
    &\equiv& f^{ab,cd} \;(T_k^a T_m^c T_l^b T_i^d + \text{h.c.} ) \;\;, \\
\label{quadD}
\qu{klmi}^{(D)} 
    &\equiv& f^{ab,cd} \;(T_k^a T_i^d T_m^c T_l^b + \text{h.c.} ) \;\;.
\eeeq
In the (`untwisted') quadrupole of Eq.~(\ref{quadU}) the colour connected
indices $a$ and $b$ of $f^{ab,cd}$ are contracted with those of two adjacent colour charges, $T_k^a$ and $T_l^b$. In the quadrupoles of Eqs.~(\ref{quadS}) and 
(\ref{quadD}) the colour charge $T_l^b$ is shifted by one position (single twist)
and two positions (double twist), respectively. Owing to the symmetry properties of 
$f^{ab,cd}$ with respect to its colour indices, other orderings of the four colour charges lead to quadrupoles that are equivalent (modulo overall signs) to
the untwisted, `single-twisted' and `double-twisted' quadrupoles
in Eqs.~(\ref{quadU})--(\ref{quadD}).

It is evident that quadrupoles operators 
can eventually lead
to colour dipole operators in various cases. For instance, by considering two pairs of distinct colour charge indices $k$ and $i$ $(k \neq i)$ and performing elementary colour algebra,
Eqs.~(\ref{quadU})--(\ref{quadD}) give
\beq
\label{qki}
\qu{kkii}^{(U)}= \qu{kkii}^{(S)} = - \qu{kkii}^{(D)} = 
-\frac{1}{2}\;C_A^2 \;\T_i\cdot \T_k \;\;,
\quad (k \neq i) \;\;.
\eeq
More generally, different orderings of the four colour charges can be 
related by using colour algebra commutation relations and, consequently, 
the quadrupole operators in Eqs.~(\ref{quadU})--(\ref{quadD}) can be related through
colour dipole operators. We are interested in identifying quadrupole operators that are `irreducible' to colour dipoles, in the sense that they do not produce dipole terms of the type $C_A^2 \;\T_i\cdot \T_k$ (i.e., a dipole $\T_i\cdot \T_k$ with the
maximally non-abelian colour factor $C_A^2$).
As stated in Sect.~\ref{s:cor123},
the operators
\begin{equation}
\label{qirr}
  \qu{klmi} \equiv 
  \frac12 f^{ab,cd} \,\left( T_k^a \{ T_m^c, T_i^d \} T_l^b + \text{h.c.} \right) \;,
\end{equation}
which are defined in Eq.~(\ref{Qbase}) (see also Eq.~(\ref{qsym})), are irreducible quadrupoles. In the following we first consider the symmetry properties of the quadrupoles
$\qu{klmi}$ and then we discuss their irreducibility and the relation with the generic quadrupole operators in Eqs.~(\ref{quadU})--(\ref{quadD}).

The quadrupole operator of Eq.~(\ref{qirr}) fulfils the symmetry properties in 
Eqs.~(\ref{qant})--(\ref{qcons}). The property in Eq.~(\ref{qsym}) directly follows from considering the hermitian conjugated term in the round brackets of 
Eq.~(\ref{qirr}) and exploiting the symmetry relation $f^{ab,cd} = f^{cd,ab}$.
Similarly, the antisymmetry of $\qu{klmi}$ with respect to the exchange 
$m \leftrightarrow i$ in the second pair of indices directly follows Eq.~(\ref{qirr})
by using the identity $f^{ab,cd} = - f^{ab,dc}$. Combining this antisymmetry with
Eq.~(\ref{qsym}), we directly obtain the antisymmetry of $\qu{klmi}$ with respect to the exchange $k \leftrightarrow l$ in the first pair of indices 
(see Eq.~(\ref{qant})). The relation (\ref{qjac}) is a consequence of the Jacobi identity in Eq.~(\ref{jacf})
for the product of two structure constants
(see also Eq.~(\ref{jusd}) and accompanying comments).
The property in Eq.~(\ref{qcons}) is a consequence of the colour conservation relation in Eq.~(\ref{csnotation}) 
for colour singlet states.
The proofs of Eqs.~(\ref{qjac}) and (\ref{qcons}) are also straightforward, although they require the use of colour algebra commutation relations to perform some reordering 
of the four colour charges of the quadrupole operator $\qu{klmi}$.

The quadrupole operator of Eq.~(\ref{qirr}) can be directly related to the quadrupoles in Eqs.~(\ref{quadU})--(\ref{quadD}). We first notice that $\qu{klmi}$
can be expressed as an antisymmetric combination of double-twisted quadrupoles: 
\begin{equation}
  \qu{klmi} = \frac12\left( \qu{klmi}^{(D)} - \qu{klim}^{(D)} \right) \;\;.
\end{equation}
The corresponding symmetric combination, $\qu{klmi}^{(D)} + \qu{klim}^{(D)}$, of quadrupoles can be expressed in terms of dipole operators and, therefore, we obtain the following general relation between the double-twisted quadrupole and $\qu{klmi}$:
\begin{equation}
\label{Q3Qdiff}
   \qu{klmi}^{(D)} = \qu{klmi} + \frac12 \,C_A^2 \;\delta_{mi}\left[ \;
    \delta_{kl} \; \T_k\cdot \T_m
- ( \delta_{km}+\delta_{lm} ) \, \T_k\cdot \T_l \;
\right] \;.
\end{equation}
The relations between $\qu{klmi}$ and the untwisted and single-twisted quadrupoles are as follows:
\beeq
\label{qsvd}
   \qu{klmi}^{(S)} &=& - \;\qu{klim}^{(D)} - \frac12 \,C_A^2 \;\delta_{li}\left[ \;
    \delta_{mk} \; \T_k\cdot \T_i
+ ( \delta_{mi}-\delta_{ki} ) \; \T_k\cdot \T_m \;
\right] \\
\label{qsvq}
   &=& \qu{klmi} + \frac12 \,C_A^2 \,\left( \delta_{kl} - \delta_{km} \right) 
       \left( \delta_{li}- \delta_{mi}\right) \;\T_l\cdot \T_m \;\;,
\eeeq
\beeq
\label{quvs}
   \qu{klmi}^{(U)} &=& - \;\qu{klmi}^{(S)} + \frac12 \,C_A^2 \;\delta_{lm}\left[ \;
    \delta_{ki} \; \T_k\cdot \T_m
- ( \delta_{km} + \delta_{mi} ) \; \T_k\cdot \T_i \;
\right] \\
   &=& \qu{klmi} + \frac12 \,C_A^2 \,\bigl\{ \;\delta_{lm}\left[ \;
    \delta_{ki} \; \T_k\cdot \T_m
- ( \delta_{km} + \delta_{mi} ) \; \T_k\cdot \T_i \;
\right] \nn \\
\label{quvq}
&& \quad \quad \quad \quad \quad \quad \;\;\;
+ \,\left( \delta_{kl} - \delta_{km} \right) 
       \left( \delta_{li}- \delta_{mi}\right) \;\T_l\cdot \T_m
\bigr\} \;\;.
\eeeq
The equality (\ref{qsvd}) is obtained by using Eq.~(\ref{quadS}) and displacing
the colour charge $T^b_l$ by one position to the right: the displacement leads to
$\qu{klim}^{(D)}$ and to the commutator $[ T^b_l , T^d_i ]$, which eventually produces the colour dipole contributions on the right-hand side of Eq.~(\ref{qsvd}).
The equality (\ref{qsvq}) is directly obtained from Eq.~(\ref{qsvd}) by using
Eq.~(\ref{Q3Qdiff}) to express $\qu{klim}^{(D)}$ in terms of $\qu{klmi}$
(note that $\qu{klmi}=-\qu{klim}$ because of Eq.~(\ref{qant})).
An analogous procedure leads to Eqs.~(\ref{quvs}) and (\ref{quvq}).
More precisely, Eq.~(\ref{quvs}) is obtained by relating Eqs.~(\ref{quadU})
and (\ref{quadS}) through the displacement of $T^b_l$, and Eq.~(\ref{quvq})
is obtained by inserting Eq.~(\ref{qsvq}) into Eq.~(\ref{quvs}).

We also notice that the three quadrupoles in Eqs.~(\ref{quadU})--(\ref{quadD})
can be directly related by using the Jacobi identity for the structure constants.
Specifically, multiplying Eq.~(\ref{jacf}) by the colour charge factor
$(T^b_k T^{a_1}_l T^{a_2}_m T^{a_3}_i + {\rm h.c.})$ and performing the sum over $\{b,a_1,a_2,a_3\}$ we obtain
\begin{equation}
\label{jusd}
\qu{klmi}^{(U)} - \qu{kmli}^{(S)} -\qu{kiml}^{(D)} =0 \;.
\end{equation}
Inserting Eqs.~(\ref{Q3Qdiff}), (\ref{qsvq}) and (\ref{quvq}) in Eq.~(\ref{jusd}),
the colour dipole contributions identically cancel, and we obtain the Jacobi identity
(\ref{qjac}) for the quadrupoles $\qu{klmi}$.

The relations (\ref{Q3Qdiff}), (\ref{qsvq}) and (\ref{quvq}) show that the quadrupoles
$\qu{klmi}^{(U)}, \qu{klmi}^{(S)}$ and $\qu{klmi}^{(D)}$ can be expressed in terms of 
$\qu{klmi}$ plus colour dipole contributions.
The quadrupoles $\qu{klmi}^{(U)}, \qu{klmi}^{(S)}$ and $\qu{klmi}^{(D)}$ are directly reducible to colour dipoles in some configurations of the hard-parton indices 
$\{ k,l,m,i\}$ (see, e.g., Eq.~(\ref{qki})). This is not the case for the quadrupole
$\qu{klmi}$, as we briefly discuss below. Owing to the symmetry properties in 
Eqs.~(\ref{qant}) and (\ref{qsym}), we first note that $\qu{klmi}$ is not vanishing only in the following configurations of hard-parton indices: four distinct indices,
three distinct indices (i.e., $\qu{klki}=-\qu{klik}=-\qu{lkki}=\qu{lkik}$ with distinct indices $k,l$ and $i$) and two pairs of distinct indices
(i.e., $\qu{kiki}= - \qu{ikki}$ with $k\neq i$). In the cases of four or three distinct indices, Eqs.~(\ref{Q3Qdiff}), (\ref{qsvq}) and (\ref{quvq}) explicitly show that the various quadrupole definitions are equivalent and, precisely, we have
\begin{equation}
\label{q34i}
\qu{klmi}^{(U)} = \qu{klmi}^{(S)} = \qu{klmi}^{(D)} = \qu{klmi} \;\;, \quad
\quad
(3 \;{\rm or}\; 4 \;{\rm distinct \;indices}) \;\;.
\end{equation}
In the case of two pairs of distinct indices, $\qu{kiki}$ vanishes onto {\em any} colour singlet states with only two hard partons (see Eq.~(\ref{q2hard}))
and, therefore, $\qu{kiki}$ is not reducible (proportional) to the colour dipole
$\T_k\cdot \T_i$ (the colour dipole is indeed not vanishing onto the colour singlet configuration of the two hard partons $k$ and $i$).

We specifically examine the action of the irreducible quadrupoles onto a colour singlet state formed by three distinct (and generic) hard partons $k, l$ and $i$.
The only non-vanishing quadrupoles have either three distinct indices or two pairs of distinct indices. Owing to the symmetries in Eqs.~(\ref{qant}) and (\ref{qsym}),
all quadrupoles with three distinct indices are proportional (through an overall sign)
to $\qu{klki}$. Then we have
\begin{equation}
\label{3ics}
\qu{klki} \,\eqcs \,- \qu{kiki} \;\;, \quad
\quad \quad
(3 \;{\rm hard \; partons}) \;\;,
\end{equation}
where we have used the colour conservation relation (\ref{qcons}) to replace the index $l$ with the sum over the indices $k$ and $i$ (note that $\qu{kkki}=0$).
By repeated use of Eq.~(\ref{qcons}) we also have
\begin{equation}
\label{2ics}
\qu{kiki} \,\eqcs \, \qu{klkl} \,\eqcs \, \qu{ilil}\;\;, \quad
\quad \quad
(3 \;{\rm hard \; partons}) \;\;.
\end{equation}
In summary, Eqs.~(\ref{3ics}) and (\ref{2ics}) show that all the non-vanishing quadrupole operators are eventually proportional, through corresponding overall signs, to a {\em single} quadrupole operator $\qu{kiki}$ with two pairs of distinct indices.
By using relations for the $SU(N_C)$ colour algebra, we have explicitly evaluated the action of this quadrupole operator onto colour singlet states formed by quarks, antiquarks and gluons. To present the results we use the notation of 
Sect.~\ref{sec:3hard},
and we consider the colour singlet states $\ket{ABC \,}$ with
$\{ ABC \} = \{ gq{\bar q} \}$ and $\{ ABC \} = \{ ggg \}$. We find
\beeq
\label{qqqg}
\!\!\!\!\!\!
\qu{BCBC} \;\ket{ABC \,} &=& \frac12 \,N_c \;\ket{ABC \,} 
\equiv N_c \,\lambda_F \;\ket{ABC \,} \;, 
\quad \quad \;\;\; \{ ABC \} = \{ gq{\bar q} \} \;, \\
\label{qgggf}
\!\!\!\!\!\!
\!\!\! \qu{BCBC} \;\ket{(ABC)_f \,} &=& 3 \,N_c \;\ket{(ABC)_f \,} 
\equiv N_c \,\lambda_A \;\ket{(ABC)_f \,} \;, 
\;\; \{ ABC \} = \{ ggg \} \;, \\
\label{qgggd}
\!\!\!\!\!\!
\!\!\! \qu{BCBC} \;\ket{(ABC)_d \,} &=& 0 \;, 
\quad \quad \quad \quad \quad \quad \quad \quad \quad \quad \quad \quad \quad \quad
\;\;\;\; \{ ABC \} = \{ ggg \} \;, 
\eeeq
where $\ket{(ABC)_f \,}$ and $\ket{(ABC)_d \,}$ are the colour antisymmetric and colour symmetric states of Eq.~(\ref{gggstates}), and we have also defined the colour
coefficients $\lambda_F$ ($\lambda_F=1/2$) and $\lambda_A$ ($\lambda_A=3$) in the fundamental and adjoint representation, respectively.

As we have just discussed, the evaluation of the action of the irreducible quadrupoles
$\qu{klmi}$ onto a colour singlet state formed by three hard partons eventually requires the explicit computation of a single quadrupole operator. It is of interest to count the number of independent quadrupole operators that have to be explicitly computed while dealing with a colour singlet state with four or more hard partons.

Considering a generic colour singlet state with $N_h$ ($N_h \geq 4$) hard partons,
we recall that the non-vanishing quadrupoles $\qu{klmi}$ can have two pairs of distinct indices, three distinct indices and four distinct indices. In the case of 
two pairs of distinct indices $k$ and $i$ ($k \neq i$) we can use the colour conservation relation (\ref{qcons}) to write
\beq
\label{2qvs3q}
\qu{kiki} \;\eqcs \; \sum_{\substack{l \\ l\,\neq\,i,k}} \qu{klki} \;\;,
\quad \quad (k \neq i) \;\;.
\eeq
This relation shows that a quadrupole with two pairs of distinct indices can always be replaced by a linear combination of quadrupoles with three distinct indices when it acts onto a colour singlet state. Therefore, we can consider the quadrupoles 
$\qu{kiki}$ as linearly dependent quadrupoles, and we can move to count the number of remaining linearly independent quadrupoles with three and four distinct indices.

To proceed in our counting it is convenient to order the $N_h$ hard partons in the colour singlet state and, correspondingly, to consider ordered sets of distinct indices in the quadrupoles $\qu{klmi}$.

In the case of irreducible quadrupoles with three distinct indices, one of the quadrupole index is repeated. Owing to Eq.~(\ref{qant}), the repeated index is placed
in the first and second pair of the quadrupole indices and we can always (modulo the overall sign) assign the first position in each pair to the repeated index.
Then, due to Eq.~(\ref{qsym}), the position of the non repeated indices does not matter. In summary, considering the ordered indices $k < l < i$, we have to deal with the following independent quadrupoles:
\beq
\label{q3d}
\qu{klki} \,, \;\qu{lkli} \,, \;\qu{ikil} \,,
\quad \quad (k < l < i ) \;\;.
\eeq 
We can also exploit 
Eq.~(\ref{qcons}), and we have
\beeq
\label{q3d1}
\qu{lkli} &=& \qu{klki} \;- \sum_{\substack{m \\ m \,\neq\, k,l,i}} \qu{lkmi} \;,
\quad \quad (k < l < i ) \;\;,\\
\label{q3d2}
\qu{ikil} &=& \qu{klki} \;- \sum_{\substack{m \\ m \,\neq\, k,l,i}} \qu{ikml} \;,
\quad \quad (k < l < i ) \;\;.
\eeeq
The identity in Eq.~(\ref{q3d1}) (Eq.~(\ref{q3d2})) is obtained by applying 
the colour conservation relation in Eq.~(\ref{qcons}) to the quadrupole index 
$l$ ($i$) in the third position and by using the properties in Eqs.~(\ref{qant})
and (\ref{qsym}).  The relations (\ref{q3d1}) and (\ref{q3d2}) show that we can select
a single independent quadrupole (e.g., $\qu{klki}$) in Eq.~(\ref{q3d}) and consider the other two quadrupoles as obtained from the selected quadrupole with three distinct indices and quadrupoles with four distinct indices. In summary, among the quadrupoles with three distinct indices we can limit ourselves to the explicit evaluation of a {\em single set} of quadrupoles with the ordered indices 
$k < l < i$. There are $N_h(N_h-1)(N_h-2)$ ways to select three indices among those of the $N_h$ hard partons, and they lead to $3!$ unordered permutations of the selected indices. Considering only one ordered permutation, we eventually select the number
\beq
\label{n3}
\frac{N_h\,(N_h-1)\,(N_h-2)}{3!}
\eeq
of independent quadrupoles with three distinct indices.

To complete our counting of independent quadrupoles we have to consider quadrupoles with four distinct indices. We order the distinct indices as $k < l < m < i$.
Owing to Eq.~(\ref{qant}) we can always select (modulo overall signs) a single ordering of the two indices in each adjacent pair of indices of the quadrupole. 
Owing to Eq.~(\ref{qsym}) we can moreover select some relative ordering among the indices of the two adjacent pairs. Eventually, we can limit ourselves to considering the following independent quadrupoles:
\beq
\label{q4d}
\qu{klmi} \,, \;\qu{kmli} \,, \;\qu{kilm} \,,
\quad \quad (k < l < m < i ) \;\;.
\eeq 
We can also exploit the Jacobi identity in Eq.~(\ref{qjac}) (and Eq.~(\ref{qant}))
to write
\beq
\label{q4d1}
\qu{kilm} = - \;\qu{klmi} \;+  \qu{kmli} \;,
\eeq
and to express one of the quadrupoles in Eq.~(\ref{q4d}) in terms of only two independent quadrupoles with four distinct indices (e.g., $\qu{klmi}$ and
$\qu{kmli}$). In summary, among the quadrupoles with four distinct indices
we can limit ourselves to the explicit computation of {\em two sets} of quadrupoles with four ordered indices $k < l < m < i$. There are 
$N_h\,(N_h-1)\,(N_h-2)\,(N_h-3)/4!$ ways to select four ordered indices among the set of the $N_h$ hard partons and, therefore, we obtain the number
\beq
\label{n4}
2 \;\frac{N_h\,(N_h-1)\,(N_h-2)\,(N_h-3)}{4!}
\eeq
of independent quadrupoles with four distinct indices.

Combining Eqs.~(\ref{n3}) and (\ref{n4}) we end up with
\beq
\label{ntot}
\frac{N_h\,(N_h-1)^2\,(N_h-2)}{12}
\eeq
{\em independent} quadrupole operators whose action onto  hard-parton scattering amplitudes has to be explicitly computed in the context of soft-gluon factorization formulae. For instance, if $N_h=4~(5)$ the number of independent quadrupole is 
$6~(20)$. Soft-gluon factorization formulae also involve colour dipole operators
$\T_k\cdot \T_i$. The number of independent colour dipoles (after using dipole symmetries and colour conservation) for a colour singlet state of $N_h$ ($N_h \geq 4$) hard partons was computed in Appendix~A of Ref.~\cite{csdip},
and it is equal to $N_h\,(N_h-3)/2$. If $N_h=4~(5)$ the number of independent dipoles
is $2~(5)$. Obviously this number of independent quadrupoles and dipoles refers to a {\em sole} colour singlet state of the $N_h$ hard partons. Depending on the colour representations of the $N_h$ hard partons, the number of colour singlet states varies,
and the actual number of independent quadrupole and dipole terms to be explicitly computed by using $SU(N_c)$ colour algebra increases accordingly.

\section{Momentum dependence of quadrupole and dipole correlations for triple soft-gluon radiation
\label{a:ee}}

In this Appendix we write down explicit expressions for the momentum dependence
of the dipole and quadrupole correlations in Eqs.~(\ref{3cordip}) and 
(\ref{3corquad}). We consider generic masses of the hard partons,
and we define $p_i^2 \equiv m_i^2$ ($m_i^2=0$ for the massless case).

The quadrupole correlation $W^{(3) {\rm quad.}}$ in Eq.~(\ref{3corquad})
is controlled by the function $\cS_{imkl}$. The explicit form of $\cS_{imkl}$
in terms of momenta can be obtained from Eq.~(\ref{simkl}) by simply inserting the kinematical expressions of $j_i(q_\ell)$ (Eq.~(\ref{j1})), 
$\gamma_i(q_1,q_2)$ (Eq.~(\ref{gammaC})) and 
$\gamma_i(q_1,q_2,q_3)$ (Eq.~(\ref{gamma123})).
This leads to a long algebraic expression for $\cS_{imkl}$. Such expression can be partly simplified by exploiting the symmetry properties in 
Eqs.~(\ref{qant})--(\ref{qcons}) of the quadrupoles $\qu{imkl}$ and the fact that we are actually interested in the computation of $W^{(3) {\rm quad.}}$.
For instance, we can subtract from $\cS_{imkl}$ any term that is symmetric 
in the first ($\{ i m\}$) or second ($\{ k l\}$) pair of indices
without affecting the value of $W^{(3) {\rm quad.}}$.
The value of $W^{(3) {\rm quad.}}$ is also unchanged by subtracting from 
$\cS_{imkl}$ any term that depends on three (rather than four) hard-parton momenta
(such term leads to a vanishing contribution because of the colour singlet relation
in Eq.~(\ref{qcons})). Moreover, we can also limit ourselves to considering the unsymmetrized component
of $\cS_{imkl}$ with respect to the momenta $q_1, q_2$ and $q_3$. Eventually,
we rewrite  $\cS_{imkl}$ in the following form:
\beq
\label{sbar}
\cS_{imkl}(q_1,q_2,q_3) = \left[ \;{\overline{\cS}}_{imkl}(q_1,q_2,q_3)
+  {\rm perms.}\{1,2,3\} \:\right] + \dots \;\;,
\eeq
where the dots in the right-hand side denote terms that give a vanishing contribution to $W^{(3) {\rm quad.}}$. Equivalently, we can rewrite Eq.~(\ref{3corquad}) as
\beq
\label{wbar}
W^{(3) {\rm quad.}}(q_1,q_2,q_3) \eqcs \sum_{i,m,k,l}  \;\qu{imkl} \;
 \left[ \;{\overline{\cS}}_{imkl}(q_1,q_2,q_3)
+  {\rm perms.}\{1,2,3\} \:\right] \;\;.
\eeq
We note that Eq.~(\ref{sbar}) can also be used for the computation of the quadrupole function $\w_{ABC}^{(3) {\rm quad.}}$ in Eq.~(\ref{w3abcquad}) (i.e., the terms denoted by dots in Eq.~(\ref{sbar}) give a vanishing contribution to 
$\w_{ABC}^{(3) {\rm quad.}}$).

The expression of ${\overline{\cS}}_{imkl}$ in terms of scalar products 
($k_i \cdot k_m \equiv k_ik_m$) of soft and hard momenta is
\begin{footnotesize}
\begin{align}
&{\overline{\cS}}_{imkl}(q_1,q_2,q_3) =
-\frac{7}{24} \per \frac{\pp{k}{l} \per \pp{i}{m} \per \pp{l}{m}}{\pq{k}{1} \per \pq{l}{1} \per \pq{i}{2} \per \pq{m}{2} \per \pq{l}{3} \per \pq{m}{3}}+\frac{1}{\pq{k}{1} \per \pq{l}{2} \per \pq{m}{3} \per \pq{i}{123}} \nonumber\\
&\times \sg \frac{\pp{i}{k} \per \pp{i}{l} \per \pp{i}{m} \per p_i(3 q_{3}-q_{12})}{12 \per \pq{i}{2} \per \pq{i}{3} \per \pq{i}{12}}+\frac{\pp{i}{m} \per p_i( q_{3}-q_{12})}{4 \per \qq{1}{2} \per \pq{i}{12} \per \pq{i}{3}} \per (\pp{k}{l} \per \pq{i}{1}+2 \per \pp{i}{l} \per \pq{k}{2})-\frac{\pp{i}{l} \per \pp{k}{m}}{q_{123}^2}\nonumber\\
&+\frac{1}{q_{123}^2 \per \qq{1}{2}} \psq 2 \per \pp{i}{l} \per \pq{m}{12} \per \pq{k}{2}+2 \per \pp{k}{l} \per \pq{i}{1} \per \pq{m}{2}+2 \per \pp{m}{l} \per \pq{i}{3} \per \pq{k}{2}-\pp{i}{m} \per (\qq{1}{3} \per \pp{k}{l}+2 \per \pq{k}{2} \per \pq{l}{3})\pdq\dg\nonumber\\
&+\frac{1}{8} \per \frac{\pp{i}{m}}{\pq{i}{3} \per \pq{m}{3} \per \pq{k}{1}} \sg \frac{1}{\pq{i}{12} \per \pq{l}{2}} \per \left[\frac{p_i(q_{2}-q_{1}) \per \pp{i}{k} \per \pp{i}{l}}{\pq{i}{1} \per \pq{i}{2}}+\frac{1}{\qq{1}{2}} \pst 4 \per \pp{i}{k} \per \pq{l}{1}+\pp{k}{l} \per p_i(q_{2}-q_{1})\pdt \right]\nonumber\\
&+\frac{2}{\pq{l}{12} \per \pq{m}{2}} \per \left[\frac{p_l(q_{1}-q_{2}) \per \pp{k}{l} \per \pp{m}{l}}{\pq{l}{1} \per \pq{l}{2}}+\frac{1}{\qq{1}{2}} \pst \pp{k}{m} \per p_l(q_{1}-q_{2})+2 \per (\pp{l}{m} \per \pq{k}{2}-\pp{k}{l} \per \pq{m}{1})\pdt\right]\dg\nonumber\\
&+\frac{1}{2 \per \pq{k}{1} \per \pq{m}{3} \per \pq{l}{12} \per \pq{i}{23}} \sg \frac{1}{\qq{1}{2} \per \qq{2}{3}} \psq 2 \per \pq{l}{1} \per (\pp{i}{k} \per \pq{m}{2}-\pp{m}{i} \per \pq{k}{3})+\pp{k}{m} \per \pq{l}{1} \per \pq{i}{3}\nonumber\\
&+\pp{k}{l} \per \pp{m}{i} \per \qq{1}{3}+\pp{i}{l} \per \pq{k}{2} \per \pq{m}{2}-2 \per \pp{k}{l} \per \pq{i}{1} \per \pq{m}{2}\pdq\nonumber\\
&+\frac{\pp{i}{m} \per p_i(q_{3}-q_{2})}{\qq{1}{2} \per \pq{i}{2} \per \pq{i}{3}} \per (\pp{i}{k} \per \pq{l}{1}+\pp{i}{l} \per \pq{k}{2}-\pp{k}{l} \per \pq{i}{1})+\frac{\pp{k}{l} \per \pp{l}{i} \per \pp{i}{m} \per p_l(q_{1}-q_{2}) \per p_i(q_{3}-q_{2})}{4 \per \pq{l}{1} \per \pq{l}{2} \per \pq{i}{2} \per \pq{i}{3}}\dg\;. \label{quadEsp}
\end{align}
\end{footnotesize}

The momentum dependence of the dipole correlation $W^{(3) {\rm dip.}}$
(see Eqs.~(\ref{3cordip}), (\ref{w3new}) and~(\ref{wikdef3}))
is controlled by the function $\cS_{ik}(q_1,q_2,q_3)$ in Eq.~(\ref{osik3g}).
The explicit expression of $\cS_{ik}(q_1,q_2,q_3)$ in terms of scalar products of momenta is very long. Analogously to the case of $\cS_{imkl}$, this expression can be slightly simplified by exploiting the symmetry properties of the colour dipole operators. We rewrite $\cS_{ik}(q_1,q_2,q_3)$ in the following form:
\begin{equation}
\label{sdbar}
  \cS_{ik}(q_1,q_2,q_3) = \left[ \; \oS_{ik}(q_1,q_2,q_3) + {\rm perms.}\{1,2,3\}
\, \right] + \dots \;\;,
\end{equation}
where the dots in the right-hand side denote terms that give a vanishing contribution to $W^{(3) {\rm dip.}}$.
We note that the symmetry properties of the colour dipole operators are directly embodied by the momentum function $\w_{ik}^{(3)}$ in Eq.~(\ref{wikdef3}). Therefore,
this function can also be rewritten as
\beeq
\label{cwbar}
\w_{ik}^{(3)}(q_1,q_2,q_3) &=& \left\{ \;
{\oS}_{ik}(q_1,q_2,q_3) + {\oS}_{ki}(q_1,q_2,q_3)
- {\oS}_{ii}(q_1,q_2,q_3) - {\oS}_{kk}(q_1,q_2,q_3) \, \right\} \nn \\
&+& {\rm perms.}\{1,2,3\} \; .
\eeeq

The explicit expression of ${\oS}_{ik}(q_1,q_2,q_3)$ is still very long,
and we write it as a sum of various terms
\begin{equation}
\label{dipEsp}
  \oS_{ik}(q_1,q_2,q_3) = \sum_{t\in\{a,b,c,d\}} \oS_{ik}^{(t)}(q_1,q_2,q_3)  \;.
\end{equation}
The term $\oS_{ik}^{(a)}$, which has no singularities if two soft gluons are collinear, is
\begin{footnotesize}
\begin{align}
&\oS_{ik}^{(a)}=\frac{31}{144} \per \frac{(\pp{i}{k})^3}{\pq{i}{1} \per \pq{k}{1} \per \pq{i}{2} \per \pq{k}{2} \per \pq{i}{3} \per \pq{k}{3}}+\frac{m_i^2 \per \pp{i}{k}}{48 \per (\pq{i}{1})^2 \per \pq{i}{2} \per \pq{k}{2}} \per \left(\frac{8 \per m_i^2}{(\pq{i}{3})^2}+\frac{3 \per m_k^2}{(\pq{k}{3})^2}-\frac{20 \per \pp{i}{k}}{\pq{i}{3} \per \pq{k}{3}}\right)\nonumber\\
&+\frac{m_i^2 \per \pp{i}{k}}{16 \per \pq{i}{12} \per \pq{i}{1}} \per \left(\frac{1}{\pq{i}{1} \per \pq{k}{2}}-\frac{1}{\pq{i}{2} \per \pq{k}{1}}\right) \per \left(\frac{7 \per \pp{i}{k}}{\pq{i}{3} \per \pq{k}{3}}-\frac{3 \per m_i^2}{(\pq{i}{3})^2}-\frac{2 \per m_k^2}{(\pq{k}{3})^2}\right)\nonumber\\
&+\frac{1}{48 \per \pq{k}{123}} \per \frac{m_k^2 \per \pp{i}{k}}{\pq{k}{1} \per \pq{i}{2}} \per \left(\frac{3}{\pq{k}{12}}-\frac{1}{\pq{k}{3}}\right) \per \left(\frac{1}{\pq{k}{2}}-\frac{1}{\pq{k}{1}}\right) \per \left(\frac{\pp{i}{k}}{\pq{i}{3}}+\frac{m_k^2}{\pq{k}{3}}\right)\nonumber\\
&+\frac{(\pp{i}{k})^3}{32 \per \pq{k}{12} \per \pq{k}{3} \per \pq{i}{1} \per \pq{i}{3}} \per \left(\frac{1}{\pq{k}{1}}-\frac{1}{\pq{k}{2}}\right) \per \left(\frac{6}{\pq{i}{12}}+\frac{p_i(q_{3}-q_{1})}{\pq{i}{13} \per \pq{i}{2}}\right)\nonumber\\
&+\frac{m_i^2 \per \pp{i}{k}}{16 \per \pq{i}{12} \per \pq{i}{1}} \per \sg \frac{2 \per \pp{i}{k}}{\pq{k}{12} \per (\pq{i}{3})^2} \per \left(\frac{1}{\pq{k}{2}}-\frac{1}{\pq{k}{1}}\right)+\left(\frac{1}{\pq{i}{2}}-\frac{1}{\pq{i}{1}}\right) \per\nonumber\\
&\times  \per \sq\frac{m_i^2}{\pq{k}{3}} \per \left(\frac{2}{\pq{i}{12} \per \pq{i}{3}}+\frac{1}{\pq{i}{23}} \per \pst\frac{1}{\pq{i}{3}}-\frac{1}{\pq{i}{2}}\pdt\right)+\frac{1}{\pq{k}{23}} \per \left(\frac{m_k^2}{\pq{k}{3}}-\frac{\pp{i}{k}}{\pq{i}{3}}\right) \per \left(\frac{1}{\pq{k}{2}}-\frac{1}{\pq{k}{3}}\right)\dq\dg\nonumber\\
&+\frac{m_k^2 \per \pp{i}{k}}{48 \per \pq{k}{123}} \per \left(\frac{1}{\pq{k}{3}}-\frac{3}{\pq{k}{12}}\right) \per \left(\frac{1}{\pq{k}{1}}-\frac{1}{\pq{k}{2}}\right) \per \sg\frac{2 \per \pp{i}{k}}{\pq{i}{12} \per \pq{k}{3} \per \pq{i}{1}}+\frac{\pp{i}{k}}{\pq{i}{13} \per \pq{k}{2}} \per \left(\frac{1}{\pq{i}{1}}-\frac{1}{\pq{i}{3}}\right)\nonumber\\
&+\frac{m_k^2}{\pq{k}{12} \per \pq{i}{3}} \per \left(\frac{1}{\pq{k}{2}}-\frac{1}{\pq{k}{1}}\right)+\frac{m_k^2}{\pq{k}{13} \per \pq{i}{2}} \per \left(\frac{1}{\pq{k}{3}}-\frac{1}{\pq{k}{1}}\right)\dg+\frac{(\pp{i}{k})^3}{288 \per \pq{k}{123} \per \pq{i}{123}} \per \left(\frac{1}{\pq{k}{1}}-\frac{1}{\pq{k}{2}}\right) \per \nonumber\\
&\times  \per \sg\frac{2}{\pq{i}{1}} \per \left(\frac{1}{\pq{i}{3}}-\frac{3}{\pq{i}{12}}\right) \per \left(\frac{1}{\pq{k}{3}}-\frac{3}{\pq{k}{12}}\right)+\left(\frac{1}{\pq{i}{2}}-\frac{3}{\pq{i}{13}}\right) \per \left(\frac{1}{\pq{i}{1}}-\frac{1}{\pq{i}{3}}\right) \per \left(\frac{1}{\pq{k}{3}}-\frac{3}{\pq{k}{12}}\right)\dg \;.\label{sika}
\end{align}
\end{footnotesize}
The term $\oS_{ik}^{(b)}$ includes the soft-gluon propagator $(q_1 q_2)^{-1}$,
and it is
\begin{footnotesize}
\begin{align}
&\oS_{ik}^{(b)}=\frac{1}{16 \per \qq{1}{2} \per \pq{i}{12}} \per \sg\frac{7 \per \pp{i}{k}}{\pq{i}{1} \per \pq{i}{3} \per \pq{k}{2} \per \pq{k}{3}} \per (2 \per m_i^2 \per \pq{k}{1}-\pp{i}{k} \per \pq{i}{12})\nonumber\\
&+\left(\frac{3 \per m_i^2}{(\pq{i}{3})^2}+\frac{2 \per m_k^2}{(\pq{k}{3})^2}\right) \per \left[\pp{i}{k} \per \left(\frac{1}{\pq{k}{2}}+\frac{\pq{i}{1}}{\pq{i}{2} \per \pq{k}{1}}\right)-2 \per \frac{m_i^2 \per \pq{k}{1}}{\pq{i}{1} \per \pq{k}{2}}\right]\nonumber\\
&+\frac{(\pp{i}{k})^2}{\pq{i}{3}} \per \left[\frac{12 \per p_i(q_{1}-q_{2})}{\pq{k}{12} \per \pq{i}{1} \per \pq{k}{3}}+\frac{1}{\pq{k}{13}} \per \left(\frac{1}{\pq{k}{3}}-\frac{1}{\pq{k}{1}}\right) \per \left(3+\frac{\pq{i}{1}}{\pq{i}{2}}-\frac{2 \per \pq{k}{1}}{\pq{k}{2}}\right)\right]\nonumber\\
&+m_i^2 \per \sq\frac{4 \per \pp{i}{k} \per p_i(q_{2}-q_{1})}{\pq{i}{12} \per \pq{i}{1} \per \pq{i}{3} \per \pq{k}{3}}+\frac{1}{\pq{i}{13}} \per \left(\frac{1}{\pq{i}{1}}-\frac{1}{\pq{i}{3}}\right) \per \left(\frac{\pp{i}{k} \per \pq{i}{12}-2 \per m_i^2 \per \pq{k}{1}}{\pq{k}{2} \per \pq{i}{3}}+\frac{\pp{i}{k} \per p_i(q_{2}-q_{1})}{\pq{k}{3} \per \pq{i}{2}}\right)\nonumber\\
&+\frac{1}{\pq{k}{12} \per \pq{i}{3}} \per \left(\frac{2 \per p_k(q_{2}-q_{1})}{\pq{i}{1}} \per \pst\frac{3 \per \pp{i}{k}}{\pq{k}{3}}-\frac{m_i^2}{\pq{i}{3}}\pdt+\frac{4 \per \pp{i}{k}}{\pq{i}{3}} \per \pst\frac{\pq{i}{2}}{\pq{i}{1}}+\frac{\pq{k}{2}}{\pq{k}{1}}-2\pdt+\frac{2 \per m_k^2 \per p_i(q_{1}-q_{2})}{\pq{k}{1} \per \pq{i}{3}}\right)\dq\nonumber\\
&+\frac{1}{\pq{k}{13}} \per \left(\frac{1}{\pq{k}{1}}-\frac{1}{\pq{k}{3}}\right) \per \sq m_k^2 \per \left(\frac{p_i(q_{1}-q_{2})}{\pq{k}{2}} \per \pst\frac{m_k^2}{\pq{k}{3}}-\frac{\pp{i}{k}}{\pq{i}{3}}\pdt+\frac{2 \per \pp{i}{k}}{\pq{k}{3}} \per \pst 2+\frac{\pq{i}{1}}{\pq{i}{2}}-\frac{\pq{k}{1}}{\pq{k}{2}}\pdt -\frac{4 \per m_i^2 \per \pq{k}{2}}{\pq{i}{2} \per \pq{k}{3}}\right)\nonumber\\
&+\frac{2 \per m_i^2 \per \pp{i}{k} \per \pq{k}{2}}{\pq{i}{2} \per \pq{i}{3}}\dq\dg+\frac{1}{48 \per \qq{1}{2} \per \pq{k}{123}} \per \sg \frac{3}{\pq{k}{1} \per \pq{i}{2}}\left(\frac{1}{\pq{k}{12}}-\frac{1}{\pq{k}{3}}\right) \per \left(\frac{\pp{i}{k}}{\pq{i}{3}}+\frac{m_k^2}{\pq{k}{3}}\right) \per (\pp{i}{k} \per \pq{k}{12}-2 \per m_k^2 \per \pq{i}{1})\nonumber\\
&+\frac{m_k^2}{\pq{i}{12} \per \pq{k}{3}} \per \sq\st\pst\frac{1}{\pq{k}{3}}-\frac{3}{\pq{k}{12}}\pdt \per \frac{2}{\pq{k}{1}}+\pst\frac{1}{\pq{k}{2}}-\frac{3}{\pq{k}{13}}\pdt \per \pst\frac{1}{\pq{k}{1}}-\frac{1}{\pq{k}{3}}\pdt\dt\nonumber\\
&\times  \per \pst 2 \per \pp{i}{k} \per p_k(q_{1}-q_{2})+m_k^2 \per p_i(q_{2}-q_{1})\pdt+\frac{6}{\pq{i}{1}} \per \left(\frac{1}{\pq{k}{3}}-\frac{1}{\pq{k}{12}}\right) \per \pst 2 \per \pp{i}{k} \per p_i(q_{1}-q_{2})+m_i^2 \per p_k(q_{2}-q_{1})\pdt\dq\nonumber\\
&+\frac{4 \per m_k^2 \per \pp{i}{k}}{\pq{k}{12} \per \pq{i}{3}} \per \left(1-\frac{\pq{k}{2}}{\pq{k}{1}}\right) \per \left(\frac{3}{\pq{k}{12}}-\frac{2}{\pq{k}{3}}\right)+\frac{m_k^2 \per \pp{i}{k}}{\pq{k}{12} \per \pq{i}{3}} \per \left(\frac{3}{\pq{k}{13}}-\frac{1}{\pq{k}{2}}\right) \per \left(\frac{1}{\pq{k}{1}}-\frac{1}{\pq{k}{3}}\right) \per p_k(q_{1}-q_{2})\nonumber\\
&+3 \per \left(\frac{1}{\pq{k}{3}}-\frac{1}{\pq{k}{12}}\right) \per (\pp{i}{k} \per \pq{k}{12}-2 \per m_k^2 \per \pq{i}{1}) \per \left[\frac{\pp{i}{k}}{\pq{i}{23} \per \pq{k}{1}} \per \left(\frac{1}{\pq{i}{2}}-\frac{1}{\pq{i}{3}}\right)+\frac{m_k^2}{\pq{k}{13} \per \pq{i}{2}} \per \left(\frac{1}{\pq{k}{1}}-\frac{1}{\pq{k}{3}}\right)\right]\dg\nonumber\\
&+\frac{\pp{i}{k}}{48 \per \qq{1}{2} \per \pq{k}{123} \per \pq{i}{123}} \per \left(\frac{1}{\pq{i}{12}}-\frac{1}{\pq{i}{3}}\right) \per \pst 2 \per \pp{i}{k} \per p_k(q_{2}-q_{1})+m_k^2 \per p_i(q_{1}-q_{2})\pdt \nonumber\\
&\times  \per \left[\left(\frac{1}{\pq{k}{3}}-\frac{3}{\pq{k}{12}}\right) \per \left(\frac{1}{\pq{k}{1}}-\frac{1}{\pq{k}{2}}\right)+\left(\frac{1}{\pq{k}{2}}-\frac{3}{\pq{k}{13}}\right) \per \left(\frac{1}{\pq{k}{1}}-\frac{1}{\pq{k}{3}}\right)\right]\;.\label{sikb}
\end{align}
\end{footnotesize}
\newpage
\noindent
The term $\oS_{ik}^{(c)}$, which contains the soft-gluon propagators $(q_1 q_2)^{-2}$
and $(q_1 q_2)^{-1} (q_1 q_3)^{-1}$,~is
\begin{footnotesize}
\begin{align}
&\oS_{ik}^{(c)}=\frac{1}{8 \per (\qq{1}{2})^2 \per \pq{i}{12} \per \pq{k}{3}} \per \sg \pst (4-d) \per \pq{i}{1}+d \per \pq{i}{2}\pdt \per \sq\frac{\pp{i}{k} \per \pq{k}{1}}{2 \per \pq{k}{123} \per \pq{i}{123}} \per \left(\frac{\pq{k}{3}}{\pq{k}{12}}-1\right) \per \left(\frac{\pq{i}{12}}{\pq{i}{3}}-1\right)\nonumber\\
&+\frac{\pq{k}{1}}{\pq{k}{12}} \per \left(\frac{m_k^2}{\pq{k}{3}}+\frac{m_i^2 \per \pq{k}{3}}{\pq{i}{3}} \per \pst\frac{2}{\pq{i}{123}}-\frac{1}{\pq{i}{3}}\pdt\right)+\frac{\pp{i}{k}}{\pq{i}{3}} \per \left(\frac{\pq{i}{1}}{\pq{i}{12}}-\frac{3}{2} \per \frac{\pq{k}{1}}{\pq{k}{12}}\right)\dq\nonumber\\
&+\frac{\pp{i}{k}}{\pq{i}{123}} \per \left(\frac{1}{\pq{i}{3}}-\frac{1}{\pq{i}{12}}\right) \per \pst (4-d) \per (\pq{i}{1})^2+d \per \pq{i}{1} \per \pq{i}{2}\pdt\dg\nonumber\\
&+\frac{1}{32 \per \qq{1}{2} \per \qq{1}{3} \per \pq{i}{12}} \per \sg\frac{\pp{i}{k}}{\pq{k}{2}} \per \sq\frac{4 \per p_k(q_{2}-q_{1})}{\pq{k}{3}}+\frac{2 \per \pq{i}{12}}{\pq{i}{3}}+2 \per \frac{\pq{i}{2} \per \pq{i}{3}+\pq{i}{1} \per \pq{i}{123}}{\pq{i}{13} \per \pq{i}{3}}\nonumber\\
&+\frac{1}{\pq{k}{13} \per \pq{i}{3}} \per \pst\pq{i}{1} \per p_k (5\per q_1 -8\per q_2 +2\per q_3)-3 \per \pq{i}{2} \per \pq{k}{3}+4 \per \pp{i}{k} \per \qq{2}{3}\pdt\dq\nonumber\\
&+m_i^2 \per \sq 4 \per \frac{\pq{i}{2} \per \pq{k}{3}+\pq{i}{3} \per \pq{k}{1}-\pq{i}{1} \per \pq{k}{13}-2 \per \pp{i}{k} \per \qq{2}{3}}{\pq{i}{13} \per \pq{k}{2} \per \pq{i}{3}}+\frac{1}{\pq{k}{13}}\left(4 \per \frac{\pq{k}{1} \per p_k(q_{3}-q_{1})}{\pq{k}{2} \per \pq{i}{3}}\right.\nonumber\\
&\left.+8 \per \frac{m_k^2 \per \qq{2}{3}-\pq{k}{13} \per \pq{k}{2}}{\pq{i}{2} \per \pq{k}{3}}+2 \per \frac{\pq{i}{1} \per p_k(q_1+4 q_2-q_3)+\pq{i}{2} \per p_k(q_{3}-q_{1})-4 \per \pp{i}{k} \per \qq{2}{3}}{\pq{i}{2} \per \pq{i}{3}}\right)\dq\nonumber\\
&+\frac{2}{\pq{i}{123} \per \pq{k}{3}} \left(\frac{1}{\pq{i}{2}}-\frac{1}{\pq{i}{13}}\right) \pst \pp{i}{k} \per \pq{i}{12} \per \pq{i}{13}+2 \per m_i^2 \per (\pq{k}{2} \per \pq{i}{3}+\pq{k}{1} \per \pq{i}{2}-\pq{i}{1} \per \pq{k}{12}-2 \per \pp{i}{k} \per \qq{2}{3})\pdt\nonumber\\
&+\frac{2}{\pq{k}{123} \per \pq{k}{3}} \per \left(\frac{1}{\pq{k}{13}}-\frac{1}{\pq{k}{2}}\right) \psq m_k^2 \pst 4 \per \pq{k}{1} \per \pq{i}{3}+p_k(q_{1}-q_{3}) \per p_i(q_{1}-q_{2})-4 \per \pp{i}{k} \per \qq{2}{3}\pdt\nonumber\\
&+2 \per \pp{i}{k} \per \pq{k}{13} \per p_k(q_{2}-q_{1})\pdq+\frac{1}{\pq{k}{123} \per \pq{i}{123}} \per \left(1-\frac{\pq{i}{12}}{\pq{i}{3}}\right) \per \left(\frac{1}{\pq{k}{13}}-\frac{1}{\pq{k}{2}}\right) \nonumber\\
&\times  \per \psq 4 \per (\pp{i}{k})^2 \per \qq{2}{3}+\pp{i}{k} \pst\pq{i}{1} \per p_k (5\per q_1-8\per q_2+2\per q_3)-3 \per \pq{i}{2} \per \pq{k}{3}\pdt +4 \per m_i^2 \per \pq{k}{1} \per p_k(q_{3}-q_{1})\pdq\dg \;. \label{sikc}
\end{align}
\end{footnotesize}
The remaining term $\oS_{ik}^{(d)}$ of Eq.~(\ref{dipEsp}) includes all the other types of soft-gluon propagators, and it is
\begin{footnotesize}
\begin{align}
&\oS_{ik}^{(d)}=\frac{m_i^2 \per m_k^2-(\pp{i}{k})^2}{4 \per q_{123}^2 \per \pq{k}{123} \per \pq{k}{1} \per \pq{i}{2} \per \pq{i}{3}}+\frac{1}{4 \per q_{123}^2 \per \pq{k}{123}} \per \sg\frac{1}{\qq{1}{2}} \per \sq \frac{1}{\pq{k}{1} \per \pq{i}{2}} \per \st \pp{i}{k} \per (\qq{2}{3}-\qq{1}{3}) \per \pst\frac{m_k^2}{\pq{k}{3}}+\frac{\pp{i}{k}}{\pq{i}{3}}\pdt\nonumber\\
&+2 \per \frac{\pp{i}{k}}{\pq{i}{3}} \per (\pq{i}{12} \per \pq{k}{123}-\pq{i}{3} \per \pq{k}{2}-\pq{i}{2} \per \pq{k}{3})+\frac{m_i^2}{\pq{i}{3}} \per \pq{k}{2} \per p_k(q_3-q_{12})\nonumber\\
&+\frac{m_k^2}{\pq{k}{3}} \pst \pq{i}{1} \per p_k(q_{3}-q_{12})-2 \per \pq{i}{3} \per \pq{k}{2}\pdt -2 \per m_k^2 \per \frac{\pq{i}{1} \per \pq{i}{12}}{\pq{i}{3}}\dt+\frac{1}{2 \per \pq{i}{13} \per \pq{k}{2}} \psg\pst\frac{1}{\pq{i}{1}}-\frac{1}{\pq{i}{3}}\pdt\nonumber\\
&\times \psq \pp{i}{k} \pst \pq{k}{1} \per p_i(4\per q_{3}-3\per q_{12})-3 \per \pq{k}{2} \per \pq{i}{12}+\pq{k}{3} \per p_i(q_{1}-3 q_{2})+2 \per \pp{i}{k} \per (\qq{2}{3}-\qq{1}{3})\pdt\nonumber\\
&+4 \per m_k^2 \per \pq{i}{12} \per \pq{i}{2}-2 \per m_i^2 \per \pq{k}{1} \per p_k(q_{3}-q_{12})\pdq +4 \per \pp{i}{k} \per p_k(q_{3}-q_{1})+2 \per m_k^2 \per p_i(q_{1}-q_{3})\pdg\nonumber\\
&+\frac{1}{\pq{i}{12} \per \pq{k}{3}} \per \psg\pst \frac{1}{\pq{i}{1}}-\frac{1}{\pq{i}{2}}\pdt \per \pst 4 \per m_k^2 \per \pq{i}{1} \per \pq{i}{3}+m_i^2 \per (\pq{k}{1} \per \pq{k}{13}-2 \per m_k^2 \per \qq{1}{3})-2 \per \pp{i}{k} \per \pq{i}{1} \per \pq{k}{123}\pdt\nonumber\\
&+\pp{i}{k} \per p_k(q_{1}-3\per q_{2}-q_{3})+m_k^2 \pst (d-5) \per p_i(q_{1}-q_{2})+4 \per \pq{i}{3}\pdt\pdg +\frac{1}{\pq{k}{13} \per \pq{i}{2}} \psg \pp{i}{k} \per \pq{k}{13}-2 \per m_k^2 \per \pq{i}{3}\nonumber\\
&+\pst\frac{1}{\pq{k}{3}}-\frac{1}{\pq{k}{1}}\pdt \per \psq m_k^2 \per \pst \pq{i}{1} \per p_k(q_{3}-q_{12})-2 \per \pq{k}{2} \per \pq{i}{3}+\pp{i}{k} \per (\qq{2}{3}-\qq{1}{3})\pdt\nonumber\\
&+\frac{\pp{i}{k}}{2} \per \pq{k}{123} \per \pq{k}{12}\pdq\pdg +\frac{1}{\pq{k}{12} \per \pq{i}{3}} \psg\pp{i}{k} \per p_k(3 \per q_{1}+q_{2}-3 \per q_{3})+2 \per m_k^2 \per p_i(q_{1}-q_{2})\nonumber\\
&+\pst\frac{1}{\pq{k}{2}}-\frac{1}{\pq{k}{1}}\pdt \per \psq 2 \per \pp{i}{k} \per \pq{k}{1} \per \pq{k}{13}+m_k^2 \pst \pq{i}{1} \per p_k(q_{23}-3 \per q_{1})-2 \per \pp{i}{k} \per \qq{1}{3}\pdt\pdq -2 \per m_k^2 \per \pq{i}{1}\pdg +\dots \nonumber
\end{align}
\begin{align}
&\dots+\frac{2}{\pq{i}{123}} \psg\pst\frac{\pq{k}{13}}{\pq{k}{2}}-1\pdt \pst\frac{\pp{i}{k}}{2}-\frac{m_k^2 \per \pq{i}{1}}{\pq{k}{13}}\pdt + \pst\frac{1}{\pq{k}{12}}-\frac{1}{\pq{k}{3}}\pdt \pst m_k^2 \per \pq{i}{2}-\pp{i}{k} \per p_k (q_{2}+2 \per q_{3})\pdt\nonumber\\
&+\frac{1}{3} \per \pst\frac{1}{\pq{k}{3}}-\frac{3}{\pq{k}{12}}\pdt \pst\frac{1}{\pq{k}{1}}-\frac{1}{\pq{k}{2}}\pdt \pst 2 \per \pp{i}{k} \per \pq{k}{2} \per p_k(q_{12}-q_{3})+m_k^2 \per (\pp{i}{k} \per \qq{2}{3}+2 \per \pq{k}{1} \per \pq{i}{1})\pdt\nonumber\\
&+\frac{1}{12} \per \pst\frac{1}{\pq{k}{2}}-\frac{3}{\pq{k}{13}}\pdt \pst\frac{1}{\pq{k}{1}}-\frac{1}{\pq{k}{3}}\pdt \psq 4 \per \pp{i}{k} \per p_k(q_{1}-q_{2}) \per p_k(q_{3}-q_{12})\nonumber\\
&+m_k^2 \per \pst 2 \per \pp{i}{k} \per (\qq{2}{3}-\qq{1}{3})+4 \per \pq{k}{1} \per \pq{i}{1}-4 \per \pq{k}{2} \per \pq{i}{2}\pdt\pdq\pdg\dq +\frac{1}{(\qq{1}{2})^2} \sq \frac{2}{\pq{i}{123}} \per \left(\frac{1}{\pq{k}{3}}-\frac{1}{\pq{k}{12}}\right)\nonumber\\
&\times  \per \psq\pp{i}{k} \per \qq{1}{3} \pst(d-4) \per \pq{k}{1}-d \per \pq{k}{2}\pdt +2 \per (d-2) \per \pq{i}{2} \per (\pq{k}{1})^2+\pq{k}{1} \per \pq{k}{2} \per \pst (4-d) \per \pq{i}{1}+\frac{d}{2} \per \pq{i}{3}\pdt\pdq\nonumber\\
&+\frac{1}{\pq{i}{12} \per \pq{k}{3}} \per \psg (d-2) \per p_i(q_{2}-q_{1}) \per \pq{k}{1} \per \pq{k}{13}+2 \per m_k^2 \per \pq{i}{1} \pst (d-4) \per \qq{1}{3}-d \per \qq{2}{3}\pdt\pdg\nonumber\\
&+\frac{1}{\pq{k}{12} \per \pq{i}{3}} \psg 2 \per \pp{i}{k} \per \qq{1}{3} \per \pst (4-d) \per \pq{k}{1}+d \per \pq{k}{2}\pdt +(d-2) \per \pq{i}{1} \per p_k(q_{1}-q_{2}) \per p_k(q_{13}-3 \per q_{2})\pdg\dq\nonumber\\
&+\frac{1}{2 \per \qq{1}{2} \per \qq{1}{3}} \sq \frac{1}{\pq{i}{12} \per \pq{k}{3}} \psg 4 \per \pp{i}{k} \per \qq{2}{3} \per p_k(2 \per q_{1}+q_{3})+2 \per m_k^2 \per \qq{2}{3} \per p_i(q_{2}-5 \per q_{1}-4 \per q_{3})\nonumber\\
&+(\pq{k}{1})^2 \per \pst(7-2 \per d) \per \pq{i}{1}+(2 \per d+1) \per \pq{i}{2}-4 \per \pq{i}{3}\pdt + 2 \per \pq{k}{1} \per \pq{k}{3} \per \pst(5-d) \per \pq{i}{1}+(d-5) \per \pq{i}{2}-2 \per \pq{i}{3}\pdt\nonumber\\
&+\pq{k}{1} \per \pq{k}{2} \per \pst(2 \per d-3) \per \pq{i}{1}+(9-2 \per d) \per \pq{i}{2}-2 \per \pq{i}{3}\pdt+3 \per (\pq{k}{3})^2 \per p_i(q_{1}-q_{2})\nonumber\\
&+\pq{k}{2} \per \pq{k}{3} \per p_i(9 \per q_{1}-3 \per q_{2}+2 \per q_{3})+ 2 \per(\pq{k}{2})^2 \per p_i(q_{3}-q_{1})\pdg\nonumber\\
&+\frac{1}{\pq{k}{12} \per \pq{i}{3}} \per \psg 2 \per \pp{i}{k} \per \qq{2}{3} \per p_k(q_{12}+2 \per q_{3})-4 \per m_k^2 \per \qq{2}{3} \per \pq{i}{2}+(\pq{k}{1})^2 \pst (2 \per d-7) \per \pq{i}{1}+2 \per \pq{i}{2}+\pq{i}{3}\pdt\nonumber\\
&+\pq{k}{1} \per \pq{k}{3} \pst (2 \per d-7) \per \pq{i}{1}+4 \per \pq{i}{2}-3 \per \pq{i}{3}\pdt +2 \per \pq{k}{1} \per \pq{k}{2} \per (\pq{i}{2}-2 \per d \per \pq{i}{1})+2 \per (\pq{k}{3})^2 \per p_i(2 \per q_{1}-q_{2})\nonumber\\
&+\pq{k}{2} \per \pq{k}{3} \pst 2 \per \pq{i}{2}+3 \per \pq{i}{3}-(2 \per d+1) \per \pq{i}{1} \pdt +(\pq{k}{2})^2 \pst (2 \per d-9) \per \pq{i}{1}-\pq{i}{3}\pdt\pdg\nonumber\\
&+\frac{4}{\pq{i}{123}} \left(\frac{1}{\pq{k}{12}}-\frac{1}{\pq{k}{3}}\right) \psg \qq{2}{3} \psq \frac{\pp{i}{k}}{2} \per p_k ( 5\per q_{1}-3 \per q_{2}+4 \per q_{3} ) +m_k^2 \per p_i (2 \per q_{2}-q_{1})\pdq +(\pq{k}{2})^2 \per \pq{i}{3}\nonumber\\
&+(\pq{k}{1})^2 \pst \pq{i}{1}+(3-d) \per \pq{i}{2}\pdt +(d-2) \per \pq{k}{1} \per \pq{k}{2} \per \pq{i}{2}+\pq{k}{1} \per \pq{k}{3} \per p_i (q_{1}-3 \per q_{2})+\pq{k}{2} \per \pq{k}{3} \per p_i(3 \per q_{1}-q_{2})\pdg\dq\dg\nonumber\\
&+\frac{1}{2 \per (q_{123}^2)^2 \per \pq{i}{123} \per \pq{k}{123}} \sg (3 \per d-10) \per \pp{i}{k} +\frac{2 \per \pp{i}{k} \per \qq{1}{3}}{(\qq{1}{2})^2} \pst (d-4) \per \qq{1}{3}-d \per \qq{2}{3}\pdt\nonumber\\
&+\frac{1}{\qq{1}{2}} \per \left[\pq{k}{1} \pst (8-3 \per d) \per \pq{i}{1}+(16-7 \per d) \per \pq{i}{2} \pdt -\frac{d}{2} \per \pq{k}{3} \per \pq{i}{3}-2 \per \pp{i}{k} \per (2 \per \qq{1}{3}+3 \per \qq{2}{3})\right]\nonumber\\
&+\frac{\qq{2}{3}}{\qq{1}{2} \per \qq{1}{3}} \sq\pp{i}{k} \per \qq{2}{3}+4 \per (d-4) \per \pq{k}{1} \per \pq{i}{1}-16 \per \pq{k}{1} \per \pq{i}{2}+4 \per (2-d) \per \pq{k}{2} \per \pq{i}{2}\dq\dg\;. \label{sikd}
\end{align}
\end{footnotesize}

The quadrupole function $\cS_{imkl}$ does not depend on $\epsilon$. The dipole function $\oS_{ik}(q_1,q_2,q_3)$ includes contributions that have a linear dependence on $\epsilon$ (see the terms proportional to $d=4 -2\epsilon$ in Eqs.~(\ref{sikc}) and (\ref{sikd})).
Such dependence originates from the use of the CDR scheme for the dimensional regularization procedure. The result for $|\J(q_1,q_2,q_3)|^2$ in the DR and 
4DH schemes is obtained by setting $\epsilon=0$ in the expression of 
$\oS_{ik}(q_1,q_2,q_3)$.


\end{document}